\documentclass[aps,prx,notitlepage,superscriptaddress,longbibliography,nofootinbib,twocolumn,floatfix,10pt]{revtex4-2}
\usepackage[utf8]{inputenc}
\usepackage[T1]{fontenc}
\usepackage{braket}
\usepackage{amsmath,amsthm,amssymb,amsfonts}
\usepackage{mathtools}
\usepackage{graphicx}
\usepackage{comment}
\usepackage{color}
\usepackage{makecell}
\usepackage{enumitem}
\usepackage{mathdots}
\usepackage{tikz}
\usepackage{tikz-3dplot}
\usepackage{epsfig}
\usepackage{blkarray}
\usepackage{colortbl}
\usetikzlibrary{patterns,patterns.meta}
\usepackage{xcolor}
\usepackage{bbm}
\usepackage{bm}
\usetikzlibrary{arrows.meta}
\usepackage[bb=boondox]{mathalfa}
\usepackage{adjustbox}

\usepackage{pgfplots}
\usetikzlibrary{shapes,arrows,patterns}
\usetikzlibrary{3d}
\usepackage{tikz-cd}
\usepgfplotslibrary{groupplots} %
\usetikzlibrary{calc} %

\usepackage{hyperref}
\usepackage{cleveref} %

\newcounter{panel}[figure]

\newcommand{\panellabel}[1]{\refstepcounter{panel}\label{#1}}
\crefname{panel}{figure}{figures}
\Crefname{panel}{Figure}{Figures}

\usepackage{accents} %
\usepackage{multirow}
\usepackage{booktabs}

\newcounter{levelfour}[subsubsection]
\renewcommand{\thelevelfour}{\roman{levelfour}}
\renewcommand{\paragraph}[1]{\par\refstepcounter{levelfour}\textit{\thelevelfour.~#1.}}

\newtheorem{definition}{Definition}
\newtheorem{proposition}{Proposition}
\newtheorem{lemma}{Lemma}

\usepackage{graphicx}
\graphicspath{ {./images/} }

\newcommand{\ii}{\operatorname{i}}
\newcommand{\ie}{{\it i.e.},\ }
\newcommand{\eg}{{\it e.g.},\ }

\newcommand{\dd}{\mathrm{d}}
\newcommand{\loc}{\mathrm{loc}}

\definecolor{WM1}{RGB}{94,129,181}
\definecolor{WM2}{RGB}{225,156,36}
\definecolor{WM3}{RGB}{143,176,50}
\definecolor{WM4}{RGB}{235,98,53}

\newcommand{\nAbar}{\bar{n}_A}
\newcommand{\nBbar}{\bar{n}_B}
\newcommand{\nXbar}{\bar{n}_X}

\newcommand{\nAbc}{\bar{n}_{A,\mathrm{c}}}
\newcommand{\nBbc}{\bar{n}_{B,\mathrm{c}}}

\newcommand{\ncrit}{n^{\rm crit}}

\newcommand{\nAsc}{n_{A}^{\rm sc}}
\newcommand{\nBsc}{n_{B}^{\rm sc}}
\newcommand{\nXsc}{n_{X}^{\rm sc}}

\newcommand{\Nat}{\mathbb{N}}
\newcommand{\Hil}{\mathcal{H}}

\newcommand{\R}{\mathbb{R}}
\newcommand{\ncritset}{\mathcal{N}}
\newcommand{\nmax}{n^{\mathrm{max}}}
\newcommand{\nAmax}{n_A^{\mathrm{max}}}
\newcommand{\nXmax}{n_X^{\mathrm{max}}}
\newcommand{\nBmax}{n_B^{\mathrm{max}}}
\newcommand{\betamin}{\bar{\beta}_{\mathrm{min}}}

\newcommand{\betaA}{\bar{\beta}_{A,\mathrm{c}}}
\newcommand{\betaB}{\bar{\beta}_{B,\mathrm{c}}}
\newcommand{\bbA}{\bar{\beta}_{A}}
\newcommand{\bbB}{\bar{\beta}_{B}}
\newcommand{\rAc}{r_{A,\mathrm{c}}}
\newcommand{\rBc}{r_{B,\mathrm{c}}}
\newcommand{\rsc}{r_{*,\mathrm{c}}}
\newcommand{\kappac}{\kappa_{\mathrm{c}}}
\newcommand{\dzg}{\delta_{0,\mathrm{g}}}%
\newcommand{\dzf}{\delta_{0,\mathrm{f}}}%

\pgfplotsset{
  pedagogyaxis/.style={
    width=0.55\columnwidth, height=0.42\columnwidth,
    axis line style={black!70, line width=0.5pt},
    tick align=inside, tick pos=both, minor tick num=1,
    tick style={black!70, line width=0.4pt},
    label style={font=\scriptsize},
    tick label style={font=\scriptsize},
    legend style={font=\scriptsize, draw=none, fill=none,
                  row sep=-1.5pt, inner sep=1.5pt},
    legend cell align=left,
    every axis plot/.append style={line width=0.7pt},
  }
}
\colorlet{VB}{WM1}\colorlet{VC}{WM2}\colorlet{VD}{WM3}\colorlet{VE}{WM4}

\newcommand{\ee}{\operatorname{e}} %
\newcommand{\erf}{\operatorname{erf}}
\newcommand{\erfc}{\operatorname{erfc}}
\newcommand{\sgn}{\operatorname{sgn}}
\newcommand{\ord}{\operatorname{ord}}
\newcommand{\lcm}{\operatorname{lcm}}

\pgfplotsset{compat=1.18}

\begin{document}
	\title{Typical volume-law entanglement between particle species}

    \author{Barbod Vahedian Ghaffari}
    \email{bvah0003@student.monash.edu}
    \affiliation{School of Mathematics and Statistics, The University of Melbourne, Parkville, VIC 3010, Australia}
    \affiliation{School of Mathematics, Monash University, Clayton, VIC 3168, Australia}
    \affiliation{School of Physics and Astronomy, Monash University, Clayton, VIC 3168, Australia}

    \author{Zhexing Kang}
	\email{zhexing3@illinois.edu}
	\affiliation{Department of Physics, University of Illinois at Urbana-Champaign, Illinois 61801, USA}
    \affiliation{Department of Mathematics, University of Illinois at Urbana-Champaign, Illinois 61801, USA}
	\affiliation{School of Physics, The University of Melbourne, Parkville, VIC 3010, Australia}
    
	\author{Kirtik Singh}
	\email{kirtiks@student.unimelb.edu.au}
	\affiliation{School of Mathematics and Statistics, The University of Melbourne, Parkville, VIC 3010, Australia}

    \author{Yale Yauk}
    \email{yale.yauk@mpq.mpg.de}
    \affiliation{Max-Planck-Institut für Quantenoptik, 85748 Garching, Germany}
    \affiliation{Munich Center for Quantum Science and Technology (MCQST), 80799 München, Germany}
    
	\author{Lucas Hackl}
	\email{lucas.hackl@unimelb.edu.au}
	\affiliation{School of Mathematics and Statistics, The University of Melbourne, Parkville, VIC 3010, Australia}
	\affiliation{School of Physics, The University of Melbourne, Parkville, VIC 3010, Australia}
    
\begin{abstract}
    We calculate the typical bipartite entanglement entropy $\braket{S_A}_N$ for subsystems of different particle species of any reasonable kind, as a function of the fixed total particle number $N$, the volume $V$, and the characteristics of the particle species $A$ and $B$. We express our result as an asymptotic expansion $\braket{S_A}_N=aV+b\sqrt{V}+c+o(1)$, and determine how $a$, $b$, and $c$ can be obtained from two generating functions characterizing the local Hilbert space dimensions of each particle species, with $c$ containing a particle-species cross term, not previously seen for geometric bipartitions. We demonstrate the generality of our findings by studying four particularly interesting systems, including the Jaynes--Cummings--Hubbard model and geometric bipartitions with a single particle species in our species bipartition framework.
\end{abstract}

\maketitle

\section{Introduction}

Quantum entanglement is one of the defining features of quantum theory, describing correlations between subsystems that have no classical counterpart. Among the many entanglement measures that have been introduced, bipartite entanglement entropy has emerged as one of the most prominent owing to its broad range of applications. It serves as a fundamental resource for quantum information processing~\cite{zhengQIProcessing}, provides a characterization of phases of matter and critical phenomena~\cite{PhysRevB.81.064439}, is useful in studying the black hole information paradox~\cite{pageblackholeinfoparadox} and holography~\cite{Ryu_2006}.

A particularly fruitful direction has been the study of typical entanglement. Beginning with Page's seminal calculation of the average entropy of a subsystem of a Haar-random pure state~\cite{page1993}, a large body of work has investigated the entanglement properties of random-state ensembles~\cite{page1993, FoongKanno1994, SanchezRuiz1995, HaydenLeungWinter2006, PopescuShortWinter2006, PhysRevLett.119.220603, Vidmar2017Quadratic, Vidmar2018VolumeLaw, Garrison_2018,hackl2019average, Bianchi_2021, _yd_ba_2020, PRXQuantum.3.030201, Murciano_2022}. Such ensembles provide analytically tractable models of highly entangled many-body states and often capture the leading entanglement structure observed in quantum-chaotic systems. More recently, these ideas have been extended to ensembles constrained by Abelian $\mathrm{U}(1)$ symmetries associated with particle number~\cite{PhysRevLett.119.220603,Bianchi2019, Bianchi_2021,Yauk:2023wbu}, magnetization, or other conserved charges~\cite{bianchi2026typicalentanglemententropycharge}. Additionally, non-Abelian $\mathrm{SU}(2)$ symmetries~\cite{PhysRevB.108.245101,chakraborty2025randommatrixpredictionaverage, SciPostPhys.17.5.127, PhysRevB.107.045102, PhysRevLett.130.140402,RussottoAresCalabrese2024,bianchi2026typicalentanglemententropycharge} and non-Lie group symmetries, such as in the setting of quantum groups~\cite{yauk2026typicalentanglementanyonchains}, have recently been studied. These studies have shown that different symmetries can substantially modify the Page picture and produce nontrivial thermodynamic corrections to the average entanglement entropy.

\begin{figure}[t]
\centering
\includegraphics{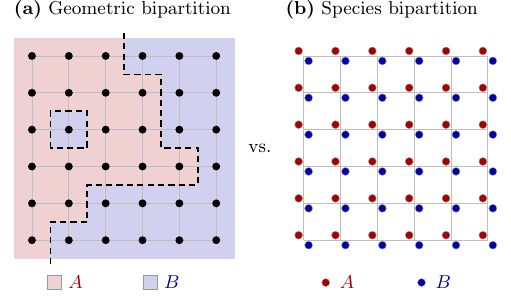}
\caption{\textit{Geometric versus species bipartition.} \textbf{(a)}~Geometric bipartition~\cite{Yauk:2023wbu}: every site hosts the same particle species, and the subsystems $A$ and $B$ are two regions of the lattice. \textbf{(b)}~Species bipartition: every site hosts both species, and the subsystems are the two species. In both cases only the total particle number $N=N_A+N_B$ is fixed.}
\label{fig:bipartition-types}
\end{figure}

Most existing studies of typical entanglement consider a geometric bipartition, in which the system is divided into spatial regions. In this setting, the entanglement structure of unconstrained and constrained random states is now understood in considerable detail. In particular,~\cite{Yauk:2023wbu} developed a general framework for fixed-particle-number random states and derived an asymptotic expansion for the corresponding average entanglement entropy in the thermodynamic limit. That work demonstrated how only information about a particle's local multiplicities for all particle-number macrostates is sufficient, along with the system size $V$ and total particle number $N$, to calculate the average and variance of the entanglement entropy. Additionally, in the thermodynamic limit, that work identified critical structure uniquely at half-subsystem fraction.

However, a spatial bipartition is not the only physically meaningful way to split a quantum system. Many experimentally relevant platforms naturally decompose into distinct particle species or degrees of freedom. Examples include hybrid light--matter systems such as cavity-QED arrays~\cite{Hartmann_2006}, circuit-QED architectures~\cite{Petersson_2012}, ultracold atomic mixtures containing multiple bosonic or fermionic species~\cite{Bloch_2008, Cazalilla_2003}, spin--boson models~\cite{RevModPhys.59.1}, and Jaynes--Cummings--Hubbard systems describing coupled matter and radiation degrees of freedom~\cite{JaynesCummings1963, Angelakis_2007, Makin_2008, Hartmann_2006, Greentree_2006}. In such systems, measurements and control operations may naturally address one species while remaining insensitive to another. There, the relevant notion of bipartite entanglement is thus the entanglement between particle species rather than between spatial regions.

Despite the prevalence of such systems, comparatively little is known about the typical entanglement structure associated with species bipartitions. %
Despite its apparent similarity to the geometric setting, the species problem is conceptually distinct: a geometric bipartition divides a system built from a single type of local degree of freedom, whereas a species bipartition generally involves two physically different species, see figure~\ref{fig:bipartition-types}. Consequently, the entanglement structure is determined by the interplay between the microscopic properties of both species rather than by a single underlying local structure. As we will see, this yields a significantly richer structure in the asymptotic expansion of the typical entanglement entropy with different classes of critical particle densities.

The goal of this work is to develop a general framework for the typical entanglement entropy between two particle species or collective degrees of freedom in lattice systems with fixed total particle number. Our analysis generalizes the fixed-particle-number framework introduced in~\cite{Yauk:2023wbu} by replacing the geometric bipartition with a species bipartition. The resulting theory applies to an extensive class of particle species, including but not limited to bosonic, fermionic, spin, and hybrid systems; only particularly pathological mathematical formulations are exempt. 

We derive an explicit asymptotic expansion for the average entanglement entropy in the thermodynamic limit at fixed particle density. The resulting expansion depends on the microscopic properties of both species and contains several qualitatively distinct regimes separated by special particle densities. We characterize the associated critical structure, determine the corresponding finite-size behavior, and establish typicality through an exponentially vanishing variance in the thermodynamic limit. We further show that the geometric bipartition result of~\cite{Yauk:2023wbu} is recovered as a special case of the general framework developed here.\\

The remainder of this manuscript is organized as follows. In section~\ref{sec:AnalyticalResults}, we introduce the general framework and derive the asymptotic scaling of the relevant Hilbert-space dimensions. We also compute the average entanglement entropy and its variance. In section~\ref{sec:case studies}, we illustrate the framework through several examples, including a simple family of testbed models to demonstrate all aspects of the critical structure, the Jaynes--Cummings--Hubbard model and geometric bipartitions as studied in~\cite{Yauk:2023wbu}. We conclude in section~\ref{sec:discussion and outlook} with a discussion of our results and provide an outlook. In appendix~\ref{ap:saddlepointanalysis} we provide technical results to derive the asymptotic scaling of the Hilbert space dimensions for an extensive class of particle species. Finally, in appendix~\ref{ap:ResolvingKdelta} we resolve the critical phenomena at finite size.

\begin{figure*}[t]
\centering
\includegraphics{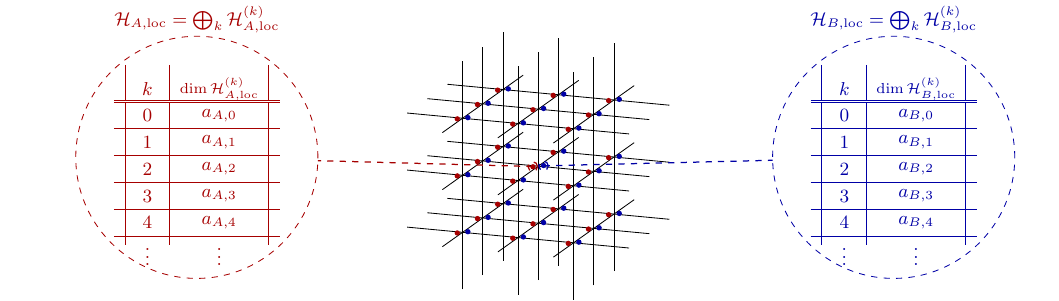}

\vspace{-0.35cm}
\caption{\textit{Local and graded-local Hilbert spaces of two species.} Each lattice site hosts two local Hilbert spaces, $\mathcal H_{A,\mathrm{loc}}$ and $\mathcal H_{B,\mathrm{loc}}$; each decomposes into sectors $\mathcal H^{(k)}_{X,\mathrm{loc}}$ of fixed onsite particle number $k$, with dimensions $a_{X,k}$ for $X\in\{A,B\}$.}
\label{fig:two_species_local_hilbert}
\end{figure*}

\section{Analytical Results: Average and Variance}\label{sec:AnalyticalResults}
In this section we derive our main analytical results, namely, an asymptotic expansion for the average entanglement entropy and its corresponding variance in the thermodynamic limit, $V\to \infty$ where $N=Vn$ with filling density $n$ at constant order, given in~\eqref{eq:bvgEE} and~\eqref{eq:BVGvar}, respectively; we show that the discontinuities of this expansion at critical densities are either Kronecker deltas or branch transitions, and we resolve both in a \emph{double scaling limit}. We also show that the entanglement entropy is \emph{typical} or \emph{self-averaging}, as its variance vanishes quickly enough in the thermodynamic limit.

\subsection{Setup and dimensional scaling: two particle species with fixed total particle number}\label{sec:setup}

We consider a general setting of $V$ sites indexed by the set $S = \set{1, \ldots, V}$. We keep the geometry of the lattice completely general; in particular, the results apply to any graph with $V$ vertices. We assign to each site $j\in S$ a copy of a local Hilbert space $\Hil_j\cong \Hil_{AB,\loc}$. In this local Hilbert space, we allow for the presence of two particle species, $A$ and $B$, so that we have the factorization
\begin{align}\label{eq:Hloc-as-tensor-product}
    \Hil_{AB,\loc} = \Hil_{A,\loc} \otimes \Hil_{B, \loc}\,.
\end{align}
So we also have at each site $j \in S$ the Hilbert spaces $\Hil_{A, j}$ and $\Hil_{B, j}$ which are copies of $\Hil_{A, \loc}$ and $\Hil_{B, \loc}$, respectively. For each species $X \in \{A,B\}$, the local Hilbert space decomposes according to the number of particles of species $X$ as
\begin{align}\label{eq:HXloc-into-particle-number-sectors}
    \Hil_{X, \loc} = \bigoplus_{k=0}^\infty \Hil_{X, \loc}^{(k)}\,,
\end{align}
where $\Hil_{X, \loc}^{(k)}$ denotes the subspace containing states with $k$ particles of species $X$, see figure~\ref{fig:two_species_local_hilbert}. This decomposition along with \eqref{eq:Hloc-as-tensor-product} allows us to obtain the decomposition
\begin{align}
    \Hil_{AB, \loc} = \bigoplus_{\ell,m = 0}^\infty \Hil_{A, \loc}^{(\ell)} \otimes \Hil_{B, \loc}^{(m)}\,.
\end{align}
So we may write
\begin{align}\label{eq:Hloc-into-particle-number-sectors}
    \Hil_{AB, \loc} = \bigoplus_{k = 0}^\infty \Hil_{AB, \loc}^{(k)}\,,
\end{align}
where $k = \ell + m$ and
\begin{align}\label{eq:fixedKDecomp}
    \Hil_{AB, \loc}^{(k)} = \bigoplus_{\ell=0}^k \Hil_{A, \loc}^{(\ell)} \otimes \Hil_{B, \loc}^{(k-\ell)}
\end{align}
denotes the subspace containing $k$ particles at a site. The decompositions~\eqref{eq:HXloc-into-particle-number-sectors} and~\eqref{eq:Hloc-into-particle-number-sectors} allow us to define the local particle-number operators $\hat{N}_{X,j}$ for $X \in \{A, B, AB\}$, by
\begin{align}
    \hat{N}_{X, j} \ket{\psi} = k \ket{\psi}, \quad \ket{\psi} \in \Hil_{X, \loc}^{(k)}\,.\label{eq:number-operator}
\end{align}
Due to~\eqref{eq:fixedKDecomp} and~\eqref{eq:number-operator}, we have
\begin{align}
    \hat{N}_{AB, j} = \hat{N}_{A, j} + \hat{N}_{B, j}\,,
\end{align}
where any local operator is extended to tensor products of Hilbert spaces by acting as the identity on all other Hilbert spaces in the tensor product.

The Hilbert space of the full lattice is obtained by taking the tensor product of the local Hilbert spaces over each site 
\begin{align}\label{eq: total hilbert space tensor product construction}
    \Hil_{AB} = \bigotimes_{j \in S} \Hil_j\,.
\end{align}
From~\eqref{eq:Hloc-as-tensor-product} and the above construction, we get
\begin{align}
    \Hil_{AB} = \Hil_A \otimes \Hil_B\,,
\end{align}
where
\begin{align}
    \Hil_A = \bigotimes_{j \in S} \Hil_{A, \loc}, \quad \Hil_B = \bigotimes_{j \in S} \Hil_{B, \loc}\,.
\end{align}
This tensor product factorization defines the \emph{particle species bipartition} of the lattice. This is the bipartitioning with respect to which the entanglement entropy is computed. 

As a notational aside, throughout the paper---unless stated otherwise---$X$ denotes either a single species $A$ or $B$, or the \emph{combined species} $AB$ characterized by~\eqref{eq:alseqcoeff}; the species $AB$ accounts for the combinatorics of the bipartite system $AB$ in the same sense that the species $A$ and $B$ do for subsystems $A$ and $B$ respectively. If the species label is omitted, it is understood to refer to the combined species $AB$.

\subsubsection{Fixing the total particle number}
Having established the species bipartition, we now specify the ensemble of states used to compute the average entanglement entropy. To do so, we first introduce the fixed-particle-number sectors and their associated particle-number operators. Using the decompositions in the previous subsection, we can decompose the Hilbert spaces $\Hil_X$ into fixed-particle-number sectors,
\begin{align}
    \Hil_{X} = \bigoplus_{N_X=0}^\infty \Hil_{X}^{(N_X)}\,
\end{align}
 where
\begin{align}
    \Hil_{X}^{(N_X)} = \bigoplus_{N_X = \sum_{j=1}^V N_{X, j}} \bigotimes_{j=1}^V \Hil_{X, j}^{(N_{X, j})}\,.
\end{align}
As a consequence of \eqref{eq:fixedKDecomp}, a state with total particle number $N$ may contain $N_A$ particles of species $A$ and $N-N_A$ particles of species $B$, giving the decomposition
\begin{align}\label{eq:fixedNdecomp}
    \Hil_{AB}^{(N)} = \bigoplus_{N_A=0}^N \Hil_{A}^{(N_A)} \otimes \Hil_{B}^{(N-N_A)}\,.
\end{align}
We can then define the number operators $\hat{N}_X$ as
\begin{align}
    \hat{N}_X = \sum_{j=1}^V \hat{N}_{X, j}\,.
\end{align}
In particular, we have
\begin{align}
    \hat{N}_{AB} = \hat{N}_A + \hat{N}_B\,.
\end{align}
It follows from the definition of the local particle-number operators that $\hat{N}_X$ acts as multiplication by $N_X$ on $\Hil_X^{(N_X)}$. Thus $\Hil_X^{(N_X)}$ is the eigenspace of $\hat{N}_X$ corresponding to particle number $N_X$. We restrict attention to Haar-random pure states in $\Hil_{AB}^{(N)}$ and compute the average entanglement entropy with respect to this ensemble.
The computation of the average entanglement entropy requires the dimensions of the fixed particle number sectors introduced above. Taking dimensions of the above decompositions yields
\begin{align}\label{eq:dimDX}
    d_X = \dim \Hil_X^{(N_X)} = \,\sum_{N_X=\sum_{j=1}^V N_{X,j}}\ \prod_{j=1}^V \underbrace{\dim \Hil_{X,j}^{(N_{X,j})}}_{=:\,a_{X,N_{X,j}}}\,.
\end{align}
Furthermore, taking dimensions of \eqref{eq:fixedNdecomp} gives
\begin{align}\label{eq:dNsumdAdB}
    d_{AB}=\sum_{N_A=0}^Nd_Ad_B\,.
\end{align}
The above formulas for the dimensions $d_A$, $d_B$ and $d_{AB}$ show that these quantities depend only on $V$, $N$ and the integers $\dim \Hil_{X, j}^{(N_{X,j})}$ which we label as $a_{X,N_{X,j}}$ for $X=A,B$ and properly define in the next subsection.

\subsubsection{Dimensional sequences}
We now define the \emph{dimensional sequences},
\begin{align}
    (a_{X,k})_k\subset \Nat\,,
\end{align}
where $a_{X,k}:= \dim\Hil_{X, \loc}^{(k)}$.
Then by \eqref{eq:fixedKDecomp}, we have the Cauchy product identity of two sequences
\begin{align}\label{eq:alseqcoeff}
    a_\ell=\sum_{k=0}^\ell a_{A,k}a_{B,\ell-k}\,,
\end{align}
where the dimension $a_\ell = \dim \Hil_{AB,\loc}^{(\ell)}$ is the number of ways $\ell$ particles can be placed among the two species at the site. For some $a_{X,k}\in (a_{X,n})_{n}$ we may think of $k$ -- the number of particles -- as being the macrostate, while $a_k$ is its corresponding maximum multiplicity. Some examples are laid out in table \ref{tab:akExamples}. We restrict ourselves to \emph{admissible sequences} $(a_{X,n})_n\subset \mathbb N$ satisfying the following conditions:
\begin{enumerate}[label=(\arabic*)]
        \item The radius of convergence satisfies $R_X=(\limsup_n a^{1/n}_{X,n})^{-1}>0$, \ie the multiplicities grow at most exponentially.%
        \item There is a vacuum: $a_{X,0}>0$.
        \item The support is aperiodic: $g_X=\gcd\set{k|a_{X,k}\neq0}=1$, \ie the greatest common divisor of $k$ for non-vanishing $a_{X,k}$ is $1$.%
        \item If $\nXmax=\infty$, we require $\sum_k a_{X,k}kz^k\xrightarrow{z\to R^-_X}\infty$.
    \end{enumerate}
Aperiodicity implies that 
\begin{align}
    \nXmax:=\min\set{n_X|a_{X,k}\!=\!0\,,\forall k\!>\!n_X}\neq0\,;
\end{align}
we also allow for $\nXmax=\infty$. We note here that virtually all physically interesting particle species are covered by this model, and a wide variety of non-admissible species can be mapped to admissible species. Otherwise, the only species which we are unable to cover are those whose multiplicities grow super-exponentially, preventing the generating function, 
\begin{align}\label{eq:genX}
    \zeta_X(z):=\sum^\infty_{k=0}a_{X,k}z^k\ \quad  0\leq |z|<R_X\,,
\end{align}
from being analytic at the origin, those species that do not support the existence of particles, $\nXmax=0$, and a particularly pathological case where the \emph{supercritical density}, defined by
\begin{align}
    \nXsc=\lim_{x\to R^-}\frac{\sum_k a_{X,k}kx^k}{\sum_k a_{X,k}x^k}\,,
\end{align}
remains finite when $\nXmax=\infty$. We discuss counterexamples to these axioms and a basic non-admissible to admissible mapping in appendix~\ref{ap:NonAdmissible}, while in appendix~\ref{ap:admissible} we show how these axioms arise naturally from the hypotheses of the saddle point method. Note that for an admissible species $X$, one finds $N^{\max}_X:=V\nXmax$ is the maximum number of such particles the system can hold.
\begin{table}[t]
\centering
\renewcommand{\arraystretch}{1.2}
\caption{\textit{Examples of particle species.} A species is characterized by the multiplicities $a_{X,k}$ of $k$ particles at a single site. The two-level atom and the one-mode boson are the two species of the Jaynes--Cummings--Hubbard model.}
\label{tab:akExamples}
\begin{tabular}{@{\hspace{0.5em}} l @{\hspace{2em}} l @{\hspace{1.5em}} c @{\hspace{1.5em}} c @{\hspace{0.5em}}}
\hline 
\hline
Species $X$ & State & $k$ & $a_{X,k}$ \\
\hline 
\multirow{3}{*}{\shortstack[l]{Spinless fermions  or \\two-level atoms}}
  & $\ket{0}$    & $0$ & $1$ \\
  & $\ket{1}$    & $1$ & $1$ \\
  & other        & ---   & $0$ \\[8pt]
\multirow{4}{*}{Spin-$\tfrac{1}{2}$ fermions}
  & $\ket{0}$                            & $0$ & $1$ \\
  & $\ket{\uparrow},\,\ket{\downarrow}$  & $1$ & $2$ \\
  & $\ket{\uparrow\downarrow}$           & $2$ & $1$ \\
  & other                                & ---   & $0$ \\[8pt]
\multirow{3}{*}{\shortstack[l]{Massless spin-$1$ bosons \\ \hspace{1em}a. two-mode}}
  & $\ket{0}$             & $0$         & $1$   \\
  & $\ket{+},\,\ket{-}$  & $1$         & $2$   \\
  & $\ket{n_+,n_-}$      & $n_+ + n_-$ & $k+1$ \\[6pt]
\hspace{1em}b. one-mode
  & $\ket{k}$  & $k$         & $1$   \\[2pt]
  
\hline
\hline
\end{tabular}
\end{table}
Finally, by \eqref{eq:alseqcoeff} we have the Cauchy product identity inside the common radius of convergence
\begin{align}\label{eq:zetaIDs}
    \zeta(z)=\zeta_A(z)\zeta_B(z)\,.
\end{align}

\subsubsection{Asymptotic scaling of dimensions in \texorpdfstring{$V$}{V}} 
The asymptotic analysis of the average entanglement entropy and its variance will require the thermodynamic scaling of the dimensions $d_A, d_B$ and $d_{AB}$. Specifically, we study the asymptotics of $d_X$ in the thermodynamic limit $V \gg1$ with large $N_X$ such that
\begin{align}
    n_X =\frac{N_X}{V}=O(1)\,.
\end{align}
We therefore derive these asymptotics here, noting that a similar approach was used in~\cite{Yauk:2023wbu}. A comprehensive treatment is laid out in appendix~\ref{ap:saddlepointanalysis}, while here we provide a brief summary of the technique. 

One has the combinatorial identity
\begin{align}
    \zeta_X(z)^V=\sum^\infty_{N_X=0}d_{X} z^{N_X}\,,
\end{align}
by taking a $V$-fold Cauchy product of $(a_{X,k})_k$ and comparing the result to \eqref{eq:dimDX}. So we have a Cauchy coefficient problem
\begin{align}
\label{eq:dimension_saddle_integral}
    d_{X}( N_X,V)=[z^{N_X}]\zeta_X(z)^V=\frac{1}{2\pi\ii}\oint \zeta_X(z)^V\frac{\dd z}{z^{N_X+1}}\,,
\end{align}
where integration is performed around a circular contour centered at the origin. For an admissible species $X$, we find for $n_X$ in any compact interval of $(0,\nXmax)$ there is a unique solution $z_{X,0}(n_X)\in (0,R)$ of the saddle point equation
\begin{align}\label{eq:zetaequivsaddle}
    z_{X,0}\frac{\zeta'_X(z_{X,0})}{\zeta_X(z_{X,0})}=n_X\,.
\end{align}
We can define the following, where $-\beta_X$ is a Legendre transform,
\begin{align}\label{eq:betapsi}
    \psi_X(z):={}&\ln \zeta_X(z)-n_X\ln z\,,\\
    \beta_X(n_X):={}&\psi_X(z_{X,0})\,.
\end{align}
We find that the saddle point method gives
\begin{align}
    d_X= \frac{\zeta_X(z_{X,0})^V}{z_{X,0}^{N_X+1}\sqrt{2\pi V\psi''_X(z_{X,0})}}(1+O(1/V))\,,
\end{align}
where 
\begin{align}
    \psi''_X(z_{X,0})=-\frac{1}{z_{X,0}^2\beta_X''(n_X)}\,.
\end{align}
Thus we may simplify
\begin{align}\label{eq:dXasymptotic}
    d_X= \sqrt\frac{-\beta_X''(n_X)}{{2\pi V}}\ee^{V \beta_X(n_X)}(1+O(1/V))\,.
\end{align}

\subsection{Average entanglement entropy}
Consider the bipartition of our system into species $A$ and species $B$. It was shown in~\cite{Bianchi2019} that, for a system with an algebra of observables $\mathcal A$ and subsystem $A$ defined by picking a subalgebra $\mathcal A_A$ (and with subsystem $B$ defined as the complement), in the presence of a non-trivial center $\mathcal A_A\cap \mathcal A_B\neq \mathbb C\mathbbm{1}$, \ie a center containing more than multiples of the identity, one can calculate the moments of the bipartite entanglement entropy $S_A$ exactly, by simply considering the dimensions of the subsystems. By the decomposition in \eqref{eq:fixedNdecomp}, our system is able to use the results of Bianchi and Donà. In this subsection, we care about the expectation value
\begin{align}\label{eq:EEanalytic}
    \braket{S_A}_N=\sum_{N_A}\varrho_{N_A}\varphi_{N_A}\,,
\end{align}
where we have for our setup
\begin{align}
    \varrho_{N_A} ={}& \frac{d_A d_B}{d_{AB}}\,,\\
    \varphi_{N_A}={}&\Psi(d_{AB} + 1) - \Psi(d_> + 1)-\frac{d_<-1}{2d_>}\,,
\end{align}
where $d_<=\min\set{d_A,d_B}$ and $d_>=\max\set{d_A,d_B}$ are the smaller and larger of the dimensions of the sector $N_A$, and $\Psi=\Gamma'/\Gamma$ is the digamma function.

Our general strategy is to use the asymptotic forms of the dimensions \eqref{eq:dXasymptotic} and convert this sum into an integral to find $\braket{S_A}_N$ up to an $o(1)$ remainder. 

\subsubsection{The Bianchi-Donà formula}\label{sec:BianchiDona}
Having established the thermodynamic scaling of the dimensions, we now use these asymptotics to study the quantities that will enter the computation of the average entanglement entropy in the thermodynamic limit. It is useful to obtain the asymptotic forms of $\varrho_{N_A}$ and $\varphi_{N_A}$. Throughout, we treat $n_X\equiv N_X/V$ as a continuous variable since we take the thermodynamic limit. By \eqref{eq:dXasymptotic}, we have 
\begin{align}
    \begin{split}
        \varrho_{N_A} = \sqrt\frac{\beta''_A(n_A) \beta''_B(n-n_A)}{-\beta''(n)2\pi V} \ee^{V\gamma(n_A)}(1+O(1/V))
    \end{split}\,,
\end{align}
where
\begin{align}
    \gamma(n_A) :={}&\beta_A(n_A) + \beta_B(n-n_A) - \beta(n)\,.
\end{align}
Since $\gamma(n_A)$ is multiplied by $V$, values of $n_A$ away from the maximum of $\gamma$ are exponentially suppressed as $V \to \infty$. We therefore define the quantity $\nAbar$ to be the value of $n_A$ that maximizes $\gamma$. By lemma~\ref{lem:combinedsaddle} this maximizer always exists and is unique. Equivalently, $\bar{n}_A$ is the quantity such that
\begin{align}\label{eq:transcedentalrelation}
    \beta_A'(\nAbar) = \beta_B'(n-\nAbar)\,.
\end{align}
To reduce notational clutter, we introduce the notation
\begin{align}\label{eq:betashorthand}
    \hspace{-1mm}\bar{\beta}_X^{(m)} := \beta_X^{(m)}\bm(\bar{n}_X(n)\bm)\,,
\end{align}
with $(m)$ indicating the $m$-th derivative and $X \in \{A, B\}$ where we define
\begin{align}
    n_B := n - n_A\,.
\end{align}
We call $\bar\beta_X=\bar\beta_X^{(0)}$ the \emph{saddle-point entropy density} of species $X$. In the thermodynamic limit, the scaling of $\varrho_{N_A}$ is determined by the behavior of $\gamma$ around $\nAbar$. So Taylor expanding $\gamma$ around $\nAbar$ yields
\begin{align}
    \gamma(n_A) ={}& \gamma(\nAbar) - \frac{\kappa}{2} (n_A - \nAbar)^2\nonumber\\
    &+ \frac16(\bbA'''-\bbB''')(n_A-\nAbar)^3+O\bm((n_A-\nAbar)^4\bm)\,,
\end{align}
where we used \eqref{eq:betashorthand} and proved in appendix~\ref{ap:sharedsaddle} that 
\begin{align}
    \kappa ={}&-\gamma''(\nAbar) > 0\,,\label{eq:kappa}\\
    \gamma(\nAbar)={}&0\,.
\end{align}
Thus we can write using~\eqref{eq:betashorthand}
\begin{align}
    \varrho_{N_A} ={}& \sqrt\frac{\beta''_A(n_A) \beta''_B(n_B)}{-\beta''(n)2\pi V}\,(1+O(1/V))\nonumber\\
    &\times\exp\!\Big[-\frac{V \kappa}{2} (n_A - \nAbar)^2\nonumber\\
    &+\frac{V}{6}(\bbA'''-\bbB''')(n_A-\nAbar)^3+O\bm(V(n_A-\nAbar)^4\bm)\Big]\,.
\end{align}
To resolve the probability distribution on its natural width, we introduce the coordinate
\begin{align}
    \delta = \sqrt{V}(n_A-\bar{n}_A)\,.
\end{align}
Then after expanding we find using \eqref{eq:betashorthand}
\begin{subequations}\label{eq:varrhodelta}
\begin{align}
    \varrho(\delta)={}&\varrho_{\mathrm G}(\delta)\left(1+\frac{\mathcal{E}_1(\delta)}{\sqrt V}+\frac{\mathcal{E}_2(\delta)}{V}+O(V^{-3/2})\right)\,,\\
    \varrho_{\mathrm G}(\delta):={}&\sqrt{\frac{\kappa}{2\pi }}\ee^{-\delta^2\kappa/2}\,,\\
    \mathcal{E}_1(\delta):={}&\frac{\delta}{2}\left(\frac{\bbA'''}{\bbA''}-\frac{\bbB'''}{\bbB''}\right)+\frac{\delta^3}{6}(\bbA'''-\bbB''')\,,
\end{align}
\end{subequations}
where we define $\varrho(\delta):=\sqrt V\varrho_{N_A}$ and find this is asymptotically equivalent to the Gaussian $\varrho_{\mathrm G}$ with vanishing mean and variance $1/\kappa$ in the $\delta$ coordinate. This will become useful when we turn the sum \eqref{eq:EEanalytic} into an integral. We call $\mathcal{E}_1/\sqrt V$ and $\mathcal{E}_2/V$ the \emph{first-} and \emph{second-order Edgeworth corrections} to the Gaussian. The first is odd in $\delta$, while $\mathcal{E}_2$ requires higher-order corrections to the asymptotic expansion~\eqref{eq:dXasymptotic} for an explicit form, but we shall see that an explicit form is not needed. Since $\sum_{N_A}\varrho_{N_A}=1$ holds exactly, and converting this sum of the smooth summand $\varrho_{N_A}$ into an integral over $\delta$ incurs at most an $o(1)$ error (appendix~\ref{ap:SumIntegralError}), both corrections have vanishing Gaussian mean,
\begin{align}\label{eq:edgeworth-mean-zero}
    \int_\R\varrho_{\mathrm G}(\delta)\mathcal{E}_1(\delta)\dd\delta=\int_\R\varrho_{\mathrm G}(\delta)\mathcal{E}_2(\delta)\dd\delta=0\,.
\end{align}

The asymptotic form of $\varphi_{N_A}$ can be found by again using the asymptotic scaling of the dimensions $d_X$ and the digamma function scaling $\Psi(x+1) = \ln(x)+O(1/x)$ for $x\gg1$. Thus we have
\begin{align}\label{eq:varphi scaling}
    \begin{split}
        \varphi_{N_A}\! =\! \ln\!\left( \!\frac{d_{AB}}{d_>} \!\right)\! -\! \frac{d_<}{2d_>}+\!O(\ee^{-cV}) \,.
    \end{split}
\end{align}
for some $c>0$. Calculating, we find
\begin{align}
    \varphi_{N_A}=\begin{cases}
    \begin{aligned}
    & V(\beta(n)-\beta_A(n_A))+\tfrac12\ln \tfrac{\beta''(n)}{\beta''_A(n_A)}\\
    & -\tfrac12\ee^{-VY}+O(1/V)\,,
    \end{aligned} & d_A>d_B\\[6mm]
    \begin{aligned}
    &V(\beta(n)-\beta_B(n_B))+\tfrac12\ln \tfrac{\beta''(n)}{\beta''_B(n_B)}\\
    &-\tfrac12\ee^{VY}+O(1/V)\,,
    \end{aligned} & d_B>d_A
    \end{cases}\,.
\end{align}

To determine which dimension is bigger at a given $n_A$ and $V$ we define
\begin{align}\label{eq:Ydefn}
    Y(n_A)=\beta_A(n_A)-\beta_B(n_B)+\frac1{2V}\ln \frac{\beta_A''(n_A)}{\beta_B''(n_B)}\,,
\end{align}
since asymptotically $d_A/d_B=\ee^{VY}(1+O(1/V))$. Then clearly the sign of $Y$ determines which dimension is larger up to an $O(V^{-2})$ error. 
We may expand around $\nAbar$ and express the result in the $\delta$ coordinate, where we recall \eqref{eq:betashorthand},
\begin{align}
    \hspace{-3mm}\varphi(\delta)={}&\begin{cases}
    \begin{aligned}
       &V\bbB-\sqrt{V}\bbA'\delta-\tfrac12\bbA''\delta^2-\tfrac12\ee^{-VY}\\
       &+\tfrac12\ln \tfrac{\beta''(n)}{\bbA''}+O(V^{-1/2})\,,
      \end{aligned} & d_A>d_B\\[6mm]
      \begin{aligned}
       &V\bbA+\sqrt{V}\bbB'\delta-\tfrac12\bbB''\delta^2-\tfrac12\ee^{VY}\\
       &+\tfrac12\ln \tfrac{\beta''(n)}{\bbB''}+O(V^{-1/2})\,,
      \end{aligned} & d_A< d_B
    \end{cases},\label{eq:varphiNAsplitdelta}\\
    \hspace{-3mm}Y(\delta)={}&\bbA-\bbB+2\bbA'\frac{\delta}{\sqrt{V}}+\frac{\delta^2}{2V}(\bbA''-\bbB'')\nonumber\\ &+\frac{1}{2V}\ln\frac{\bbA''}{\bbB''}+O(V^{-3/2})\,.\label{eq:Yinitial}
\end{align}

\subsubsection{Converting the sum into an integral}\label{sec:sum-to-integral}
We convert the average entanglement entropy sum into an integral in the thermodynamic limit $V \to \infty$, introducing rescaled coordinates $\delta = \sqrt{V}(n_A - \bar{n}_A)$ so that the resulting integral is taken against a Gaussian measure.
In the thermodynamic limit we find the asymptotics
\begin{align}
    \hspace{-2mm}\braket{S_A}_N=\sum_{N_A=0}^N\varrho_{N_A}\varphi_{N_A}
    =\!\int_{-\infty}^\infty\! \varrho(\delta) \varphi(\delta) \dd\delta +o(1),\label{eq:integral formula for EE}
\end{align}
where the continuum functions $\varrho(\delta)$ and $\varphi(\delta)$ were introduced in~\eqref{eq:varrhodelta} and~\eqref{eq:varphiNAsplitdelta}, respectively.
The sum has been replaced by a Riemann integral, where we can bound the conversion error to be at most $O(1/\sqrt{V})$ using an Euler--Maclaurin formula, as explained in appendix~\ref{ap:SumIntegralError}. The function $\varrho(\delta)$ approaches a Gaussian with corrections, while $\varphi(\delta)$ is a piecewise-analytic function.

Naively, one might think that the second-order Edgeworth correction $\mathcal{E}_2/V$ in $\varrho(\delta)$ would produce a constant-order contribution after integrating against the leading order $V\bbA$ or $V\bbB$ term. However, since the condition $\sum \varrho_{N_A}=1$ is exact, both Edgeworth corrections have vanishing Gaussian mean~\eqref{eq:edgeworth-mean-zero}. In particular, it can be shown that when integrating against the leading order term in $\varphi$, which is constant in $\delta$, the contribution is $o(1)$ if we are near nonanalytic points (in the sense of the scaling regime used in section~\ref{sec:resolving-kdeltas}) or $0$ if we are exactly at a nonanalytic point or far enough away.

\begin{figure*}[tp]
\centering
\includegraphics{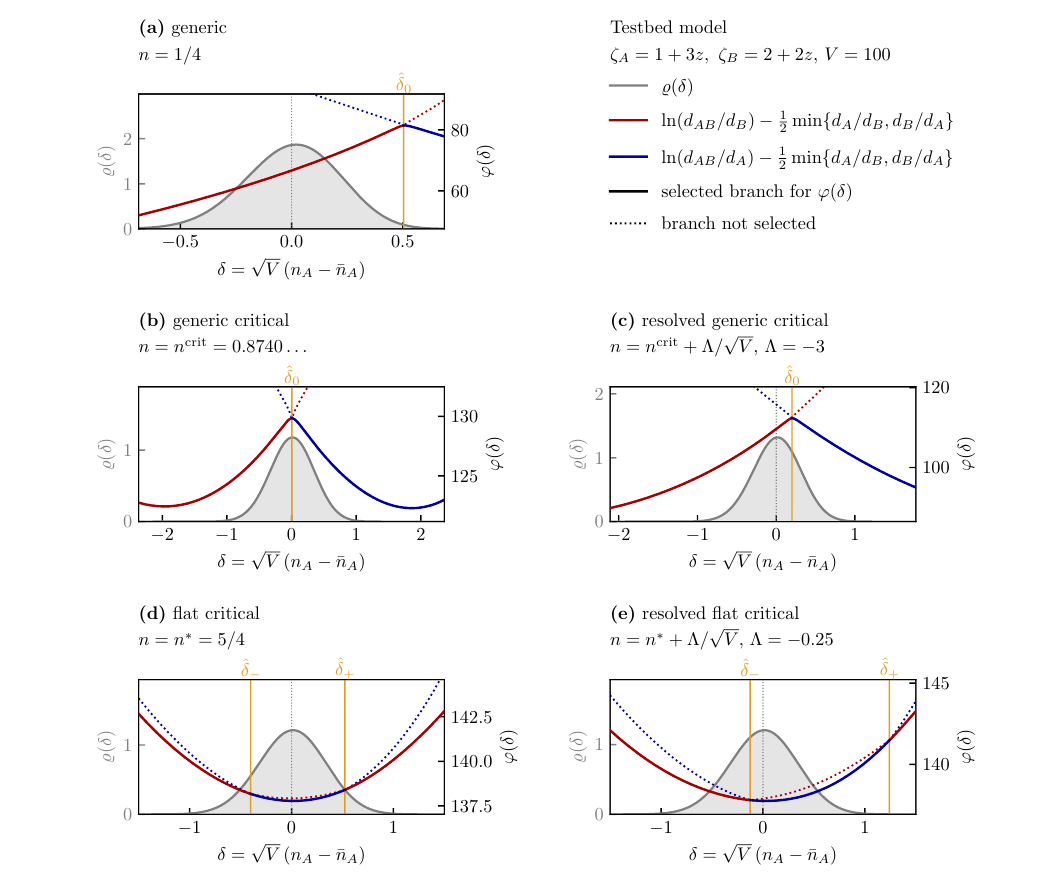}
\caption{\textit{When does the Gaussian see the kink?}
    The two factors of the integrand in~\eqref{eq:intsplit3ways} for the testbed model: the Gaussian $\varrho(\delta)$ (shaded) and the observable $\varphi(\delta)$, which follows the red branch where $d_A<d_B$ and the blue branch where $d_A>d_B$. The kinks of $\varphi$, where the branches cross, are marked in orange. \textbf{(a)}~Generic density: the kink lies at $\hat\delta_0\propto\sqrt{V}$, away from the peak, so its effect is suppressed. \textbf{(b)}~Generic critical density: the kink sits at the peak. \textbf{(c)}~Double scaling limit around the generic critical density, in which $\hat\delta_0$ is of constant order. \textbf{(d)}~Flat critical density: two kinks $\hat\delta_\pm$ at finite distances on either side of the peak. \textbf{(e)}~Double scaling limit around the flat critical density, in which the two kinks are no longer symmetric about the peak.
    }
    \label{fig:kink-vs-gaussian}
\end{figure*}

\subsubsection{Computing the integral}
We can write
\begin{align}\label{eq:intsplit3ways}
    \int_{\R} \varrho(\delta) \varphi(\delta) \dd\delta={}&\int_{\set{d_A>d_B}}\varrho(\delta)\ln\frac{d_{AB}}{d_A}\dd \delta\nonumber\\+&\int_{\set{d_A<d_B}}\varrho(\delta)\ln\frac{d_{AB}}{d_B}\dd \delta \nonumber\\-&\frac12\int_{\R}\varrho(\delta)\ee^{-V|Y|}\dd \delta\,,
\end{align}
where the integration is split between regions where one dimension is larger. Since $\varrho(\delta)$ has a constant width, we only need to look for constant-order roots $\hat\delta_0$ of \eqref{eq:Yinitial} to determine where to split the integral. We see at fixed $n$ (since $\nAbar=\nAbar(n)$) if $\bbA<\bbB$ or $\bbB<\bbA$ that one branch would dominate while the other would not contribute. The interesting case is when $\bbA=\bbB$, at which we denote $n=\ncrit$ a \emph{critical density}, where we have our integral split in a non-trivial manner. As a result, at $n=\ncrit$ we obtain contributions after integration that are discontinuous in $n$, namely the Kronecker deltas and branch transitions of section~\ref{sec:resolving-kdeltas}, as discussed in appendix~\ref{ap:criticaldensity}. Such an analysis is a special case of resolving these contributions in appendix~\ref{ap:ResolvingKdelta} where we let $n=\ncrit+\Lambda/V^s$ vary around the critical density. These critical cases can be thought of as the mean of the Gaussian $n_A=\nAbar$ coinciding with a kink of the otherwise analytic $\varphi$ function (see figure~\ref{fig:kink-vs-gaussian}). 

For the Page correction term on the last line of \eqref{eq:intsplit3ways}, we see that it contributes at constant order if and only if $\bbA=\bbB$ and $\bbA'=0$. We prove that this occurs if and only if $\zeta_A(1)=\zeta_B(1)<\infty$ and $n=n^*$ in appendix~\ref{ap:saddlepointanalysis}, where $n^*:= \zeta'(1)/\zeta(1)$. In particular, a constant-order contribution from this term is only possible if both species only allow for a finite number of particles at each local site.

We denote the set of such critical densities $\ncritset$ so that we can use the indicator function notation
\begin{align}
     1_A(x)=\begin{cases}
        1\,,\ x\in A\\
        0\,,\ \text{otherwise}
    \end{cases}\,.
\end{align}
Evaluating the integrals, with the notation in~\eqref{eq:betashorthand}, we finally obtain for $\bbA\leq\bbB$ (otherwise we exchange subsystems $A\leftrightarrow B$) 
\begin{align}
    \begin{split}
        \braket{S_A}_N ={}&
        V\bbA
        -\sqrt{\frac{2V}{\pi\kappa}}\, |\beta'(n)| \,{1}_{\ncritset}(n) \\
        &+\!\frac12\!\left[\!\frac{1}{1+r_A}\!+\!\ln\!\frac{r_A}{r_A+1}\!+\!\Xi_A\beta'(n){1}_{\ncritset^C}(n)\right.\\ &\hspace{10mm}\left.-\frac12\left(\ln r_A+\frac{1-r_A}{1+r_A}
         \right){1}_{\ncritset\setminus\set{n^*}}(n)\right]  \\
        &+\Pi_*\,\delta_{\zeta_A(1),\zeta_B(1)}\delta_{n,n^*}
        +o(1)\,.
    \end{split}
    \label{eq:bvgEE}
\end{align}
Here we have defined the \emph{concavity ratio} $r_A:=\bbA''/\bbB''$, its inverse $r_B:=1/r_A$, the minimum $r_*:=\min\!\set{r_A,r_B}$, and the \emph{particle species cross terms} $\Xi_A :=\frac{1}{ \kappa^2}(\bbB'''r_A-\bbA'''r_B)$ and $\Xi_B:=-\Xi_A$. This expression is valid for all $n\in(0,\nmax)$ constant with respect to $V$. The last term contributes only at the flat critical density $n^*$, which requires $\zeta_A(1)=\zeta_B(1)$. There $\bbA=\bbB$, and at a non-degenerate flat critical density the same species is selected on both sides of $n^*$ (appendix~\ref{ap:classify-jumps}); we label it $A$, so that $r_A(n^*)=\rsc:=r_*(n^*)$. At a degenerate flat critical density $r_A(n^*)=\rsc=1$ for either labeling. Here and below, a subscript $c$ denotes evaluation at $n=\ncrit$, as in $\betaA''$, $\nAbc'=\nAbar'(\ncrit)$, $\rAc=r_A(\ncrit)$ and $\kappac=\kappa(\ncrit)$. The Kronecker delta at $n^*$ then carries the weight
\begin{align}\label{eq:Pistar}
    \Pi_*:=\lim_{r\to\rsc}\Theta(r)-\frac12\left(\frac{1}{1+\rsc}+\ln\frac{\rsc}{1+\rsc}\right)\,,
\end{align}
with $\Theta$ defined in~\eqref{eq:Theta}. The limit is needed only at a degenerate flat critical density, where $\rsc=1$ and $\Theta(1)$ is not defined; there $\lim_{r\to1}\Theta(r)=\frac12(\frac12-\ln2)-\frac12$, so that $\Pi_*=-\frac12$ is the Page term alone. At a non-degenerate flat critical density, $\Pi_*$ also contains a logarithmic contribution besides the Page term, see appendix~\ref{ap:classify-jumps}. Hence $\braket{S_A}_{n^*V}=V\bbA+\lim_{r\to\rsc}\Theta(r)+o(1)$ at every flat critical density, and $\Pi_*$ is the Kronecker weight of the constant term at $n^*$ (proposition~\ref{prop:jumps-kdeltas}).

\begin{table*}[t]
\centering
\renewcommand{\arraystretch}{1.5}
\caption{\textit{Resolved profiles at the critical densities.} The resolved profiles of~\eqref{eq:inh-res} in the window $n=\ncrit+\Lambda V^{-s}$ with $s=s_K$ of~\eqref{eq:sK} at a generic and $s=\frac12$ at a flat critical density. In the generic column, $\ncrit<n^*$; for $\ncrit>n^*$, exchange $A\leftrightarrow B$ while keeping the same weights. In the flat column, $\betaA''\geq\betaB''$; otherwise exchange $A\leftrightarrow B$. Smooth functions are evaluated at $\ncrit$. The remaining symbols are $\kappac=\kappa(\ncrit)$~\eqref{eq:kappa}, $\delta_{0,\mathrm{g}}$ (to leading order)~\eqref{eq:deltaCgdeltaCPg}, $w_0$ and $w_1$~\eqref{eq:XiArAetc}, $\tau$~\eqref{eq:tau}, $r_*$ (defined below~\eqref{eq:bvgEE}), $w_0^*$ and $w_2^*$~\eqref{eq:flatweights}, and $P$~\eqref{eq:near-flat-page-correction} (non-degenerate) or~\eqref{eq:PageResDegen} (degenerate). The dependence on $\Lambda$ enters through $w$, $\delta_{0,\mathrm{g}}$ and $P$. We read off the resolved profile terms from~\eqref{eq:gencritres1} for the generic critical and~\eqref{eq:flatResAVGlog} for the flat critical case (where only $P$ need be resolved in the flat degenerate case).}
\label{tab:resolved-kdeltas}
\def\cell#1{\begin{tabular}[c]{@{}l@{}}#1\end{tabular}}%
\def\cellc#1{\begin{tabular}[c]{@{}c@{}}#1\end{tabular}}%
\begin{tabular}{@{\hspace{0.5em}} l @{\hspace{2em}} c @{\hspace{2em}} c @{\hspace{0.5em}}}
\hline
\hline
Term & Generic critical~\eqref{eq:gencritres1}  & Flat critical~\eqref{eq:flatResAVGlog} \\
\hline
$A(\Lambda;\ncrit)$ &  $\frac12\beta(\ncrit)$ & $\frac12\beta(\ncrit)$ \\
$B(\Lambda;\ncrit)$ & $-\sqrt{\frac{2}{\pi\kappac}}|\beta'(\ncrit)|\,w_1$ &  $0$ \\
$C(\Lambda;\ncrit)$ & $(1-w_0)\,c_A(\ncrit)+w_0\,c_B(\ncrit)-\tau\sqrt{\frac{\kappac}{2\pi}}\,\dzg\,w_1$ & $\frac12\Big[\frac{1+(\rsc-1)w_2^*}{1+\rsc}-\ln(1+1/\rsc)-w_0^*\ln\rsc\Big]+P(\Lambda)$ \\[2pt]
\hline
\hline
\end{tabular}
\end{table*}

Away from the critical densities, our result~\eqref{eq:bvgEE} simplifies considerably to
\begin{align}
    \braket{S_A}_N=V a_{<}(n)+c_{<}(n)+o(1)\,,
\end{align}
where the subscript $<$ selects the branch of the species $X\in\set{A,B}$ with the smaller saddle-point entropy density
\begin{align}
    a_X(n):={}&\beta_X\bm(\nXbar(n)\bm)=\bar{\beta}_X\,,\label{eq:aX}\\
    c_X(n):={}&\frac12\left(\frac{1}{1+r_X}+\ln\frac{r_X}{1+r_X}+\Xi_X\beta'(n)\right)\,,\label{eq:cX}
\end{align}
so that $a_{<}(n)=\min\set{a_A(n),a_B(n)}$. Note that we generally do \emph{not} have $c_{<}(n)=\min\set{c_A(n),c_B(n)}$, as the selection is done based on $a_X(n)$, which then also applies to $c_{<}(n)$.

It is important to recall that a flat critical density is only allowed for pairs of species $A$ and $B$ such that their total multiplicities match $\sum_k a_{A,k}=\sum_ka_{B,k}<\infty$. Such pairs of species can be further categorized by whether $\beta_A''\bm(\nAbar(n^*)\bm)=\beta_B''\bm(\nBbar(n^*)\bm)$, which is equivalently characterized by \eqref{eq:flatdoubledegenchar}.
For the case where species $A$ and $B$ are $p$- and $q$-fold convolutions of a single species $S$ in the sense that the generating functions satisfy $\zeta_A=\zeta_S^p,\, \zeta_B=\zeta_S^q$, this formula simplifies to~\cite[Eq.~(23)]{Yauk:2023wbu} with $f=p/(p+q)$ and a related volume (see section~\ref{sec:GeometricBipartition}). This shows that our species bipartition contains the single-species geometric bipartition as a special case.

\subsection{Resolving Kronecker deltas and branch transitions}
\label{sec:resolving-kdeltas}
At fixed density, the average entanglement entropy~\eqref{eq:bvgEE} takes the form
\begin{align}\label{eq:pointwise-abc}
    \braket{S_A}_N=a(n)V+b(n)\sqrt V+c(n)+o(1)\,,
\end{align}
where $a(n)$, $b(n)$ and $c(n)$ are the pointwise limits at fixed $n$. At the critical densities $\ncrit\in\ncritset$ this convergence is generally not uniform, and these functions, or their derivatives, may be discontinuous there. These features are a direct consequence of the non-analyticities in the Bianchi-Donà sum~\eqref{eq:EEanalytic}, which contains $\min$ and $\max$ functions. A coefficient has a \emph{branch transition} at $\ncrit$ if its one-sided limits differ, and a \emph{Kronecker delta} if its value at $\ncrit$ differs from the mean of the one-sided limits. In the present section and the accompanying appendix~\ref{ap:ResolvingKdelta} we discuss how these discontinuities can be \emph{resolved} by performing appropriate double scaling limits, where $n=\ncrit+\Lambda/V^{s}$, \ie we study $\braket{S_A}_N$ around $\ncrit$ in a window of a certain scale specified by the power $s$; a pedagogical example is given in appendix~\ref{ap:pedagogy}. Appendix~\ref{ap:ResolvingKdelta} mainly derives a calculation of $\braket{S_A}_N$ to an $o(1)$ error in the double scaling limit, from which we can easily read off the resolved profiles of the discontinuities; heuristically, indicator function terms are replaced with weights continuous in $\Lambda$: $1_{\ncritset}(n)\rightarrow w(\Lambda)$.

At critical densities $\ncrit$, we distinguish the two cases (see definition~\ref{def:critical-densities})
\begin{align}
    \text{(non-degenerate)}\quad &\Leftrightarrow\quad \betaA''\neq\betaB''\,,\\
    \text{(degenerate)}\quad &\Leftrightarrow\quad \betaA''=\betaB''\,.
\end{align}
At a non-degenerate generic critical density, $b$ has a Kronecker delta and $c$ a branch transition, while at a flat critical density only $c$ has a Kronecker delta; the general classification is given in proposition~\ref{prop:jumps-kdeltas} and table~\ref{tab:discontinuities} of appendix~\ref{ap:classify-jumps}. The difference between the two cases mostly matters for the scale $s$ that we need to use to specify our window around $\ncrit$. At the non-degenerate critical densities we have $s=\frac12$ and we can expand around criticality,
\begin{align}\label{eq:resolved-abc}
    \braket{S_A}_N\!=\!{}&a(\ncrit)V\!+\!b(\Lambda;\ncrit)\sqrt V\!+\!c(\Lambda;\ncrit)+o(1)\,,
\end{align}
where $b$ and $c$ are continuous in $\Lambda$. The coefficients $b$ and $c$ are not themselves the resolved profiles of $b(n)$ and $c(n)$ near $\ncrit$ since Taylor expansions in $(n-\ncrit)$ contribute to lower orders. We therefore split
\begin{subequations}\label{eq:inh-res}
\begin{align}
    b(\Lambda;\ncrit)={}&B_{\rm inh}^V(\Lambda)+B(\Lambda;\ncrit)\,,\\
    c(\Lambda;\ncrit)={}&C_{\rm inh}^V(\Lambda)+C_{\rm inh}^{\sqrt V}(\Lambda)+C(\Lambda;\ncrit)\,.
\end{align}
\end{subequations}
The \emph{inherited} terms $B_{\rm inh}^V$, $C_{\rm inh}^V$ and $C_{\rm inh}^{\sqrt V}$ are generated by these Taylor expansions, and their superscript records the order of the term from which they descend (appendix~\ref{ap:interplay} and table~\ref{tab:inherited}). They vanish at $\Lambda=0$ and resolve no discontinuity of~\eqref{eq:pointwise-abc}. The \emph{resolved profiles} $B$ and $C$ are the remaining coefficients of $\sqrt V$ and of constant order, listed in table~\ref{tab:resolved-kdeltas} and derived in appendix~\ref{ap:calc-avgEE-res}. They recover the pointwise coefficients $b(n)$ and $c(n)$ around $\ncrit$ for $\Lambda=0$ and as $\Lambda\to \pm\infty$, \eg for $b(n)$ we have
\begin{align}\label{eq:BC-limits}
\begin{split}
    B(0;\ncrit)&=b(\ncrit)\,,\\
    \lim_{\Lambda\to\pm\infty}B(\Lambda;\ncrit)&=\lim_{n\to(\ncrit)^\pm}b(n)\,,
\end{split}
\end{align}
and analogously for $C$ and $c(n)$. A branch transition is thus resolved by a profile whose limits as $\Lambda\to\pm\infty$ differ, and a Kronecker delta by a profile whose value at $\Lambda=0$ differs from the mean of these limits. Degenerate generic critical densities, and the switch of the branch selected at order $V$ in the degenerate flat case, require wider windows $n=\ncrit+\Lambda V^{-s_K}$ with $s_K<\frac12$, but do not affect the limit profile terms significantly---see table~\ref{tab:windows} and appendix~\ref{ap:calc-avgEE-res}.

\subsection{Variance}\label{sec:variance}
The variance $\left(\Delta S_A\right)^2_N\equiv\braket{S^{\,2}_A}_N-\braket{S_A}_N^2$ of the entanglement entropy for fixed $N$ is~\cite{Bianchi2019,Yauk:2023wbu}
\begin{align}
    \label{eq:variance_sum_definition}
    \left(\Delta S_A\right)^2_N=\frac{1}{d_{AB}+1}\!\left[\sum_{N_A=0}^N\varrho_{N_A}\!\left(\varphi_{N_A}^2\!+\!\chi_{N_A}\right)\!-\!\braket{S_A}_N^2\!\right]\,,
\end{align}
with $\braket{S_A}_N$, $\varrho_{N_A}$, and $\varphi_{N_A}$ defined in the previous sections, and
\begin{align}\label{eq:chi_definition}
    \begin{split}
    \chi_{N_A}={}&(d_A+d_B)\Psi'(d_>+1)-(d_{AB}+1)\Psi'(d_{AB}+1)\\
    &-\frac{(d_<-1)(d_<+2d_>-1)}{4d_>^2}\,,
    \end{split}
\end{align}
where $d_<=\min\set{d_A,d_B}$ and $d_>=\max\set{d_A,d_B}$ are the smaller and larger of the dimensions of the sector $N_A$, and $\Psi'$ is the trigamma function~\cite[Eq.~(26)]{Yauk:2023wbu}. Expanding with $\Psi'(x+1)=1/x-1/(2x^2)+O(x^{-3})$ gives
\begin{align}\label{eq:chi_asymptotic}
    \chi_{N_A}={}&\frac{d_<}{2d_>}-\left(\frac{d_<}{2d_>}\right)^2+O(d_>^{-2}+d_{AB}^{-1})\,,
\end{align}
so that $\chi_{N_A}\to\frac14$ where $d_A=d_B$ and $\chi_{N_A}$ is exponentially small where $d_A$ and $d_B$ differ exponentially. In any case, $\chi_{N_A}$ is bounded uniformly, so $\sum_{N_A=0}^N \varrho_{N_A}\chi_{N_A}=O(1)$.

The average entanglement entropy we calculated in the previous subsections is typical. Since $-\frac12\leq\varphi_{N_A}\leq\Psi(d_{AB}+1)<\ln(d_{AB}+1)$ for every sector, the square bracket in~\eqref{eq:variance_sum_definition} is at most $O(V^2)$. Hence
\begin{align}\label{eq:typicalityscaling}
   \frac{(\Delta S_A)_N}{\braket{S_A}_N}\lesssim \frac{1}{\sqrt{d_{AB}}}\xrightarrow{V\to\infty}0\,,
\end{align}
which by the asymptotics of $d_{AB}$ from~\eqref{eq:dXasymptotic} vanishes exponentially fast, valid given $n\in(0,\nXmax)$. In particular, this bound does not require every sector to be large; indeed, for binary species at $N=V$, the sector $N_A=0$ has $d_Ad_B=1$. The same conclusion was reached in~\cite{Bianchi2019} assuming that every sector is large, while we simply require  $n\in(0,\nXmax)$.

Having established that $\braket{S_A}_N$ is typical, we can take one further step and compute the standard deviation $(\Delta S_A)_N$ at leading order in $V$ along with some prefactors. Since the $\chi_{N_A}$ term contributes $O(1)$ to the bracket in~\eqref{eq:variance_sum_definition}, we ignore it in the following calculations. The sum $\sum_{N_A=0}^N \varrho_{N_A}\varphi_{N_A}^2$ is evaluated with the methods of the previous sections. We obtain for $\bbA\leq \bbB$ (without loss of generality)
\begin{subequations}
\begin{align}\label{eq:phisquaredint}
    \int_\R \varrho(\delta) \varphi(\delta)^2\dd \delta={}&V\beta'(n)^2\int_\R\varrho(\delta) \delta^2\dd \delta -V^2\bbA^2\nonumber\\{}&+2V\bbA\int_\R \varrho(\delta) \varphi(\delta)\dd \delta +o(V)\,,\\
    ={}&\frac{V}{\kappa}\beta'(n)^2 -V^2\bbA^2\nonumber\\{}&+2V\bbA \braket{S_A}_N+o(V)\,,
\end{align}
\end{subequations}
while we use~\eqref{eq:bvgEE} to expand
\begin{align}
    \braket{S_A}_N^2={}&\frac{2V}{\pi\kappa}\beta'(n)^2 1_{\ncritset}(n)-V^2\bbA^2\nonumber\\
    &+2V\bbA\braket{S_A}_N+o(V)\,.
\end{align}
Combining these expressions with the asymptotics of $d_{AB}$ from~\eqref{eq:dXasymptotic}, thus yields
\begin{align}\label{eq:BVGvar}
    \begin{split}
        \left(\Delta S_A\right)^2_N={}\frac{\sqrt{2\pi}\beta'(n)^2}{\kappa|\beta''(n)|^{1/2}}\left(1-\frac2\pi  1_{\ncritset}(n)\right)&V^{3/2}\ee^{-V\beta(n)}\\ +o\bm(&V^{3/2}\ee^{-V\beta(n)}\bm)\,.
    \end{split}
\end{align}
This sharpens the bound~\eqref{eq:typicalityscaling} by $V^{-1/2}$. The variance given here disappears at leading order when $n=n^*$, so one would have to go to lower orders to obtain an explicit value for the variance, where the $\chi_{N_A}$ term~\eqref{eq:chi_asymptotic} need no longer be negligible. 

\section{Case studies}\label{sec:case studies}
Our formula~\eqref{eq:bvgEE} for the average entanglement entropy $\braket{S_A}_N$ applies to general systems, once $\beta_A(n)$ and $\beta_B(n)$ are found. Here, we consider several case studies to build intuition and perform numerical checks. We first look at the simplest non-trivial systems, before considering the physical scenario of the Jaynes--Cummings--Hubbard model~\cite{Hartmann_2006,Greentree_2006,Angelakis_2007} and how to recover the case of geometric bipartitions.

\subsection{Testbed model: \texorpdfstring{$\zeta_A(z)=1+3z$ and $\zeta_B(z)=2+2z$}{zeta A(z)=1+3z and zeta B(z)=2+2z}}\label{sec:testbedmodelA}
In order to numerically verify our general formula, we wanted to come up with the simplest model characterized by $\zeta_A$ and $\zeta_B$, which shows all the different features and special cases found in the general formula~\eqref{eq:bvgEE}. To keep the numerics feasible, we focused on the smallest non-trivial particle types, where both of these polynomials are of degree $1$. We further allowed ground-state degeneracy, which may be less physical, but the goal here is to test the framework. Based on the characterization in appendix~\ref{ap:criticaldensity}, a flat critical density $n=n^*$ exists if and only if $\zeta_A(1)=\zeta_B(1)<\infty$. The smallest possible polynomials with this property are:
\begin{itemize}
    \item $\zeta_A(z)=1+z$ and $\zeta_B(z)=1+z$: Too trivial, as this is just the case of two identical species that can be related to previous work with $f=1/2$.
    \item $\zeta_A(z)=1+2z$ and $\zeta_B(z)=2+z$: This model has a degenerate flat critical density at $n^*$, with no other critical densities. An interesting family of generating functions with particle-hole-like symmetry based on this is introduced in section~\ref{sec:PH-family}. 
    \item $\zeta_A(z)=1+3z$ and $\zeta_B(z)=2+2z$: This model has a non-degenerate generic critical density $\ncrit=0.873994\dots$ and a non-degenerate flat critical density $n^*=5/4$. It is therefore an ideal testing ground.
\end{itemize}

We now analyze the model with $\zeta_A(z) = 1 + 3z$ and $\zeta_B(z) = 2 + 2z$ in detail. By \eqref{eq:dimension_saddle_integral} and the binomial formula, the combinatorics for subsystems $A$ and $B$ are
\begin{align} \label{eq:SubDimTestBed}
    \hspace{-2mm} d_A(V,N_A) = \binom{V}{N_A} 3^{N_A}\,, \,\,d_B(V,N_B) = \binom{V}{N_B} 2^V\,,
\end{align}
respectively. For the total Hilbert space dimension we can use~\eqref{eq:dNsumdAdB}
\begin{align} \label{eq:dNTotTestBed}
    d_N(V,N) = \sum_{N_A=\max\set{0,N-V}}^{\min\set{V,N}} \binom{V}{N_A}3^{N_A} \binom{V}{N-N_A}2^V,
\end{align}

\begin{figure}[tp]
\centering
\includegraphics{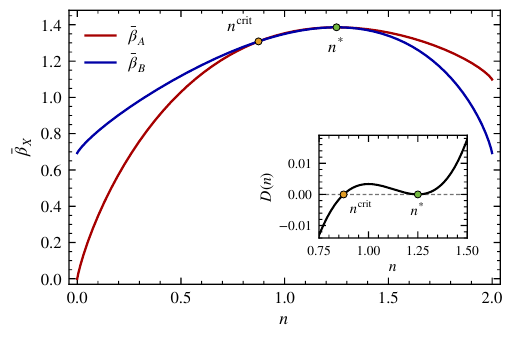}
\caption{\textit{Saddle-point entropy densities of the testbed model $\zeta_A(z)=1+3z$ and $\zeta_B(z)=2+2z$.} The smaller of $\bbA$ and $\bbB$ is the volume-law coefficient of the entanglement entropy. They cross at the generic critical density $\ncrit$, where the selected species changes, and touch at the flat critical density $n^*$, where it does not. The inset shows their difference $D(n)=\bbA-\bbB$.}
  \label{fig:CritDiaTestBed}
\end{figure}

\begin{figure*}[t]
\centering
\includegraphics{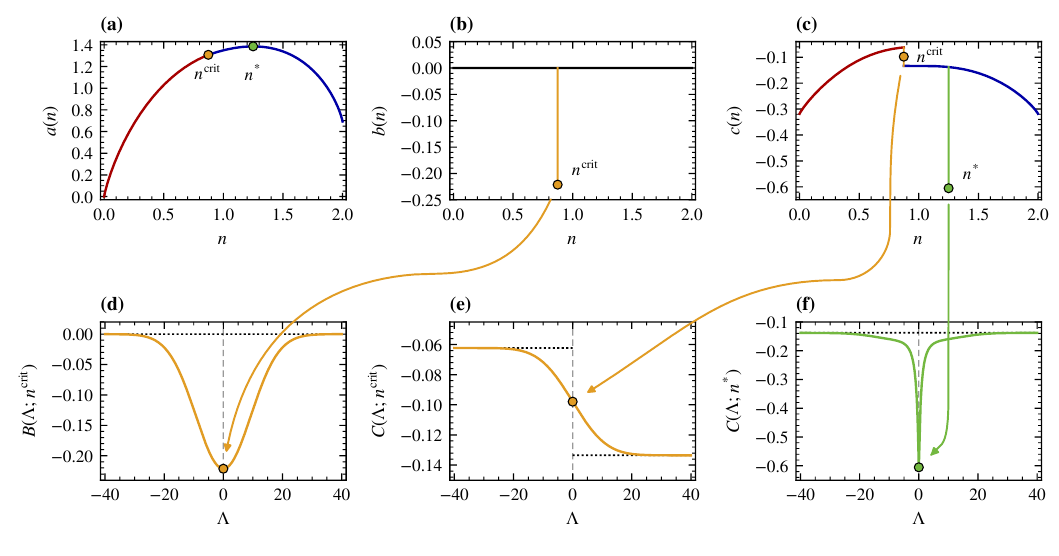}
\caption{\textit{Asymptotic coefficient functions and their resolved critical profiles for the testbed model.} \textbf{(a)}--\textbf{(c)}~Coefficients $a(n)$, $b(n)$ and $c(n)$ of the expansion $\braket{S_A}=aV+b\sqrt{V}+c+o(1)$, given by~\eqref{eq:testbedA-a},~\eqref{eq:testbedA-b} and~\eqref{eq:testbedA-c}, with the generic critical density $\ncrit$ and the flat critical density $n^*$ marked. \textbf{(d)}~Resolved profile of the Kronecker delta of $b(n)$ at $\ncrit$, in the double scaling limit $n=\ncrit+\Lambda/\sqrt{V}$. \textbf{(e)}~Resolved profile of the branch transition of $c(n)$ at $\ncrit$. \textbf{(f)}~Resolved profile of the Kronecker delta of $c(n)$ at $n^*$, in the double scaling limit $n=n^*+\Lambda/\sqrt{V}$.}
    \label{fig:asym-coeffs}
\end{figure*}

We solve the saddle point equation~\eqref{eq:zetaequivsaddle}
\begin{align}
    z_{A,0}(n_A)={}&\frac{n_A}{3-3n_A}, \\z_{B,0}(n_B)={}&\frac{n_B}{1-n_B}, \\
    z_{0}(n) ={}&\frac{2n-2+\sqrt{n^2-2n+4}}{6-3n}\,,
\end{align}
so we may use lemma~\ref{lem:combinedsaddle} to find $\nAbar$ and $\nBbar$:
\begin{align}
    \bar{n}_A(n) &= \frac{n + 2 - \sqrt{n^2 - 2n + 4}}{2}, \\
    \bar{n}_B(n) &= \frac{n - 2 + \sqrt{n^2 - 2n + 4}}{2},
\end{align}
and the entropy density functions by~\eqref{eq:betapsi},
\begin{align}
    \beta_A(n_A) &= -(1-n_A)\ln(1-n_A) - n_A \ln n_A + n_A \ln 3\,, \\
    \beta_B(n_B) &=  - (1-n_B)\ln(1-n_B) - n_B \ln n_B +\ln 2\,.
\end{align}
With these definitions, we can plot the criticality diagram as in figure~\ref{fig:CritDiaTestBed}, where we can clearly see the two intersection points of $\beta_A(\bar n_A)$ and $\beta_B(\bar n_B)$, which are two critical densities. We have a flat critical density $n^*=5/4$ and a generic critical density $\ncrit = 0.873994...$, which is the transcendental solution of the relation
\begin{align}
    \beta_A \bm(\bar n_A(n) \bm)=\beta_B\bm(\bar n_B(n)\bm)\,.
\end{align}
Substituting the saddle densities and entropy density functions above into
the general result~\eqref{eq:bvgEE}, with $\Pi_*$ of~\eqref{eq:Pistar} at the
unequal-curvature flat critical point, gives a closed asymptotic expression
for the testbed model. Away from
the two critical densities, the constant-order coefficient is $c_A(n)$ or $c_B(n)$ of~\eqref{eq:cX}. The entropy can now be written in the form
\begin{align}
    \braket{S_A}_N
    ={}&a(n)V+b(n)\sqrt{V}+c(n)+o(1)\,,
    \label{eq:testbedA-asymptotic}
\end{align}
where the volume-law coefficient is
\begin{align}
    a(n)
    ={}&
    \begin{cases}
        \beta_A \bm(\bar n_A(n) \bm), & 0<n\leq\ncrit\\
        \beta_B\bm(\bar n_B(n)\bm), & \ncrit<n<2
    \end{cases}\,,
    \label{eq:testbedA-a}
\end{align}
the square-root coefficient is
\begin{align}
    b(n)=-\sqrt{\frac{2}{\pi \kappa(\ncrit)}}|\beta'(\ncrit)|\delta_{n,\ncrit}\,,
    \label{eq:testbedA-b}
\end{align}
and the constant coefficient is
\begin{align}
    c(n)
    ={}&
    \begin{cases}
        c_A(n) & 0<n<\ncrit\\
        \begin{aligned}
            &\frac14\left(1+\ln r_A(\ncrit) \right.\\ 
            &\,\,\,\,\left.-2\ln\bm(1+r_A(\ncrit)\bm)\right)
        \end{aligned} & n=\ncrit\\
        \Theta(3/4) & n=n^*\\
        c_B(n)
        & \text{else}\\
    \end{cases}\,.
    \label{eq:testbedA-c}
\end{align}
At the flat critical density, we have $\nAbar=3/4$, $\nBbar=1/2$, and
$\bbA=\bbB=\ln 4$. Moreover,
$\beta_A''(3/4)=-16/3$ and
$\beta_B''(1/2)=-4$, so that $\rsc=3/4$ and the
unequal-curvature result~\eqref{eq:Theta} applies. We display the expansion coefficients from~\eqref{eq:testbedA-asymptotic} in
figure~\ref{fig:asym-coeffs}.

\begin{figure*}[t]
\centering
\includegraphics{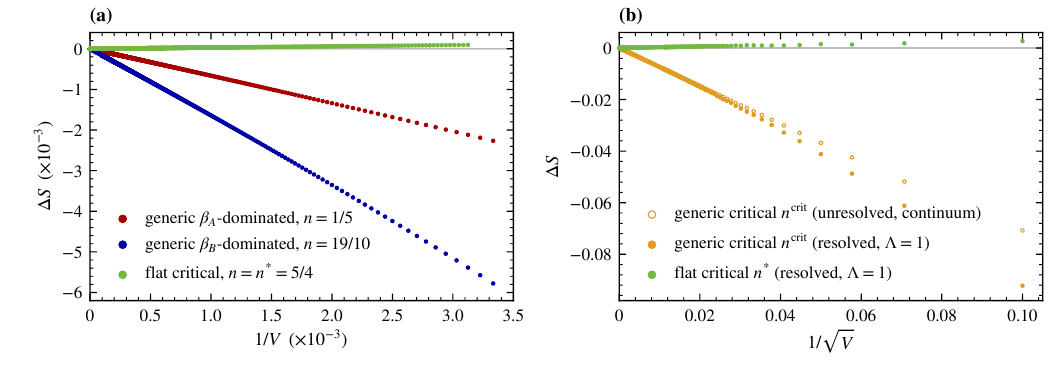}
\caption{\textit{Finite-size scaling of the error $\Delta S=S_{\mathrm{num}}-S_{\mathrm{app}}$ in the testbed model.} The numerically computed average $S_{\mathrm{num}}$ is compared with the asymptotic approximation $S_{\mathrm{app}}$ in six representative regimes, in each of which $\Delta S$ extrapolates linearly to zero. \textbf{(a)}~Fixed generic and flat-critical densities, against $1/V$. \textbf{(b)}~Generic critical density with and without resolving its Kronecker delta and branch transition, and the resolved flat critical density, against $1/\sqrt{V}$. The resolved cases are taken at $\Lambda=1$ in the window $n=\ncrit+\Lambda/\sqrt{V}$.}
    \label{fig:fssa-summary}
\end{figure*}

At the generic critical density $\ncrit\neq n^*$, we see a Kronecker delta at order $\sqrt{V}$ and a branch transition at constant order, and at $n^*$ a Kronecker delta at constant order (see also proposition~\ref{prop:jumps-kdeltas}). We wish to resolve these contributions by finding their limit profiles as we approach $n\to \ncrit$ or $n\to n^*$. Both critical densities are non-degenerate, $\betaA''\neq\betaB''$, so the scale that resolves the Kronecker deltas and the branch transition is set by letting $n=\ncrit+\Lambda/\sqrt{V}$ for both $\ncrit\in \set{0.873994\dots, 5/4}$ by~\eqref{eq:sK} and table~\ref{tab:windows}. On this scale, the average entanglement given by~\eqref{eq:gencritres1} and~\eqref{eq:flatResAVGlog} can be expressed as
\begin{align}
    \braket{S_A}_N=V a(\ncrit) + b( \Lambda;\ncrit)\sqrt{V}+c(\Lambda;\ncrit)\,.
\end{align}
As in~\eqref{eq:inh-res}, $b(\Lambda;\ncrit)$ and $c(\Lambda;\ncrit)$ contain, besides the resolved profiles, inherited terms obtained from Taylor expanding the terms of order $V$ and $\sqrt V$ around the critical density. Following table~\ref{tab:resolved-kdeltas}, the resolved profiles are
\begin{align}\label{eq:BCCtestbed}
    B(\Lambda;\ncrit)=&{}-\sqrt{\frac{2}{\pi \kappac}}\,|\beta'(\ncrit)|\,w_1\,,\displaybreak[1]\\
    C(\Lambda;\ncrit)=&{}(1-w_0)\,c_A(\ncrit)+w_0\,c_B(\ncrit)\nonumber\\
    &-\tau\sqrt{\frac{\kappac}{2\pi}}\,\dzg\,w_1\,,\displaybreak[1]\\
    C(\Lambda;n^*)=&{} \frac12\left(\frac{1+(\rsc-1)w_2^*(\Lambda)}{1+\rsc}    \right.\nonumber\\
    &\quad -\ln(1+1/\rsc)\nonumber \\
    &\quad \left. -w_0^*(\Lambda)\ln \rsc\right) +P(\Lambda)\,.
\end{align}
Here, $w_0$ and $w_1$ are the weights~\eqref{eq:XiArAetc} with $\kappa\to\kappac$ and $\dzg=\Lambda(\frac12-\nAbc')$ from~\eqref{eq:deltac1}, and $\tau=-0.0025\ldots$ is given by~\eqref{eq:tau}. This is the generic entry of table~\ref{tab:resolved-kdeltas}, and the term proportional to $\tau$ is barely visible for the testbed model. At $n^*$, the weights $w_0^*(\Lambda)$ and $w_2^*(\Lambda)$ are defined in~\eqref{eq:flatweights}, and $P(\Lambda)$, the resolved Page term, is defined in~\eqref{eq:near-flat-page-correction}. Since $\beta_A''(3/4)=-16/3<\beta_B''(1/2)=-4$, these are used with $A\leftrightarrow B$ exchanged, so that $\rsc=3/4$.

The three profiles are shown in panels (d)--(f) of figure~\ref{fig:asym-coeffs}. In the language of~\eqref{eq:jump-kronecker}, they are of the following types:
\begin{itemize}
    \item $B(\Lambda;\ncrit)$ vanishes as $|\Lambda|\to\infty$. It resolves the Kronecker delta at order $\sqrt V$, with weight $K[b]=B(0;\ncrit)=-0.2212\dots$.
    \item $C(\Lambda;\ncrit)$ interpolates between $c_A(\ncrit)$ as $\Lambda\to-\infty$ and $c_B(\ncrit)$ as $\Lambda\to\infty$, and $C(0;\ncrit)$ is their mean. It resolves the branch transition at constant order, with jump $J[c]=c_B(\ncrit)-c_A(\ncrit)=-0.0713\dots$, and carries no Kronecker delta.
    \item $C(\Lambda;n^*)$ tends to $c_B(n^*)$ as $\Lambda\to\pm\infty$. It resolves the Kronecker delta at constant order, with weight $K[c]=\Theta(3/4)-c_B(n^*)=-0.4674\dots$, of which the Page term contributes $P_{\rm flat}=-0.4492\dots$.
\end{itemize}
The remaining terms of $b$ and $c$ are the inherited terms of~\eqref{eq:inh-res}, which vanish at $\Lambda=0$.

We check our analytical predictions by performing a finite-size scaling analysis of how the difference $\Delta S=S_{\mathrm{num}}-S_{\mathrm{app}} = o(1)$ behaves in four representative regimes suggested by the criticality diagram of the testbed figure~\ref{fig:CritDiaTestBed}: a generic $\beta_A$-dominated point $n=\tfrac{1}{5}$, a generic critical point $n=\ncrit=0.873994...$, a flat critical point $n=n^*=\tfrac{5}{4}$, and a generic $\beta_B$-dominated point $n=\tfrac{19}{10}$. Here, $S_{\mathrm{num}}$ denotes the exact numerically evaluated entropy, computed from the integer-sector Haar formula, except in the unresolved generic-critical case, where we use a normalized continuum integral with saddle-point approximations of the dimensions at $n=\ncrit$. For the unresolved cases, $S_{\mathrm{app}}$ is~\eqref{eq:testbedA-asymptotic} truncated after the constant term. For the resolved generic- and flat-critical cases, it is based on~\eqref{eq:gencritres1} and~\eqref{eq:flatResAVGlog}, respectively, with the Page term~\eqref{eq:near-flat-page-correction} included in the latter. The numerical evaluation retains the finite-$V$ crossing corrections inside the Gaussian weights in the generic-critical case and uses the unexpanded Gaussian integral for the logarithmic term in the flat-critical case, thereby retaining terms beyond constant order.

For the fixed-density calculations, we choose $V$ so that $N=nV$ is an integer: multiples of $10$ for $n=1/5,19/10$, and multiples of $4$ for $n=5/4$. For the resolved critical sequences, we target $\Lambda=1$ and choose
\begin{align}
N_V&=\operatorname{Round}(n_0V+\sqrt{V}),\\
n_V&=\frac{N_V}{V},\qquad \Lambda_V=\frac{N_V-n_0V}{\sqrt{V}}\,,
\end{align}
where $n_0=\ncrit$ or $n^*$. The numerical entropy and the subtracted resolved approximation use the same actual density $n_V$, with the scaling functions evaluated at $\Lambda_V$.

For the two generic $\beta_A$ or $\beta_B$-dominated cases and the flat critical case $n=n^*$ without resolving its Kronecker delta, the residuals are plotted against $\frac{1}{V}$, and in each case they extrapolate linearly to the origin, as shown in figure~\ref{fig:fssa-summary}(a). For the generic critical point $n=\ncrit$, both the unresolved and resolved treatments of the Kronecker delta and branch transition are plotted against $\frac{1}{\sqrt{V}}$, and both again approach the origin under linear extrapolation, as shown in figure~\ref{fig:fssa-summary}(b). The resolved flat-critical case is also plotted against $\frac{1}{\sqrt{V}}$ in figure~\ref{fig:fssa-summary}(b), and it is approaching the origin as expected. Together, these checks therefore confirm that the asymptotic formula in~\eqref{eq:bvgEE} captures the expected behavior for large $V$, \ie $\Delta S = o(1)$.

\subsection{Particle-hole-like symmetry family: \texorpdfstring{$\zeta_A(z) = a+bz$ and $\zeta_B(z) = b+az$}{zeta A(z)=a+bz and zeta B(z)=b+az}}\label{sec:PH-family}

\begin{figure*}[t]
\centering
\includegraphics{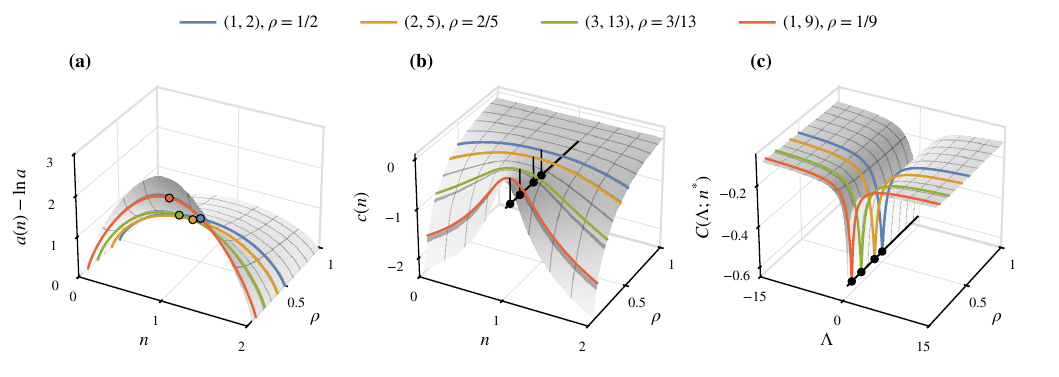}
\caption{\textit{Asymptotic coefficient functions and their resolved flat-critical profile in the particle-hole-like family.} For $\zeta_A(z)=a+bz$ and $\zeta_B(z)=b+az$, the functions $a(n)-\ln a$, $c(n)$ and $C(\Lambda;n^*)$ depend on $(a,b)$ only through the asymmetry ratio $\rho=a/b$, so each is drawn as a surface over $\rho$, on which the pairs $(a,b)$ of the legend are curves. \textbf{(a)}~Volume-law coefficient~\eqref{eq:PHfam-a}; the markers identify the flat critical density $n^*=1$. \textbf{(b)}~Constant-order coefficient~\eqref{eq:PHfam-cgen}, whose value at $n^*$ (black line) lies $\frac12$ below its limit as $n\to n^*$: a Kronecker delta. \textbf{(c)}~Resolved profile~\eqref{eq:PHfam-C} of this Kronecker delta in the double scaling limit $n=n^*+\Lambda/\sqrt{V}$, which narrows slowly as $\rho$ decreases.}
    \label{fig:PHasym-coeffs}
\end{figure*}
In the following, we investigate the family of generating functions,
\begin{align}
    \zeta_A(z)=az^k+bz^l\,,\qquad
    \zeta_B(z)=bz^k+az^l\,,
\end{align}
where $a,b\in\mathbb{Z}_{+},a< b$ and $k,l \in \Nat, k \neq l$. This family of generating functions has the special property that at the flat critical density $n^*$ it has $\betaA''=\betaB''$, so that $n^*$ is a degenerate flat critical density.

Without loss of generality we take $a<b$, since $a>b$ is obtained by exchanging $A\leftrightarrow B$. The case $a=b$ corresponds to two identical species and reduces to the $f=\frac{1}{2}$ geometric-bipartition case discussed in section~\ref{sec:GeometricBipartition}. We can also assume that $k=0$ and $l=1$, since we require the existence of a vacuum and aperiodicity, respectively. The exact fixed-particle-number dimensions are
\begin{align}
    d_A(V,N_A)={}&\binom{V}{N_A}a^{V-N_A}b^{N_A}\,,\\
    d_B(V,N_B)={}&\binom{V}{N_B}b^{V-N_B}a^{N_B}\,.
\end{align}
The combined generating function according to~\eqref{eq:zetaIDs} is
\begin{align}
    \zeta(z)=\zeta_A(z)\zeta_B(z)=ab+(a^2+b^2)z+abz^2\,,
\end{align}
and satisfies
\begin{align}
    \zeta_B(z)=z\,\zeta_A(z^{-1})\,,\qquad
    \zeta(z)=z^2\zeta(z^{-1})\,,
\end{align}
which generate the particle-hole-like symmetry of this model: exchanging particles and holes maps species $A$ onto species $B$.

Introducing the asymmetry ratio $\rho=\frac{a}{b}\in(0,1)$, we solve the saddle point equation~\eqref{eq:zetaequivsaddle} as
\begin{align}
    z_{A,0}(n_A)={}&\frac{\rho n_A}{1-n_A}\,,\\
    z_{B,0}(n_B)={}&\frac{n_B}{\rho(1-n_B)}\,,\\
    z_0(n)={}&\frac{(\rho^2+1)(n-1)+\mathcal{R}(n)}{2\rho(2-n)}\,,\\
    \mathcal{R}(n):={}&\sqrt{4\rho^2+(1-\rho^2)^2(n-1)^2}\,,
\end{align}
where we have introduced $\mathcal{R}(n)$ to shorten $z_0$. We can use lemma~\ref{lem:combinedsaddle} to find $\nAbar$ and $\nBbar$ as
\begin{align}
    \nAbar(n)={}&\frac12\left(n+\frac{\rho^2+1-\mathcal{R}(n)}{1-\rho^2}\right),\\
    \nBbar(n)={}&\frac12\left(n-\frac{\rho^2+1-\mathcal{R}(n)}{1-\rho^2}\right),
\end{align}
and the entropy density functions follow from~\eqref{eq:betapsi} as
\begin{align}
    \beta_A(n_A)={}&-(1-n_A)\ln(1-n_A)-n_A\ln n_A\nonumber\\
    &-n_A\ln\rho+\ln a\,,\\
    \beta_B(n_B)={}&-(1-n_B)\ln(1-n_B)-n_B\ln n_B\nonumber\\
    &+n_B\ln\rho+\ln b\,.
\end{align}
These satisfy $\beta_A(x)=\beta_B(1-x)$. Together with the palindromic symmetry of $\zeta$, this gives
\begin{align}
    z_0(2-n)=z_0(n)^{-1}\,,\quad
    \nAbar(2-n)=1-\nBbar(n)\,,
\end{align}
and hence
\begin{align}
    \hspace{-2mm}\beta_A\bm(\nAbar(2-n)\bm)\!=\!\beta_B\bm(\nBbar(n)\bm)\,,\,
    D(2-n)\!=\!-D(n)\,,
\end{align}
where $D(n)=\bbA-\bbB$ is the difference~\eqref{eq:Ddef}.

The symmetry guarantees a crossing at $n=1$. It is also the unique critical density. Since $\zeta_A(1)=\zeta_B(1)=a+b$, it is the flat density $n^*=1$ defined in appendix~\ref{ap:saddlepointanalysis}, so that the family has a flat critical density and no generic critical density. We may also compute $\betaA''=\betaB''=-\frac{(\rho+1)^2}{\rho}$, so this family therefore realizes the degenerate flat critical case discussed in appendix~\ref{ap:calc-flat-bvgEE-res}.

\begin{figure*}[t]
\centering
\includegraphics{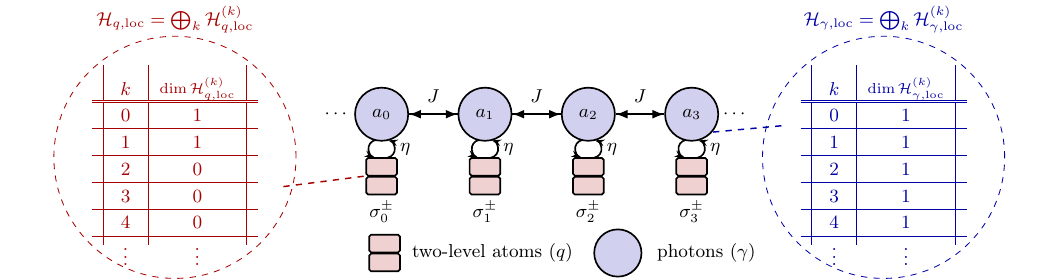}

\vspace{-0.10cm}
\caption{\textit{Local degrees of freedom and excitation sectors of the Jaynes--Cummings--Hubbard chain.} Each site $j$ hosts a cavity mode $\hat a_j$ (species $\gamma$) and a two-level atom $\hat\sigma_j^\pm$ (species $q$). The photon hopping $J$ and the atom--cavity coupling $\eta$ of the JCH Hamiltonian conserve the total excitation number, which is the only property of the dynamics that we use. The side tables give the local multiplicities $\dim\mathcal H_{X,\mathrm{loc}}^{(k)}$ of $k$ excitations.}
\label{fig:jch-schematic-gamma-q}
\end{figure*}

We now focus on the general entropy result~\eqref{eq:bvgEE}. The volume-law coefficient is simply
\begin{align}
    a(n)={}&
    \begin{cases}
        \beta_A\bm(\nAbar(n)\bm), & 0<n<1\\
        \ln(a+b), & n=1\\
        \beta_B\bm(\nBbar(n)\bm), & 1<n<2
    \end{cases}\,,
    \label{eq:PHfam-a}
\end{align}
for $a<b$, and satisfies $a(2-n)=a(n)$. Away from $n=1$, the constant-order coefficient follows directly from the non-critical part of~\eqref{eq:bvgEE},
\begin{align}
    c(n)={}&-\frac12\delta_{n,1}+
    \begin{cases}
        c_A(n)\,, & n\leq 1 \\
        c_B(n)\,, & 1<n<2 
    \end{cases}\,.
    \label{eq:PHfam-cgen}
\end{align}
The Kronecker-delta contribution comes from the Page correction term, so we use the scaling of the inner window discussed in appendix~\ref{ap:kdeltares-pert-flat},
\begin{align}
    n=1+\frac{\Lambda}{\sqrt{V}}\,.
\end{align}
For the equal-curvature case, the resolved Page term is given by~\eqref{eq:PageResDegen}. Using
\begin{align}
    \kappac=\frac{2(\rho+1)^2}{\rho}\,,\qquad
    \beta''(1)=-\frac{(\rho+1)^2}{2\rho}\,,
\end{align}
we obtain the resolved Page term
\begin{align}
    P(\Lambda)={}&-\frac12\operatorname{erfcx}\!\left(\frac{1+\rho}{2\sqrt{\rho}}|\Lambda|\right)\,,\label{eq:PHfam-P}
\end{align}
and define the total resolved profile
\begin{align}
    C(\Lambda;n^*):={}&\frac12\left(\frac12-\ln 2\right)+P(\Lambda)\,.
    \label{eq:PHfam-C}
\end{align}
This is the flat-critical entry of table~\ref{tab:resolved-kdeltas} at $\rsc=1$; it does not include the inherited terms of~\eqref{eq:inh-res}.
It interpolates between
\begin{align}
    C(0;1)={}&-\frac14-\frac12\ln2\,,\\
    \lim_{|\Lambda|\to\infty}C(\Lambda;1)={}&\frac14-\frac12\ln2\,.
\end{align}

In summary, we have shown that $\beta_X-\ln a$ and $\nXbar$ depend on $(a,b)$ only through $\rho$, and thus so do $a(n)-\ln a$, $c(n)$ (since $c$ consists of derivatives of $\beta_X$ evaluated at the saddle $\nXbar$), and $C(\Lambda;n^*)$. This is clearly seen in figure~\ref{fig:PHasym-coeffs} as a surface plot over $(n,\rho)$ with particular configurations $(a,b)$ plotted as lines on the surface.

\subsection{Jaynes--Cummings--Hubbard model/coupling of two-level atoms and bosons}\label{sec:JCH}

The Jaynes--Cummings--Hubbard (JCH) model describes the interaction between two-level systems and quantized cavity modes (\ie atoms and bosons) on a lattice. Figure~\ref{fig:jch-schematic-gamma-q} shows the schematic. This model is widely applied in atomic physics, quantum optics, and solid-state quantum information circuits~\cite{Hartmann_2006,Greentree_2006,Angelakis_2007}. With $\hbar=1$, the Hamiltonian for the JCH model is
\begin{align}
    \begin{split}
    \hat H ={}& \sum_{j=1}^V \omega_c \hat a_j^\dagger \hat a_j 
    + \sum_{j=1}^V \omega_a \hat\sigma_j^+ \hat\sigma_j^- \\
    & + J \sum_{j=1}^{V-1} \left( \hat a_{j+1}^\dagger \hat a_j + \hat a_j^\dagger \hat a_{j+1} \right)  \\
     &  + \eta \sum_{j=1}^V \left( \hat a_j \hat\sigma_j^+ + \hat a_j^\dagger \hat\sigma_j^- \right),
    \end{split}
\end{align}
where $\hat\sigma_j^\pm$, $\hat a_j^\dagger$ and $\hat a_j$ are creation and annihilation operators for the two-level atom and photonic cavity in the $j$-th cavity respectively. The tunneling rate between adjacent cavities is $J$, while the photon--atom coupling strength is given by the vacuum Rabi frequency $\eta$. The cavity and atomic transition frequencies are denoted by $\omega_c$ and $\omega_a$, respectively. We denote the atomic subsystem by $q$ (atom excitation) and the bosonic subsystem by $\gamma$ (photons). The total excitation number of the JCH model is conserved, since the total number operator $\hat{N}=\hat{N}_q+\hat{N}_{\gamma}=\sum_j (\hat{\sigma}_j^+\hat{\sigma}^-_j+\hat{a}_j^\dagger\hat{a}_j)$ commutes with the Hamiltonian. For our statistical consideration, we only use the Hilbert space decomposition and notions of particle numbers to study the typical entanglement entropy between atoms and photons in a Haar-random state with fixed total excitation number $\hat{N}$.

We have the subsystem Hilbert-space dimensions
\begin{align}
    d_q(V,N_q)=\binom{V}{N_q},\quad d_\gamma(V,N_\gamma)
    =\binom{V+N_\gamma-1}{N_\gamma}.
    \label{eq:JCH-subsystem-dimension}
\end{align}
The exact Hilbert-space dimension of the system is 
\begin{align}
    d_N(V, N) = \sum_{m=0}^{\min\set{V, N}} \binom{V}{m} \binom{V + N - m - 1}{N - m}.
\end{align}
We obtain the generating functions
\begin{align}
    \zeta_q(z)=1+z, \quad \zeta_\gamma(z)=\frac{1}{1-z}, \quad
    \zeta(z)=\frac{1+z}{1-z}
\end{align}
and from the saddle point equation~\eqref{eq:zetaequivsaddle}, we compute the saddle points for the subsystems,
\begin{align}
    z_{q,0}(n_q) \!=\! \tfrac{n_q}{1-n_q}\,,\,
    z_{\gamma,0}(n_\gamma) \!=\! \tfrac{n_\gamma}{1+n_\gamma}\,,\,
    z_0(n)\!=\!\tfrac{s-1}{n}\,,
\end{align}
where we have introduced $s=\sqrt{1+n^2}$. We may use lemma~\ref{lem:combinedsaddle} to find $\nAbar$ and $\nBbar$ as
\begin{align}
    \bar{n}_{q}={}&\frac{n-s+1}{2}\quad\text{and}\quad \bar{n}_{\gamma}={}&\frac{n+s-1}{2}\,.
\end{align}
Then, the entropy density functions are
\begin{align}
    \beta_q(n_q) &= (n_q-1) \ln(1-n_q)-n_q \ln n_q\,,\\
    \beta_\gamma(n_\gamma) &= (n_\gamma+1) \ln(1+n_\gamma)-n_\gamma \ln n_\gamma\,.
\end{align}

\begin{figure}[tp]
\centering
\includegraphics{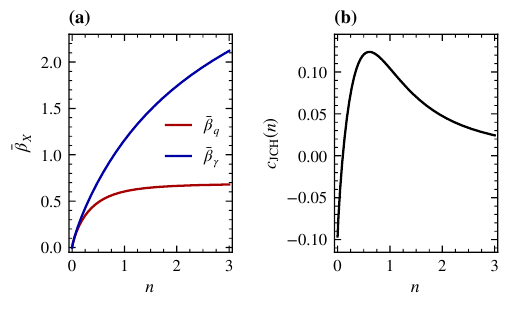}
\caption{\textit{Saddle-point entropy densities and asymptotic coefficients of the JCH model.} \textbf{(a)}~Saddle-point entropy densities $\bar\beta_q$ of the atoms and $\bar\beta_\gamma$ of the photons at total excitation density $n$. Since $\bar\beta_q<\bar\beta_\gamma$ at every $n$, the atomic branch gives the volume-law coefficient and there are no critical densities. \textbf{(b)}~Constant-order coefficient $c_{\mathrm{JCH}}(n)$ of~\eqref{eq:JCH-average-final}.}
  \label{fig:beta-both}
\end{figure}

Note that with the physical constraint on local excitation levels of atoms and photons, at the shared saddle point, the photonic subsystem has a larger entropy density and hence an exponentially larger Hilbert-space dimension in the thermodynamic limit, as can be seen in figure~\ref{fig:beta-both}. Specifically, initially the entropy densities of the two subsystems behave similarly because for each individual site, the first excitation added to the site provides the same total local dimensions. However, as the particle filling ratio $n$ continues to increase, the individual sites of the atomic subsystem are capped at one excitation, and its corresponding entropy density function begins to behave asymptotically. In comparison, the entropy density for the photonic subsystem grows without bound.

In fact, we can find by direct computation that $\beta_q\bm(\bar{n}_{q}(n)\bm)<\beta_\gamma\bm(\bar{n}_{\gamma}(n)\bm)$ for every density $n$. Consequently, the JCH model has no critical structure and consists only of generic densities $n$. Furthermore, since the photonic subsystem can support an infinite number of particles, there is no flat density $n^*$ for this system. Substituting the JCH saddle-point data into the generic branch of \eqref{eq:bvgEE}, we obtain
\begin{align}
    \kappa_{\rm JCH} &= \frac{4s}{n},\\
    r_{q} &= 1 + 2n\left(n + s\right),\\
    \Xi_{q} &= \frac{n}{2s^2}.
\end{align}
We obtain the final expression for the average entanglement entropy of the JCH model
\begin{align}
    \braket{S_q}_N
    ={}&V \bar{\beta}_q+c_{\rm JCH}(n) +o(1)\,.
    \label{eq:JCH-average-final}
\end{align}
with $c_{\rm JCH}(n)=\frac12[\frac{1}{1+r_q}+\ln\frac{r_q}{1+r_q}+\Xi_q\beta'(n)]$, shown in figure~\ref{fig:beta-both}.

\subsection{Geometric bipartitions}\label{sec:GeometricBipartition}
\begin{figure}[t]
\centering
\includegraphics{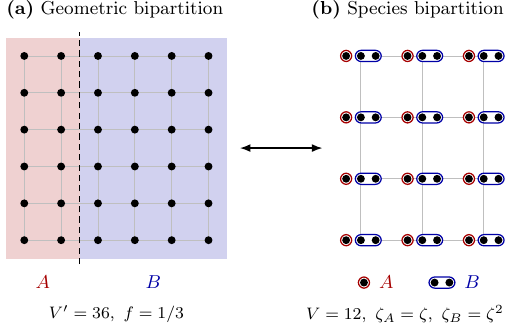}
\caption{\textit{Correspondence between geometric and species bipartitions.} \textbf{(a)}~Geometric bipartition of $V'$ sites, each hosting a single species with generating function $\zeta$, into a fraction $f=p/(p+q)$ of sites in $A$ and the rest in $B$, here with $p=1$ and $q=2$. \textbf{(b)}~Equivalent species bipartition of $V=V'/(p+q)$ sites, each hosting the composite species $\zeta_A=\zeta^p$ and $\zeta_B=\zeta^q$, which group $p$ sites of $A$ and $q$ sites of $B$ from (a).}
\label{fig:geometric-vs-species}
\end{figure}

We want to show that a geometric bipartition of a system containing a single particle species considered in~\cite{Yauk:2023wbu} is a special case of a species bipartition. Although the bipartitions are different in nature, as one sees when looking at figure~\ref{fig:geometric-vs-species}, they can be related through a correspondence. 

Consider a system with $S'=\set{1,\dots,V'}$ local sites containing a single particle species defined by $\zeta$; let the bipartition be made geometrically, \ie $S'_A\subset S'$ sites belonging to subsystem $A$ and $S'\setminus S'_A$ to subsystem $B$; and let $f=V'_A/V'\equiv |S'_A|/|S'|$. Such a bipartite system corresponds to a bipartite system of $S=\set{1,\dots,V}$ local sites, each containing particle species $A$ and $B$ defined by $\zeta_A=\zeta^p$ and $\zeta_B=\zeta^q$ respectively, where $f=p/(p+q)$, $V'=(p+q)V$ and $p,q\in \mathbb{Z_+}$ are coprime.

We can show this explicitly by constructing the correspondence in both directions and then comparing the dimensions of the subsystem Hilbert spaces. Without loss of generality we only consider systems with $f\leq 1/2$. It is clear that given any values for $p,q,V$ we can map it to a geometric bipartition model with $V'=(p+q)V$ and $f=p/(p+q)$. The converse is also true from the following argument: given $V'\in \mathbb Z_+$ and $f\in \mathbb Q\cap{[0,1/2]}$ such that $V'f\in \mathbb Z_+$, we find $p,q\in \mathbb Z_+$ such that $f=p/(p+q)$ in reduced form, \ie $\gcd(p,q)=1$, and set $V=V'/(p+q)$, which is valid since $V'f\in \mathbb Z_+ $ implies $(p+q)\,| V'$.

Now we show how we are able to have $p$- and $q$-fold copies of the original species, in a sense. If the generating function for an admissible species is given by $\zeta$, we can write the $p$-fold composite species with $\zeta_A=\zeta^p$, and similarly $\zeta_B=\zeta^q$. By the Cauchy product identity, these are $p$- and $q$-fold convolutions of the dimensional sequences of the species; induction on proposition~\ref{prop:ABisadmissible} preserves admissibility. For counting purposes, it means the combinatorics of each copy of the species is accounted for at fixed particle number: 
    \begin{align}
        \dim \mathcal H_A^{(N_A)}=&{}[z^{N_A}]\zeta_A(z)^V\,,\\
        =&{} [z^{N_A}]\zeta(z)^{pV}\,,\\
        =&{}[z^{N_A}]\zeta(z)^{fV'} = \dim {\mathcal H_A'}^{(N_A)}\,,
    \end{align}
by the exact identity~\eqref{eq:dimension_saddle_integral}. A similar argument follows for species $B$, and thus the combined species $AB$ using~\eqref{eq:dNsumdAdB}. Since the dimensions of the subsystems are the same for both systems at fixed particle number, the average entanglement entropy $\braket{S_A}$ (and fluctuation measures such as $(\Delta S_A)_N$)~\eqref{eq:EEanalytic} are thus the same. This correspondence is illustrated in figure~\ref{fig:geometric-vs-species} for $f=1/3$.

We can now calculate the mean and variance explicitly. It was shown for such composite systems that $\beta^{\mathrm {tot}}_{m}(n)=m\beta(n/m)$ for $\zeta^{\mathrm {tot}}_m=\zeta^m$ by~\cite[Eq.~(51)]{Yauk:2023wbu}. Without loss of generality, we let $p\leq q$, \ie $f\leq 1/2$. We first pick some $n\in (0,\nmax)$ for the particle species, then $n_{AB}=(p+q)n$ by computing the saddle point equation for $\zeta_{AB}=\zeta^{p+q}$. We can now solve the saddle point equation for the species $n=z_0\zeta'(z_0)/\zeta(z_0)$ for $z_0$; we can plug such a $z_0$ into the saddle point equation for subsystem $A$ to find $\nAbar=z_0\zeta'_A(z_0)/\zeta_A(z_0)=pn$ and similarly $\nBbar=qn$ (since $z_0(n)=z_{AB,0}(n_{AB})$ by $n_{AB}=(p+q)n$ and the saddle point equation). We now consider the critical densities: for $\beta_A(\nAbar)=\beta_B(\nBbar)$ we can compute and find criticality occurs if and only if $p=q$, \ie at $f=1/2$ every $n$ is critical while at $f<1/2$ no $n$ is critical. And clearly $n_{AB}=n_{AB}^*$ if and only if $n=n^*$. Using this information, we rediscover~\cite[Eqs.~(23) and (32)]{Yauk:2023wbu}
\begin{align}\label{eq:EEgeometric}
    \braket{S_A}_N={}&V'f\beta(n)-\frac{|\beta'(n)|}{\sqrt{2\pi |\beta''(n)|}}\sqrt{V'}\delta_{f,1/2}\nonumber\\&+\frac12\left(f+\ln(1-f)-\delta_{f,1/2}\delta_{n,n^*}\right)+o(1)\,,\\
    \left(\Delta S_A\right)^2_N={}&\frac{\sqrt{2\pi}\beta'(n)^2}{|\beta''(n)|^{3/2}}\left[f(1-f)-\frac1{2\pi}  \delta_{f,1/2}\right]\nonumber\\&\times V'^{3/2}\ee^{-V'\beta(n)}+o\bm(V'^{3/2}\ee^{-V'\beta(n)}\bm)\,.
\end{align}
We see that the universal constant-order term for the entanglement entropy is recovered since we can compute $r_A=q/p$ and $\Xi_A=\beta'''(n)(\frac{1}{q^2}\frac{q}{p}-\frac{1}{p^2}\frac{p}{q})/\kappa^2=0$. This also shows that the branch transitions at constant order of section~\ref{sec:resolving-kdeltas} are a feature of species bipartitions. 

\section{Discussion and Outlook} \label{sec:discussion and outlook}
In this work, we have calculated the average and variance of the entanglement entropy of Haar-random states with fixed total particle number in systems where two particle species are placed at each local site of a $V$-site lattice and the bipartition is made between the species. Our setup covers a broad class of particle species, including fermions and bosons, with and without spin.

Our framework only depends on the volume $V$, total particle number $N$ and information about the multiplicities $a_{X,k}=\dim \mathcal H_{X,\loc}^{(k)}$ of each particle number macrostate at a local site. In particular, the particle-specific information given by the sequence $(a_{X,k})_{k\in\Nat}$ is encoded in a generating function $\zeta_X(z)=\sum_{k=0}^\infty a_{X,k}z^k$, which is in turn encoded into the negative Legendre transform $\beta_X$ of $\ln \zeta_X$. We derive the average entanglement entropy $\braket{S_A}_N$ up to $O(1)$ in the thermodynamic limit $V\to\infty$ and the variance $(\Delta S_A)_N^2$ to leading order where we find it vanishes exponentially fast, thus showing typicality. All terms in this expansion can be written in terms of $\beta_X$ and its derivatives for $X=A,B$. Our quantities also admit an interesting thermodynamic interpretation~\cite{bianchi2026typicalentanglemententropycharge}, where $\beta_X$ has the interpretation of the \emph{local entropy}.

It should be emphasized that physically interesting transitions show up at $O(\sqrt V)$ and $O(1)$ for $\braket{S_A}_N$ and at leading order for the variance. The existence of these transitions (isolated Kronecker-delta-like spikes and branch transitions) depends only on the particle species and filling density $n=N/V$. These occur since the exact formula for $\braket{S_A}_N$~\eqref{eq:EEanalytic} contains nonanalytic points where $d_A=d_B$. We classified these critical densities $\ncrit$ according to whether $\ncrit= n^*$, which exists if and only if the total local site multiplicities of the species match finitely, \ie $\dim \mathcal H_{A,\loc}=\dim \mathcal H_{B,\loc}$ and $\nmax<\infty$. In particular, such a requirement forces the Hilbert spaces of the two subsystems to be the same $\mathcal H_A\cong \mathcal H_B$ and only in such systems do we see a Page correction term contribution.

The existence of critical points has been previously identified in studies of systems with geometric bipartitions~\cite{Yauk:2023wbu,page1993} exclusively at half subsystem fraction $f=1/2$ for every density $n$. In contrast, our broader framework allows for a much richer critical structure. Interestingly, we find if the critical densities are not discrete in $(0,\nmax)$, \ie if there exists a continuum of critical densities, then the system can be described by a geometric bipartition at half subsystem fraction. Thus, the remaining critical densities are isolated points. 

Although~\cite{bianchi2026typicalentanglemententropycharge} considered systems with $\mathrm{U}(1)$ or $\mathrm{SU}(2)$ symmetries, they computed the average entanglement entropy and variance for systems with a geometric bipartition. A natural extension would be to generalize these results with a species bipartition, or with vector valued charges in the sense that each component charge is fixed independently as considered in the present manuscript for the total particle number.

Although our framework considers bipartitions between species, it does not directly apply to systems such as the $t$-$J$ model, where the Hilbert space under consideration is some physical subspace $\mathcal H_{\mathrm{phys}}\subset \mathcal H$ that does not itself factorize between the species. In such systems, additional constraints break the Cauchy product factorization $a_k=(a_{A,\cdot}*a_{B,\cdot})_k$ which we assume here. An interesting direction of future work would be to consider how such constraints could modify our results.

There exist also several natural next steps of what was considered in the present manuscript. First, we could also consider decompositions that combine geometric regions and types of species, \ie a subsystem could be a species in a specific region. Second, we could select Haar random states in a subspace of Hilbert space determined by more than one constraint, which could be several $\mathrm{U}(1)$ conserved quantities, as considered in~\cite{Langlett2025}, but also a combination of Abelian and non-Abelian charges. Third, it is a natural question to go beyond typical bipartite entanglement entropy by considering tri-partitions~\cite{hackl2025average} and multi-partitions and then compute quantities, such as typical mutual information, negativity or other measures of multipartite entanglement. 

\section*{Acknowledgments}
We thank Mario Kieburg for identifying the issue of periodic sequences after a talk given by BVG at the ANZAMP 2026 annual meeting. BVG acknowledges support from the Vacation Scholars Program of the School of Mathematics and Statistics at The University of Melbourne. ZK acknowledges the International Programs in Engineering at the University of Illinois Urbana-Champaign and The University of Melbourne for its Study Abroad Program. YY acknowledges support from the International Max Planck Research School for Quantum Science and Technology (IMPRS-QST). This research is part of the Munich Quantum Valley (MQV), which is supported by the Bavarian state government with funds from the Hightech Agenda Bayern Plus. LH acknowledges support through an Australian Research Council Discovery Early Career Researcher Award (DECRA) DE230100829 funded by the Australian Government.

\textbf{Author contribution statement.} BVG carried out most of the analytical work: he derived the average entanglement entropy and its variance, and derived and numerically verified the resolution of the Kronecker deltas and branch transitions at generic and flat critical densities. He wrote sections~\ref{sec:AnalyticalResults} and~\ref{sec:discussion and outlook} and appendices~\ref{ap:saddlepointanalysis} and~\ref{ap:ResolvingKdelta}, with the exception of the parts below, and wrote section~\ref{sec:GeometricBipartition}. ZK did the initial analysis of the Jaynes--Cummings--Hubbard model, identified and analyzed the particle-hole-like family, and developed the finite-size scaling analysis and high-precision numerics across the generic-density, generic-critical and flat-critical regimes. He also produced the case-study figures, wrote most of sections~\ref{sec:testbedmodelA}--\ref{sec:JCH}, and contributed proofs of saddle-point dominance that guided the treatment of periodically supported sequences. KS developed the initial setup and asymptotic strategy of the project, wrote the Introduction and the first draft of appendix~\ref{ap:pedagogy}, and drafted sections~\ref{sec:setup} and~\ref{sec:BianchiDona} jointly with BVG. YY contributed the example of species with finite spread in appendix~\ref{ap:nonadmissibleAxiom5}, re-examined the framework of \cite{Yauk:2023wbu} on which this work builds, and took part in discussions. LH conceived and supervised the project, derived the $\Theta$ term, proposed the testbed model $\zeta_A(z)=1+3z$, $\zeta_B(z)=2+2z$, identified the problem of saddle-point dominance for periodically supported sequences, and carried out numerical checks of the asymptotic formulas with BVG and ZK. All authors discussed the results and reviewed the manuscript.

Large language models (LLMs), including Claude (Anthropic) and ChatGPT (OpenAI), were used throughout this project as a tool and collaborator. Proposals made by an LLM were treated like proposals made by a coauthor: they were discussed and checked by the authors, and adopted, revised or discarded. No result was included solely because an LLM produced it. LLMs were used in the following ways. (i)~Derivations and proofs: to propose and check intermediate steps of the saddle-point analysis of appendix~\ref{ap:saddlepointanalysis} and of the resolutions in appendix~\ref{ap:ResolvingKdelta}, to find errors such as sign mistakes and missing assumptions, and to test claims against explicit counterexamples. The resulting asymptotic formulas were checked against exact finite-size evaluations of the Bianchi-Donà sum, see figure~\ref{fig:fssa-summary}. (ii)~Appendix~\ref{ap:SumIntegralError}, which bounds the error of converting the Bianchi-Donà sum into an integral, was drafted by an LLM and then checked and revised by the authors. (iii)~Code: Mathematica code for symbolic and numerical calculations, plotting code, and TikZ code for figures was written with LLM assistance and checked by the authors, numerical code in particular by comparison with independent calculations. (iv)~Text: LLMs were used to draft and edit parts of the text, and to check notation, cross-references and the bibliography. (v)~Review: before submission, an LLM-assisted internal review checked the mathematics, wording and references, and each suggestion was assessed individually by the authors. The authors take full responsibility for the work.

\clearpage

\onecolumngrid
\makeatletter
\let\set@footnotewidth\set@footnotewidth@one
\expandafter\let\csname combine@insert@1\endcsname\combine@insert@@ne
\makeatother
\appendix
\section{Saddle Point Analysis}\label{ap:saddlepointanalysis}

In this analysis, we require an estimate for the dimensions $d_X$ in~\eqref{eq:dimDX} in the thermodynamic limit, for use in the Bianchi-Donà equations for the moments of the entanglement entropy $S_A$. We begin by using a standard saddle point method to get the asymptotics of $d_X$. Then, we translate the assumptions used in the saddle point method to restrictions on the particle species we consider, and then discuss what happens when they are not met. We also consider the interplay of the saddle point analysis and the Cauchy product factorization of the generating functions $\zeta=\zeta_A\zeta_B$ to help simplify calculations in the main text. Finally, we conclude with a discussion on the critical particle densities $\ncrit$.

\subsection{Saddle point estimate}
We begin by trying to compute the dimensions $d_{X}( N_X,V)=[z^{N_X}]\zeta_X(z)^V$ when $n_X=N_X/V=O(1)$ and $N_X,V\gg1$. Fortunately, a great deal of literature exists contemplating such questions but we choose to follow~\cite[\S VIII.8]{FlajoletSedgewick2009}. We recall that our generating function $\zeta_X$, which has radius of convergence $R_X>0$ and is thus analytic at the origin, consists of non-negative coefficients $a_{X,k}$ since these represent local Hilbert space dimensions. Furthermore, without loss of generality, we restrict $\zeta_X$ to be aperiodic in the sense that $\gcd\set{k:a_{X,k}\neq0}=1$, and satisfy $\zeta_X(0)\neq0$, \ie a vacuum exists $a_{X,0}>0$. We shall show how we recover generality in section~\ref{ap:NonAdmissible}. We also define the \emph{supercritical density} by %
\begin{align}
    \nXsc:=\lim_{x\to R_X^-}\frac{x\zeta'_X(x)}{\zeta_X(x)}\,,
\end{align}
and impose the condition $n_X<\nXsc$. This requirement is sharp. Then by Proposition VIII.7 for any $n_X$ in a compact interval of $(0,\nXsc)$ there exists a unique positive solution $z_{X,0}$ of the equation
\begin{align}
    z_{X,0}\frac{\zeta'_X(z_{X,0})}{\zeta_X(z_{X,0})}=n_X\,.
\end{align}
And by Theorem VIII.8 we find our saddle point estimate
\begin{align}
    [z^{N_X}]\zeta_X(z)^V=\frac{\zeta_X(z_{X,0})^V}{z_{X,0}^{N_X+1}\sqrt{2\pi V\xi}}(1+O(1/V))\,,
\end{align}
where 
\begin{align}
    \xi:=\dfrac{\dd^2}{\dd z^2}(\ln \zeta_X(z)-n_X\ln z)\bigg|_{z=z_{X,0}}\,.
\end{align}
We can express our result to be more in line with \cite{Yauk:2023wbu} by defining
\begin{align}\label{eq:betaLegendre}
    \psi_X(z)={}&\ln \zeta_X(z)-n_X\ln z\,,\\
    \beta_X(n_X)={}&\psi_X(z_{X,0})\,,
\end{align}
in the sense that $-\beta_X(n_X)=n_Xt-f_X(t)$ is the Legendre transformation of 
\begin{align}
    f_X(t):=\ln \zeta_X(\ee^t)\,,
\end{align}
with $n_X$ and $t=\ln z_{X,0}$ as the conjugate pair. The saddle point equation then corresponds to $f_X'(t)=n_X$. By the second derivative of the Legendre transformation one can compute
\begin{align}
    \xi\equiv\psi''_X(z_{X,0})=-\frac{1}{z_{X,0}^2\beta_X''(n_X)}\,.
\end{align}
Thus we may simplify
\begin{align}
    d_X(N_X,V)= \sqrt\frac{-\beta_X''(n_X)}{{2\pi V}}\ee^{V \beta_X(n_X)}(1+O(1/V))\,.
\end{align}
A full expansion in powers of $1/V$ exists; however, we only require up to an $O(1/V)$ remainder.
\subsection{Admissible species}\label{ap:admissible}
We shall now translate these technical requirements on generating functions to restrictions on the type of particle species we admit. Firstly, for our generating function $\zeta_X$ to be analytic at the origin, \ie $R_X>0$, we require $\limsup_k a_{X,k}^{1/k}<\infty$. This means the multiplicities can grow at most exponentially $a_{X,k}=O(r^{-k})$ for some $0<r<R_X$. Secondly, without loss of generality we say that there is a vacuum $a_{X,0}>0$ and the support of the sequence is aperiodic, 
\begin{align}
    g_X:=\gcd\set{k:a_{X,k}\neq0}=1\,.
\end{align}
We also have the condition that we may only use $n_X$ in compact sub-intervals of $(0,\nXsc)$: define 
\begin{align}
    \nXmax:=\min\set{k: \forall j>k\,, a_{X,j}=0}\,,
\end{align}
then by aperiodicity we have $\nXmax>0$ (otherwise we would have $g_X=\gcd\set{0}=0$), and if $\nXmax<\infty$ we find $R_X=\infty$ and thus
\begin{align}
    \nXsc=\lim_{x\to +\infty}\frac{\sum^{\nXmax}_{k=0}a_{X,k}kx^k}{\sum^{\nXmax}_{k=0}a_{X,k}x^k}=\nXmax>0\,.
\end{align}
But if $\nXmax=\infty$ we must impose an additional condition to have $\nXsc=\infty$. Although we can deal with the case where $\nXsc<\infty$ and $\nXmax=\infty$ by only considering densities $n_X<\nXsc$, we want to be able to consider every possible density $n_X\in (0,\nXmax)$ and we shall find we do not restrict ourselves much by doing this. We talk about the relaxation of this condition in section~\ref{ap:NonAdmissible}. We can characterize this requirement on the spread with the following proposition:
\begin{proposition}\label{prop:TxChar}
    If $\nXmax=\infty$, we have $\nXsc=\infty$ if and only if $\lim_{x\to R^-_X}\sum_ka_{X,k}kx^k=\infty$.
\end{proposition}
\begin{proof}%
    ($\impliedby$) Suppose for the sake of contradiction that $\nXsc<\infty$; we will first prove that $\zeta_X(z)<\infty$. Consider that 
    \begin{align}
        (\ln \zeta_X)'(x)=\frac{\sum_ka_{X,k}kx^{k-1}}{\sum_ka_{X,k}x^k}\leq \nXsc/x\,,
    \end{align}
    by monotonicity for $x\in (0,R_X)$. If we integrate in the region $[x_0,x]\subset (0,R_X)$ we find by the fundamental theorem of calculus
    \begin{align}
        \ln\zeta_X(x)=\ln \zeta_X(x_0) +\int_{x_0}^x \frac{\sum_ka_{X,k}kt^{k-1}}{\sum_ka_{X,k}t^k} dt\leq \ln \zeta_X(x_0)+\nXsc\ln\frac{R_X}{x_0}<\infty\,
    \end{align}
    uniformly since $R_X<\infty$. We then may take the limit $\lim_{z\to R^-_X}\zeta_X(z)<\infty$. Then consider
    \begin{align}
        \lim_{z\to R^-}\sum_ka_{X,k} kz^k =\nXsc \lim_{z\to R^-_X}\zeta_X(z)<\infty\,,
    \end{align}
    which contradicts our hypothesis, hence $\nXsc=\infty$. 
    
    ($\implies$) Conversely, we see
    \begin{align}
        \nXsc= \lim_{z\to R^-_X}\frac{\sum_k a_{X,k} kz^k}{\sum_k a_{X,k} z^k}\leq a_{X,0}+\lim_{z\to R^-}\sum_k a_{X,k} kz^k\,.
    \end{align}
    So if $\nXsc=\infty$ then $\lim_{z\to R^-}\sum_k a_{X,k} kz^k=\infty$.
\end{proof}
We summarize these requirements in definition~\ref{def:admissible}:

\begin{definition}\label{def:admissible}
    For a particle species $X$ with dimensional sequence $(a_{X,n})_n\subset \mathbb N$, we say the sequence or species is \textbf{admissible} whenever the following hold:
    \begin{enumerate}[label=(\arabic*)]
        \item We require a nonzero radius of convergence, \ie $R_X=(\limsup_n a^{1/n}_{X,n})^{-1}>0$ by the Cauchy-Hadamard formula. This is equivalent to saying the multiplicities grow at most exponentially, \ie $a_{X,k}=O(r_X^{-k})$ for all $k$ for some $r_X>0$.
        \item There is a vacuum: $a_{X,0}>0$.
        \item The support is aperiodic: $g_X=1$. In particular, this implies that the particles can exist on the lattice, \ie $\nXmax\neq0$.
        \item If $\nXmax=\infty$, we require $\sum_k a_{X,k}kz^k\xrightarrow{z\to R^-_X}\infty$.
    \end{enumerate}
\end{definition}
We remark that axioms (3) and (4) were not included in \cite{Yauk:2023wbu}, leading to a large class of counterexamples. However, it should be noted that such counterexamples are unphysical. 

Recall from section~\ref{sec:setup} that we can think of the entire local site as a single particle species in the sense that we have the dimensional sequence $(a_{AB,k})_k\subset \mathbb N$ where $a_{AB,k}=(a_A*a_B)_k$ is the $k$-th element of the Cauchy product of the sequences $(a_{A,n})_n,(a_{B,n})\subset \mathbb N$. With this factorization we may prove an important proposition:
\begin{proposition}\label{prop:ABisadmissible}
    If particle species $A$ and $B$ are admissible, then the combined species $AB$ is admissible.
\end{proposition}
\begin{proof}%
    The species $AB$ is well defined, so it remains to check each part of definition~\ref{def:admissible}. As a corollary to this proposition, any species whose generating function is defined by a finite product of admissible species' generating function is admissible, by induction. \\
    Part 1: We have $\zeta_{AB}=\zeta_A \zeta_B$ with $\zeta_A$ and $\zeta_B$ having radii of convergence of $R_A>0$ and $R_B>0$ respectively. Then $\zeta_{AB}$ has a radius of convergence of $R=\min\set{R_A,R_B}$.\\
    Part 2: $a_{AB,0}=a_{A,0}a_{B,0}>0$.\\
    Part 3: Suppose $j\in \set{k\in \mathbb N:a_{A,k}\neq 0}$, then we find
        \begin{align}
            a_{AB,j}=\sum^j_{\ell=0}a_{A,j-\ell}a_{B,\ell}\geq a_{A,j}a_{B,0}>0\,,
        \end{align}
        by the existence of a vacuum for species $B$. Then $j\in\set{k\in \mathbb N:a_{AB,k}\neq 0}$. In particular, $g_{AB}=\gcd \set{k\in \mathbb N:a_{AB,k}\neq 0}$ divides each $j\in \set{k\in \mathbb N:a_{A,k}\neq 0}$ but $g_A=\gcd \set{k\in \mathbb N:a_{A,k}\neq 0}=1$ by aperiodicity of species $A$. This forces $g_{AB}=1$. \\
    Part 4: We will first prove $\nmax_{AB}=\nAmax+\nBmax$. Consider for $k\in\mathbb N$,
    \begin{subequations}
    \begin{align}
        a_{AB,(\nAmax+\nBmax+k)}={}&\sum^{\nAmax+\nBmax+k}_{\ell=0} a_{A,(\nAmax+\nBmax+k-\ell)}a_{B,\ell}\,, \\
        ={}& \sum^{\nAmax+\nBmax+k}_{\ell=\nBmax+k} a_{A,(\nAmax+\nBmax+k-\ell)}a_{B,\ell}\,,\\
        ={}&a_{A,\nAmax}a_{B,\nBmax}\delta_{k,0}\,.
    \end{align}
    \end{subequations}
    This follows from the definitions of $\nAmax$ and $\nBmax$. Therefore, we can have $\nmax_{AB}=\infty$ only if $\nmax_{A}=\infty$ or $\nmax_{B}=\infty$. Without loss of generality suppose $R=R_A\leq R_B$ with $\nmax=\infty$. We then find
    \begin{align}
        \sum_k a_{AB,k} kz^k =\sum_k \sum_{j=0}^k a_{A,k-j}a_{B,j} kz^k\geq a_{B,0}\sum_ka_{A,k}kz^k \xrightarrow{z\to R}\infty\,
    \end{align}
    by axiom 4 for species $A$.
\end{proof}

\subsection{Pathological species}\label{ap:NonAdmissible}
In this section we discuss species which are not admissible in the sense of definition~\ref{def:admissible}. We first provide a method here to extract the asymptotics for species which could violate requirements 2, 3 and 4: Suppose $(a_{X,k})_{k\in \mathbb Z}\subset \Nat_0$\footnote{We allow for negative indices so that conserved charges other than particle number can be considered.} such that there are indices $k_0<k_1$ such that $a_{X,k_0},a_{X,k_1}>0$ and $a_{X,k}=0$ for all $k<k_0$. Furthermore, we still suppose that requirement 1 is satisfied for this system\footnote{This is required for the saddle point method to go through. However, the asymptotics can sometimes be calculated in a different manner. For instance, the dimensions and average entanglement entropy were calculated for a system of $N$ distinguishable particles in a $V$-site lattice with the bipartition made geometrically in~\cite[Appendix C]{Yauk:2023wbu}. Notably, the leading-order contribution was proportional to $V\ln V$ instead of obeying a volume law.}. For such a species, we are allowed to apply modifications to the species by first translating every index such that $k_0\mapsto 0$ in section~\ref{ap:nonadmissibleAxiom2}, and then by mapping the asymptotics to an aperiodic sequence in section~\ref{ap:nonadmissibleAxiom4}. After this procedure, the saddle point method goes through for densities $n<\min\set{\nAsc,\nBsc}$. We now provide a discussion for each of the requirements in definition~\ref{def:admissible}.

\subsubsection{Fixing the normalization}\label{ap:nonadmissibleAxiom2}
We may consider the slightly more interesting case where $k=0$ is not necessarily the starting point of our generating function sum. It is a common technique to factorize as follows:
\begin{align}
    \zeta_X(z)=\sum_{k=k_0}a_k z^k=z^{k_0}\sum_{k=k_0}a_{k} z^{k-k_0}=z^{k_0}\sum_{k=0}a_{k+k_0} z^{k}\equiv z^{k_0}\hat\zeta_X(z)\,,
\end{align}
where we define $\hat\zeta_X(z)=\sum_{k=0}\hat{a}_kz^k$ with $\hat{a}_k\equiv a_{k+k_0}$. To extract the asymptotics we notice
\begin{align}
    [z^{N}]\zeta_X(z)^V=[z^{N}]z^{Vk_0}\hat\zeta_X(z)^V=[z^{V(n-k_0)}]\hat\zeta_X(z)^V\,.
\end{align}
So we can use the saddle point method for $0<n-k_0<k_{\mathrm{max}}-k_0$, where $k_{\mathrm{max}}$ is defined analogously to $\nXmax$. That is, starting with the sequence $(a_k)_{k=k_0}^\infty$, we may relabel $k\mapsto k-k_0$ to fix the normalization and work as we did before.

\subsubsection{Periodically supported sequences}\label{ap:nonadmissibleAxiom4}
Suppose the species satisfies all requirements except aperiodicity. First consider the trivial case where $g_X=0$ which occurs if and only if $\nXmax=0$, so no particles of this species are allowed to exist. Such a case is uninteresting, and was explicitly excluded in~\cite{Yauk:2023wbu}. We now consider the case where we have a periodically supported dimensional sequence, $g_X>1$. Then the support satisfies $\set{k:a_{X,k}\neq0}\subseteq g_X\Nat$, so $\zeta_X$ inherits a $\mathbb Z_{g_X}$ symmetry $\zeta_X(\omega z)=\zeta_X(z)$ for every $g_X$-th root of unity $\omega$. Consequently the positive real saddle $z_{X,0}$ is not the unique maximizer of $|\zeta_X|$ on the contour $|z|=z_{X,0}$: by the daffodil lemma~\cite[Lemma IV.1]{FlajoletSedgewick2009}, the maximum modulus is attained precisely at the $g_X$ points $z_{X,0}\,\ee^{2\pi \ii m/g_X}$, each of which is an equally dominating saddle of the integrand in~\eqref{eq:dimension_saddle_integral}. The single-saddle estimate of Theorem VIII.8 in~\cite{FlajoletSedgewick2009} therefore does not apply, and applying it regardless would produce a nonzero $d_X$ for every $N_X$, whereas $d_X$ manifestly vanishes unless $g_X\mid N_X$.

Rather than track the relative phases of the $g_X$ saddles, we remove the periodicity by a change of variable and reduce to an admissible species. Define the \emph{compressed sequence} and its generating function
\begin{align}
    \tilde a_{X,\ell}:=a_{X,g_X\ell}\,, \qquad \tilde \zeta_X(w):=\sum_{\ell=0}^\infty \tilde a_{X,\ell}w^\ell\,.
\end{align}
Since every nonzero $a_{X,k}$ has $g_X\mid k$, no information is lost and
\begin{align}\label{eq:compressionmap}
    \zeta_X(z)=\tilde \zeta_X(z^{g_X})\,.
\end{align}

\begin{proposition}\label{prop:compressedadmissible}
    If $(a_{X,k})_k$ satisfies axioms 1, 2, and 4 of definition~\ref{def:admissible}, then the compressed sequence $(\tilde a_{X,\ell})_\ell$ is admissible, with $\tilde R_X=R_X^{g_X}$ and $\tilde \nXsc=\nXsc/g_X$.
\end{proposition}
\begin{proof} %
    Axiom 1: By~\eqref{eq:compressionmap}, $\tilde R_X = (\limsup_{k}\tilde a^{1/k}_{X,k})^{-1}=(\limsup_{k} a^{1/(g_Xk)}_{X,g_Xk})^{-g_X}=R_X^{g_X}$. Axiom 2: $\tilde a_{X,0}=a_{X,0}>0$. Axiom 3: $\gcd\set{\ell:\tilde a_{X,\ell}\neq0}=\gcd\set{k/g_X:a_{X,k}\neq0}=g_X/g_X=1$. Axiom 4: substituting $w=z^{g_X}$,
    \begin{align}
        \sum_\ell \tilde a_{X,\ell}\,\ell\, w^\ell=\frac{1}{g_X}\sum_k a_{X,k}\,k\,z^k\,,
    \end{align}
    and $w\to \tilde R_X^-$ if and only if $z\to R_X^-$, so the two limits diverge together. The stated value of $\tilde \nXsc$ follows from~\eqref{eq:saddlecompression} below.
\end{proof}

Extracting coefficients from~\eqref{eq:compressionmap} and writing $M_X:=N_X/g_X$, we obtain
\begin{align}\label{eq:periodiccoeffextraction}
    d_X(N_X,V)=[z^{N_X}]\zeta_X(z)^V=\begin{cases}
        [w^{M_X}]\tilde\zeta_X(w)^V\,, & g_X\mid N_X\\
        0\,, & g_X\nmid N_X
    \end{cases}\,,
\end{align}
which is exactly the admissible problem of the previous subsections for the species $\tilde X$ at particle number $M_X$ and density
\begin{align}
    m_X:=\frac{M_X}{V}=\frac{n_X}{g_X}\,.
\end{align}
The saddle point analysis therefore applies to $\tilde \zeta_X$ without modification: for $m_X$ in a compact subinterval of $(0,\tilde n_X^{\mathrm{max}})$ there is a unique $\tilde z_{X,0}(m_X)\in(0,\tilde R_X)$ solving $\tilde z_{X,0}\tilde\zeta_X'(\tilde z_{X,0})/\tilde\zeta_X(\tilde z_{X,0})=m_X$, and
\begin{align}\label{eq:periodicdX}
    d_X(N_X,V)=\sqrt{\frac{-\tilde\beta_X''(m_X)}{2\pi V}}\,\ee^{V\tilde \beta_X(m_X)}(1+O(1/V))\,,\qquad g_X\mid N_X\,,
\end{align}
with $d_X=0$ otherwise. Differentiating~\eqref{eq:compressionmap} gives the dictionary between the two descriptions,
\begin{align}\label{eq:saddlecompression}
    \tilde z_{X,0}(m_X)=z_{X,0}(n_X)^{g_X}\,,\qquad
    \tilde\beta_X(m_X)=\beta_X(n_X)\,,\qquad
    \tilde\beta_X'(m_X)=g_X\beta_X'(n_X)\,,\qquad
    \tilde\beta_X''(m_X)=g_X^2\beta_X''(n_X)\,,
\end{align}
where $-\beta_X$ is defined as before by the Legendre transform of $f_X(t)=\ln\zeta_X(\ee^t)$. In particular, if one insists on expressing~\eqref{eq:periodicdX} in terms of the uncompressed $\beta_X$, the multiplicity of dominating saddles reappears as an explicit factor of $g_X$,
\begin{align}\label{eq:dimXgX}
    d_X(N_X,V)=g_X\sqrt{\frac{-\beta_X''(n_X)}{2\pi V}}\,\ee^{V \beta_X(n_X)}(1+O(1/V))\,,\qquad g_X\mid N_X\,,
\end{align}
in agreement with a direct summation over the $g_X$ symmetric saddles. We work with the compressed description throughout, in which no such factor appears.

Combined with section~\ref{ap:nonadmissibleAxiom2}, the full procedure for a species violating axioms 2 and 3 is: first shift $k_0\mapsto0$ to install a vacuum, then compress by $g_X$ to remove the periodicity. Provided the other admissibility requirements hold, the result is an admissible species, and the physical content of the two operations is simply that the on-site particle number is offset by $k_0$ and quantized in units of $g_X$.

\paragraph{Two periodic species}
For the bipartite problem it is the interplay of $g_A$ and $g_B$ that matters. Since both species have a vacuum, $\set{k:a_{A,k}\neq0}\cup\set{k:a_{B,k}\neq0}\subseteq \set{k:a_{AB,k}\neq0}\subseteq \set{k:a_{A,k}\neq0}+\set{k:a_{B,k}\neq0}$, so
\begin{align}
    g_{AB}=\gcd(g_A,g_B)\,.
\end{align}
If $g_A=g_B=g$, then $g_{AB}=g$ and~\eqref{eq:compressionmap} compresses all three generating functions simultaneously, preserving the Cauchy product identity~\eqref{eq:zetaIDs},
\begin{align}
    \tilde\zeta(w)=\tilde\zeta_A(w)\tilde\zeta_B(w)\,.
\end{align}
The constraints $g\mid N$ and $g\mid N_A$ mean the decomposition~\eqref{eq:fixedNdecomp} is a sum over $M_A=N_A/g$ with unit spacing, so the compressed problem is a genuine two-species admissible problem on the same $V$ sites at density $m=n/g$. Every result of this paper then applies verbatim after the substitutions $n\mapsto m$, $\beta_X\mapsto\tilde\beta_X$; in particular~\eqref{eq:bvgEE} holds with $\braket{S_A}_N$ evaluated at $m=n/g$.

If instead $g_A\neq g_B$, compressing by $g_{AB}$ removes only the common periodicity and leaves at least one species periodic, so the reduction is incomplete. To see which values of $N_A$ are valid to pick, we solve the following system of congruences with the generalized Chinese Remainder Theorem
\begin{align}
    \begin{cases}
        N_A\equiv 0 &\pmod{g_A}\\
        N_A\equiv N &\pmod{g_B}
    \end{cases} \iff N \equiv 0 \pmod{g_{AB}}\,,
\end{align}
that is, the system is solvable if and only if $g_{AB}|N$. By~\eqref{eq:periodiccoeffextraction}, $g_{AB}|N$ is necessary for $d_{AB}\neq0$, but it is not sufficient for small $N$: $N$ must also be a sum of $V$ on-site particle numbers $k$ with $a_{AB,k}\neq0$. For example, $\zeta_A(z)=1+z^4$ and $\zeta_B(z)=1+z^6$ have $g_{AB}=2$, but no sector with $N=2$. Let $F_{AB}$ be the Frobenius number of the numerical semigroup generated by $\set{k/g_{AB}:a_{AB,k}\neq0}$. For $V$ large enough, every multiple of $g_{AB}$ beyond $g_{AB}F_{AB}$ is such a sum, provided of course that $N\leq N_{\max}:=V\nmax$, so that $d_{AB}\neq0$ if and only if $g_{AB}|N$ for $N>g_{AB}F_{AB}$. If $\nmax<\infty$, the reflection $k\mapsto\nmax-k$ of the on-site particle numbers shows that gaps of the same kind can occur within a $V$-independent distance below $N_{\max}$, given by the Frobenius number of the reflected semigroup. Since these bounds do not depend on $V$, neither restriction matters in the thermodynamic limit at fixed density $0<n<\nmax$. So solving the system, we may only pick $N_A$ in a congruence class modulo $\lcm(g_A,g_B)$ with residue
\begin{align}
    N_A^{(0)}
=g_A\left[\frac{N}{g_{AB}}\left(\frac{g_A}{g_{AB}}\right)^{-1}\mod{\frac{g_B}{g_{AB}}} \right]\,,
\end{align}
where the inverse is understood in the multiplicative group modulo $g_B/g_{AB}$. So we may only consider total particle numbers $N$ that are multiples of $g_{AB}$ and we must restrict the sum over $N_A$ in~\eqref{eq:fixedNdecomp} accordingly. We note that when $g_A\neq g_B$, the results of this paper do not hold with simply using the compressed formulas; one must redo the derivations or use the uncompressed formulas while tracking factors of $g_X$ from~\eqref{eq:dimXgX}.

\subsubsection{Supercritical filling densities}\label{ap:nonadmissibleAxiom5}
Suppose we have a species where axioms 1 through 3 are satisfied, but axiom 4 is violated: we have $\nXmax=\infty$ but by proposition~\ref{prop:TxChar} we have finite spread $\nXsc<\infty$. We recall there is no solution $t$ to the saddle point equation $f'_X(t)=n_X$ for $n_X\geq\nXsc$.

We now present a concrete example of such a case: consider composite particles which are made of two parts, each containing a left and a right spin-$\frac{1}{2}$ particle. That is, one composite particle corresponds to four spin-$\frac{1}{2}$ constituents. The Hilbert space with $k$ composite particles is evidently $\tilde{\mathcal{H}}^{(k)}=(\mathbb{C}^2_L)^{\otimes 2k}\otimes (\mathbb{C}^2_R)^{\otimes 2k}$. The physical Hilbert space, which will satisfy $\nXmax=\infty$ and $\nXsc<\infty$, is given by the gauge-constrained invariant sector of the spins under $\mathrm{Spin}(4)\cong \mathrm{SU}(2)_L\times \mathrm{SU}(2)_R$. That is, the physical subspace is that where the $2k$ left spin-$\frac{1}{2}$ particles form a singlet $(\mathbf{S}_L^2=0)$ under rotations, and the $2k$ right spin-$\frac{1}{2}$ particles also independently form a singlet $(\mathbf{S}_R^2=0)$ under rotations. We write $\mathcal{H}^{(k)}=\mathrm{Inv}_{\mathrm{SU}(2)_L}\bm((\mathbb{C}^{2}_L)^{\otimes 2k}\bm)\otimes \mathrm{Inv}_{\mathrm{SU}(2)_R}\bm((\mathbb{C}^{2}_R)^{\otimes 2k}\bm)$.

As an example, for $k=2$, there are $8$ spin-$\frac{1}{2}$ particles, but we can consider the left and right parts separately. Repeated use of the fusion rule of irreducible representations of $\mathrm{SU}(2)$ given by $\frac{1}{2}\otimes\frac{1}{2}=0\oplus 1$ leads to the expression
\begin{align}
    \left(\frac{1}{2}\right)^{\otimes 4}=0\oplus 0\oplus 1\oplus 1\oplus 1\oplus 2\,,
\end{align}
from which we can read off that the invariant sector is a 2-dimensional subspace. Since the right part independently forms the invariant sector, $a_{X,2}=4$. More generally, there are $C_k=\frac{1}{k+1}\binom{2k}{k}$ ways to form a singlet from $2k$ spin-$\frac{1}{2}$ particles, so $a_{X,k}=C_k^2$.

The number $C_k$ is the $k^{\mathrm{th}}$ Catalan number, which forms an important sequence in enumerative combinatorics. Its asymptotics are well known to follow $C_k\sim \frac{4^k}{\sqrt{\pi}k^{3/2}}$. From this, we see that $\zeta_X(z)=\sum_{k=0}^\infty C_k^2 z^k$ has a radius of convergence $R_X=1/16$, and in fact the spread can be calculated exactly:
\begin{align}
    \nXsc=\lim_{z\to R_X^-}\frac{\sum_k kC_k^2z^k}{\sum_k C_k^2z^k}=\frac{\sum_{k=0}^\infty \frac{k C_k^2}{16^k}}{\sum_{k=0}^\infty \frac{C_k^2}{16^k}}=\frac{\pi-3}{4-\pi}\approx0.16494\dots\,.
\end{align}

In such systems, for densities $n_X>\nXsc$, which we refer to as the \emph{supercritical regime}, no saddle-point solution for calculating the Hilbert space dimension exists. However, its asymptotic scaling is still calculable (both numerically and analytically) and comes under condensation phenomena encountered in statistical physics. In typical particle configurations at a mean density in the supercritical regime, excess particles condense onto one site. This has the effect of causing the saddle-point to freeze at its boundary value, with subleading corrections to be computed case by case. That is,
\begin{equation}
    d_X(N_X,V)=\ee^{V\beta_X(n_X)+o(V)}\,, \qquad \beta_X(n_X)=\begin{cases}
        \ln\zeta_X(z_{X,0})-n_X\ln z_{X,0}\,, & n_X\leq \nXsc \\ 
        \ln\zeta_X(R_X)-n_X\ln R_X\,, & n_X>\nXsc \\ 
    \end{cases}.
\end{equation}
The formula for $\beta_X$ when $n_X<\nXsc$ is obtained by a Legendre transformation~\eqref{eq:betaLegendre}, which is well defined since a saddle point solution exists; for $n_X>\nXsc$ it is computed from methods in large deviation theory. This phenomenon has interesting consequences for the average entanglement entropy, because in geometric bipartitions it completely changes the Page curve~\eqref{eq:EEgeometric}. The volume-law coefficient is no longer linear in $f$ but could also be parabolic $f(1-f)$. We defer details of this phenomenon to future calculations; in particular, the interplay of supercriticality between two species is increasingly more subtle.

\subsection{Shared saddle point}\label{ap:sharedsaddle}
For an admissible species $X$ we recall that for each $n_X\in(0,\nXmax)$ there is a unique Legendre conjugate variable solving the saddle point equation, $t\in(-\infty,\ln R_X):f_X'(t)=n_X$. We may now prove some results using these structures and the Cauchy product identity. Our first point of interest is to determine the behavior of  
\begin{align}
    \gamma(n_A):=\beta_A(n_A)+\beta_B(n-n_A)-\beta(n)\,.
\end{align}
We see that if we show $\beta_X$ is concave then $\gamma$ would be concave. Therefore, if we show that there is a density $\nAbar\in(0,\nAmax)$ such that $\beta'_A(\nAbar)=\beta_B'(\nBbar)$ (recall $\nBbar:=n-\nAbar$) then $\nAbar$ would be the unique maximizer of $\gamma$ over its full domain. We shall first prove concavity and show the first two derivatives of $\beta_X$.

\begin{proposition}\label{prop:betaX}
    Let $z_{X,0}(n_X)$ denote the unique solution of the saddle-point equation $f_X'(\ln z_{X,0})=n_X$. Then the following hold,
    \begin{enumerate}[label=(\arabic*)]
        \item $\beta'_X(n_X)=-\ln z_{X,0}$ and $\beta''_X(n_X)=-\left(n_X-n_X^2+z_{X,0}^2\zeta_X''(z_{X,0})/\zeta_X(z_{X,0})\right)^{-1}<0$.
        \item If $\nXmax=\infty$, then $\beta'_X(n_X)>0$ for all $n_X\in(0,\infty)$.
        \item If $\nXmax<\infty$, then there exists a unique $n^*_X=\zeta_X'(1)/\zeta_X(1)\in (0,\nXmax)$ such that $z_{X,0}(n^*_X)=1$ and $\beta'_X(n^*_X)=0$. Moreover, $\beta'_X(n_X)>0$ for $n\in (0,n^*_X)$ and $\beta'_X(n_X)<0$ for $n\in (n^*_X,\nXmax)$.
    \end{enumerate}
\end{proposition}
\begin{proof}%
    Parts 2 and 3 of this proof are adapted from Proposition A4 of~\cite{Yauk:2023wbu}.\\
    Part 1: Recall $t_X=\ln z_{X,0}$ so we may use some basic properties of the Legendre transformation to obtain the result, namely 
    \begin{align}
    \dfrac{\dd (-\beta_X)}{\dd n_X}=t_X\quad\text{and}\quad \dfrac{\dd^2 (-\beta_X)}{\dd n_X^2}=\left(\dfrac{\dd^2 f_X}{\dd t_X^2}\right)^{-1}\,.
    \end{align}
    \\
    Part 2: If $\nXmax=\infty$ then clearly $R_X\leq1$ since $(a_{X,k})_k\in \Nat_0$, so $z_{X,0}\in[0,R_X)$ hence $\beta_X'(n_X)=-\ln z_{X,0}>0$.\\
    Part 3: If $\nXmax<\infty$ then $R_X=\infty$ so any $z_{X,0}\in [0,\infty)$ is permitted. We see that $\beta_X'(n_X)=-\ln z_{X,0}$ is positive for $z_{X,0}\in[0,1)$, zero for $z_{X,0}=1$ and negative for  $z_{X,0}>1$. In particular, since the saddle point equation $f'_X(\ln z_{X,0})=n_X$ is monotonically increasing with $z_{X,0}$, one finds $\beta_X'(n_X)$ is positive for $n\in (0,n_X^*)$, zero for $n_X=n_X^*$ and negative for $n_X\in (n_X^*,\nXmax)$ where one can compute $n_X^*=f'_X(0)=\zeta_X'(1)/\zeta_X(1)$.
\end{proof}

We now prove the existence and uniqueness of a maximizer $\nAbar$, present some important structural identities at $\nAbar$ and motivate why we refer to it as a \emph{saddle point}.

\begin{lemma}\label{lem:combinedsaddle}
    For admissible species $A$ and $B$, and the combined species $AB$ the following hold and there exists a unique saddle point $\nAbar$.
    \begin{enumerate}[label=(\arabic*)]
        \item $z_0(n)=z_{A,0}(\nAbar)=z_{B,0}(\nBbar)$.
        \item $\beta(n)=\beta_A(\nAbar)+\beta_B(\nBbar)$.
        \item $\beta'(n)=\beta_A'(\nAbar)=\beta'_B(\nBbar)$.
        \item $1/\beta''(n)=1/\beta''_A(\nAbar)+1/\beta''_B(\nBbar)$
    \end{enumerate}
\end{lemma}
\begin{proof}[Proof of lemma~\ref{lem:combinedsaddle}] We prove each part sequentially.\\
    Part 1: Consider that for some $n\in(0,\nmax)$ we have some unique $\bar{t}$ such that $f'_{AB}(t)=n$. We notice that $\beta'_A\bm(n_A(\bar t)\bm)=-\bar t=\beta'_B\bm(n_B(\bar t)\bm)$. So if we find $n=n_A(\bar t)+n_B(\bar t)$ we would have found that $n_A(\bar t)=\nAbar$ is the unique maximizer we have been looking for. Using the Cauchy product identity we may compute where $\bar t=\ln z_0$
    \begin{align}
        n=f'_{AB}(\bar{t})= \frac{z_0\zeta'(z_0)}{\zeta(z_0)}=\frac{z_0\zeta_A'(z_0)}{\zeta_A(z_0)}+\frac{z_0\zeta_B'(z_0)}{\zeta_B(z_0)}=f'_A(\bar t)+f'_B(\bar t)=n_A(\bar t)+n_B(\bar t)\,.
    \end{align}
    Hence the existence and uniqueness of $\nAbar$ have been shown. Additionally, it is known that the Legendre conjugate variable $\bar t$ is shared across the functions (equivalently the saddle points are shared).
    \\
    Part 2: Consider
    \begin{align}
        f_{AB}(t)=\ln \zeta_{AB}(\ee^{t})=\ln \zeta_{A}(\ee^{t})+\ln \zeta_{B}(\ee^{t})=f_A(t)+f_B(t)\,,
    \end{align}
    so plugging in $t=\bar t$ we obtain the result
    \begin{align}
        -\beta(n)=n\bar t-f_{AB}(\bar t)=(\nAbar+\nBbar)\bar t-f_A(\bar t)-f_B(\bar t)=-\beta_A(\nAbar)-\beta_B(\nBbar)\,.
    \end{align}
    \\
    Part 3: The result follows immediately from part 1 by plugging $t=\bar t$ into $\beta'_X(n_X)=-\ln z_{X,0}$ from proposition~\ref{prop:betaX}.\\
    Part 4: The result follows from direct computation using the form of $\beta''_X(n_X)$ from proposition~\ref{prop:betaX}. Explicitly, with $\bar t=\ln z_0$
    \begin{subequations}
    \begin{align}
        -(\beta_A''(\nAbar))^{-1}-(\beta_B''(\nBbar))^{-1}={}&\nAbar-\nAbar^2+z_0^2\zeta''_A(z_0)/\zeta_A(z_0)\nonumber\\&+{}\nBbar-\nBbar^2+z_0^2\zeta''_B(z_0)/\zeta_B(z_0)\,,\\
        ={}& n-n^2+2\nAbar \nBbar +z_0^2\frac{\zeta''_A(z_0)\zeta_B(z_0)+\zeta''_B(z_0)\zeta_A(z_0)}{\zeta(z_0)}\,,\\
        ={}&n-n^2+2\nAbar \nBbar +z_0^2{\zeta''(z_0)}/{\zeta(z_0)}-2z_0^2\frac{\zeta'_A(z_0)\zeta_B'(z_0)}{\zeta_A(z_0)\zeta_B(z_0)}\,,\\
        ={}&n-n^2+z_0^2{\zeta''(z_0)}/{\zeta(z_0)}+2\nAbar\nBbar -2f'_A(\bar t)f'_B(\bar t)\,,\\
        ={}&{}-(\beta''(n))^{-1}\,.
    \end{align}
    \end{subequations}
\end{proof}

We finally note that $\nAbar$ is a strictly increasing function of $n$.
\begin{proposition}\label{prop:monotonen}
    On the interval $0<n<{\nmax}$, the saddle point $\nXbar(n)$ is monotonically increasing with $\nXbar'(n)=\frac{1}{1+r_X}$.
\end{proposition}
\begin{proof}
    From lemma~\ref{lem:combinedsaddle} we know that for each $n\in (0,\nmax)$ there is a unique conjugate variable $\bar{t}\in(-\infty,\ln R_X)$ such that $\nXbar=f'_X(\bar{t})$. So $\nXbar$ is a function of $n$, and we may compute
    \begin{subequations}
    \begin{align}
        \nXbar'(n)={}&\dfrac{\dd}{\dd n}f'_X(\bar{t})\,,\\
        ={}& \left(\dfrac{\dd t}{\dd n}\dfrac{\dd^2 f_X}{\dd t^2}\right)\bigg|_{t=\bar{t}}=\left(\dfrac{\dd t}{\dd n}\dfrac{\dd n_X }{\dd t}\right)\bigg|_{t=\bar{t}}\\
        ={}&\frac{\beta''(n)}{\beta_X''(\nXbar)}=\frac{1}{1+r_X}\,,
    \end{align}
    \end{subequations}
    where we have used properties of the Legendre transform and part (4) of lemma~\ref{lem:combinedsaddle} in the last equality.
\end{proof}

\subsection{Critical densities}\label{ap:criticaldensity}
For the filling density $n=N/V$, we consider four different cases which affect the entanglement entropy~\eqref{eq:bvgEE}.
\begin{definition}\label{def:critical-densities}
    For $n\in(0,\nmax)$ we say
    \begin{enumerate}[label=(\arabic*)]
        \item $n$ is a \textbf{generic critical density}, denoted $n=\ncrit$, whenever $n\neq n^*$ and 
        \begin{align}
            n\in \ncritset :=\set{n\in (0,\nmax): \beta_A\bm(\nAbar(n)\bm)=\beta_B\bm(\nBbar(n)\bm)}\,.
        \end{align}
        \item $n$ is a \textbf{flat density} whenever $\nmax<\infty$ and $n=n^*\equiv \zeta'(1)/\zeta(1)$. By proposition~\ref{prop:betaX} this is equivalent to saying $n^*$ is the unique density in $(0,\nmax)$ (if it exists) where $\beta'(n^*)=0$.
        \item $n=n^*$ is a \textbf{flat critical density} if $n^*\in \ncritset$.
        \item $n$ is a \textbf{generic density} in any other case.
    \end{enumerate}
    A generic or flat critical density $\ncrit$ is \textbf{degenerate} if $\beta''_A\bm(\nAbar(\ncrit)\bm)=\beta''_B\bm(\nBbar(\ncrit)\bm)$, \ie if the concavity ratio defined below~\eqref{eq:bvgEE} satisfies $r_A(\ncrit)=1$, and \textbf{non-degenerate} otherwise.
\end{definition}
We can characterize when a density is critical by defining
\begin{subequations}\label{eq:Ddef}
\begin{align}
    D(n):={}&\beta_A\bm(\nAbar(n)\bm)-\beta_B\bm(\nBbar(n)\bm)\,,\\
        ={}& \psi_A\bm(z_{0}(n)\bm)-\psi_B\bm(z_{0}(n)\bm)\,,\\
        ={}& \ln \frac{\zeta_A\bm(z_{0}(n)\bm)}{\zeta_B\bm(z_{0}(n)\bm)}-(\nAbar(n)-\nBbar(n))\ln z_{0}(n)\,.
\end{align}
\end{subequations}
Then whenever $D(n)=0$ we find that $n$ is a critical density. In particular, we see that $z_0(n^*)=1$ so $n^*$ is a flat critical density if and only if $\zeta_A(1)=\zeta_B(1)<\infty$. Furthermore, when $n^*$ is a flat critical density, we have $\beta_A''\bm(\nAbar(n^*)\bm)=\beta_B''\bm(\nBbar(n^*)\bm)$ if and only if 
\begin{align}\label{eq:flatdoubledegenchar}
   n^*_A-{n^*_A}^2+\zeta''_A(1)/\zeta_A(1)=n^*_B-{n^*_B}^2+\zeta''_B(1)/\zeta_B(1)\,
\end{align}
by direct computation with proposition~\ref{prop:betaX}.

Here we prove an interesting result about how critical densities are spaced. If the species are distinct then the critical points must be discrete in the interval $(0,\nmax)$, equivalently, if the critical points are not discrete we must then have species $A$ and $B$ being identical.   
\begin{proposition}\label{prop:continuouscriticality}
    $D(n)=0$ in some interval containing an accumulation point $I_n\subset (0,\nmax)$ if and only if $\zeta_A=\zeta_B$.
\end{proposition}
\begin{proof}%
    Consider
    \begin{subequations}
    \begin{align}
        D(n)={}&f_A(\bar t) -f'_A(\bar t) \bar t -f_B(\bar t) +f'_B(\bar t) \bar t\,,\\
        ={}&(f_A-f_B)(\bar t)- (f_A-f_B)'(\bar t)\bar t\,.
    \end{align}
    \end{subequations}
    We then solve this first-order differential equation in some conjugate interval $I_{\bar t}\subset (-\infty,\ln R )$ and obtain for some $C\in \R$ that $f_A(\bar t)=f_B(\bar t)+C\bar t$. Equivalently, $\zeta_A(z)=z^C\zeta_B(z)$ in some interval $I_z\subset (0,R)$. Using the existence of a vacuum, $\zeta_X(0)\neq0$, we find $C=0$. Each mapping from $n$ to $t$ to $z$ space is a homeomorphism so there is an accumulation point in $I_z$; analyticity forces $\zeta_A=\zeta_B$ identically, \ie species $A=B$.
\end{proof}
Interestingly, this is exactly the case where criticality occurs uniquely at half-subsystem fraction in geometric bipartitions at fixed particle number. 

\subsection{Converting the sum into an integral}\label{ap:SumIntegralError}
We now provide a compact argument why the error of converting the sum in the Bianchi-Donà formula into an integral is of subleading order, as already stated in~\eqref{eq:integral formula for EE} of section~\ref{sec:sum-to-integral}. We consider fixed admissible species at a fixed density $0<n<\nmax$, taking $V\to\infty$ with $N=nV$ integer. Our starting point is the exact discrete sum
\begin{equation}\label{eq:SI:exactsum}
\braket{S_A}_N
=\sum_{N_A\in\mathcal I_N}\varrho_{N_A}\varphi_{N_A},
\end{equation}
where the exact sector weights $\varrho_{N_A}$ and entropy terms $\varphi_{N_A}$ are defined in the main text, and $\mathcal I_N$ is the set of allowed particle-number sectors. In particular, $\sum_{N_A\in\mathcal I_N}\varrho_{N_A}=1$. Admissibility ensures that the shared saddle at $N_A=V\nAbar$ stays strictly inside the common domain of analyticity. The sector weight has width $O(\sqrt V)$ in $X=N_A$, with rescaled variance bounded above and away from zero.

\textbf{Step 1: Remove the tails.} Let $\chi$ be a smooth bump function, such that it is $1$ in a neighborhood of $\nAbar$ and vanishes outside a slightly larger neighborhood. With $p_k:=\chi(k/V)\varrho_k$, the first replacement is
\begin{equation}\label{eq:SI:cutoff}
\sum_{N_A\in\mathcal I_N}\varrho_{N_A}\varphi_{N_A}
=\frac{\sum_{k\in\mathcal I_N}p_k\varphi_k}
{\sum_{k\in\mathcal I_N}p_k}+O(\ee^{-cV})\,,
\end{equation}
for some $c>0$. The excluded probability is exponentially small, and $\varphi_k=O(V)$, so both the entropy tail and the effect of normalizing by $\sum_kp_k=1+O(\ee^{-cV})$ are exponentially small. All cutoff-weighted functions below are extended by zero outside their support, allowing lattice sums over $\mathbb{Z}$.

\textbf{Step 2: Use normalized saddle continuations.}
For real $X$ on the cutoff support, put $n_A=X/V$ and $n_B=n-n_A$. Using 
\begin{equation}\label{eq:SI:betafunctions}
\begin{aligned}
\gamma(n_A)&=\beta_A(n_A)+\beta_B(n_B)-\beta(n)\,,\\
VY(n_A)&=V(\beta_A(n_A)-\beta_B(n_B))
 +\frac12\ln\frac{\beta_A''(n_A)}{\beta_B''(n_B)}\,,
\end{aligned}
\end{equation}
the leading dimensions in~\eqref{eq:dXasymptotic} define the approximations
\begin{equation}\label{eq:SI:continuation}
\begin{aligned}
\widehat{\varrho}_X&:=\frac{\chi(X/V)}{Z_V}
\sqrt{\frac{\beta_A''(n_A)\beta_B''(n_B)}{-2\pi V\beta''(n)}}
 \ee^{V\gamma(n_A)}\,,\\
\widehat{\varphi}_X&:=V(\beta(n)-\beta_B(n_B))
 +\frac12\ln\frac{\beta''(n)}{\beta_B''(n_B)}-f\bm(VY(n_A)\bm),\\
f(t)&=\max\set{t,0}+\tfrac12\ee^{-|t|}\,.
\end{aligned}
\end{equation}
Hats distinguish these functions from the exact sector quantities. Choose $Z_V$ so that $\int_{\R}\widehat{\varrho}_X\,\dd X=1$, and set $q_k:=\widehat{\varrho}_k/\sum_j\widehat{\varrho}_j$. With $a=\betamin=\min\set{\bbA,\bbB}$, the centered second replacement is
\begin{equation}\label{eq:SI:centering}
\frac{\sum_{k\in\mathcal I_N}p_k(\varphi_k-aV)}
{\sum_{k\in\mathcal I_N}p_k}
=\sum_{k\in\mathbb Z}q_k(\widehat{\varphi}_k-aV)
 +O(\tfrac{1}{\sqrt{V}})
=\frac{\sum_{k\in\mathbb Z}\widehat{\varrho}_k(\widehat{\varphi}_k-aV)}
{\sum_{k\in\mathbb Z}\widehat{\varrho}_k}
 +O(\tfrac{1}{\sqrt{V}})\,.
\end{equation}
The leading dimensions have uniform relative error $O(V^{-1})$ on the cutoff support. This accuracy survives normalization by the respective lattice sums; $Z_V$ cancels from $q_k$. The entropy terms differ absolutely by $O(V^{-1})$. These estimates follow from the coefficient-extraction contour integrals for individual dimensions, without replacing the sector sum by an integral. To estimate the first equality, note that $\beta(n)=\bbA+\bbB$ gives $\widehat{\varphi}_{V\nAbar}=aV+O(1)$. Its bounded $X$-derivative and the $O(\sqrt V)$ sector width give $\sum_kq_k|\widehat{\varphi}_k-aV|=O(\sqrt V)$. The total replacement error is therefore $O(V^{-1})O(\sqrt V)+O(V^{-1})=O(V^{-1/2})$, uniformly including at critical densities. The second equality is algebraic, by the definition of $q_k$. Adding back $aV$ gives the same error bound for the uncentered averages, since both sets of weights are normalized.

\textbf{Step 3: Convert the sum into an integral.}
The final replacement is
\begin{equation}\label{eq:SI:quadrature}
\frac{\sum_{k\in\mathbb Z}\widehat{\varrho}_k\widehat{\varphi}_k}
{\sum_{k\in\mathbb Z}\widehat{\varrho}_k}
=\int_{\R}\widehat{\varrho}_X\widehat{\varphi}_X\,\dd X
 +O(\tfrac{1}{\sqrt{V}})\,.
\end{equation}
For real $X$, we define
\begin{equation}\label{eq:SI:centeredfunction}
I_V:=\int_{\R}\widehat{\varrho}_X\widehat{\varphi}_X\,\dd X \quad\text{and}\quad H_V(X):=\widehat{\varrho}_X(\widehat{\varphi}_X-I_V)\,.
\end{equation}
Since $\int H_V\,\dd X=0$, the normalized quadrature error is
\begin{equation}\label{eq:SI:quadratureidentity}
\frac{\sum_k\widehat{\varrho}_k\widehat{\varphi}_k}{\sum_k\widehat{\varrho}_k}-I_V
=\frac{\sum_kH_V(k)-\int_{\R}H_V(X)\,\dd X}{\sum_k\widehat{\varrho}_k}\,.
\end{equation}
Standard saddle estimates give, with primes denoting derivatives in $X$,
\begin{equation}\label{eq:SI:weightbounds}
\begin{gathered}
\|\widehat{\varrho}'\|_1=O(V^{-1/2}),\qquad
\|\widehat{\varrho}''\|_1=O(V^{-1}),\\
\int_{\R}|X-V\nAbar|\,|\widehat{\varrho}_X''|\,\dd X=O(V^{-1/2})\,.
\end{gathered}
\end{equation}
The first-derivative quadrature bound implies $\sum_k\widehat{\varrho}_k=1+O(V^{-1/2})$, so the denominator stays bounded away from zero.
To control crossings, put $t(X)=VY(X/V)$, so $t'=O(1)$ and $t''=O(V^{-1})$. The Page-corrected $f$ is $C^2$, including at $t=0$, with $0\le f'\le1$ and $f''(t)=\tfrac12\ee^{-|t|}$. Integration by parts gives
\begin{equation}\label{eq:SI:crossingbound}
0\le\int_{\R}\widehat{\varrho}_X f''(t)(t')^2\,\dd X
=-\int_{\R}\big(\widehat{\varrho}_X'f'(t)t'
 +\widehat{\varrho}_X f'(t)t''\big)\,\dd X
=O(V^{-1/2})\,.
\end{equation}
Thus $\int\widehat{\varrho}_X|\widehat{\varphi}_X''|\,\dd X=O(V^{-1/2})$, independently of the number or slopes of crossings. Since $ |I_V-\widehat{\varphi}_{V\nAbar}|=O(\sqrt V)$, the weight bounds and bounded $\widehat{\varphi}_X'$ imply $\|H_V''\|_1=O(V^{-1/2})$ by the product rule. Euler--Maclaurin now gives
\begin{equation}\label{eq:SI:EM}
\left|\sum_{k\in\mathbb Z}H_V(k)-\int_{\R}H_V(X)\,\dd X\right|
\le\frac1{12}\|H_V''\|_1=O(V^{-1/2})\,,
\end{equation}
which proves~\eqref{eq:SI:quadrature} via~\eqref{eq:SI:quadratureidentity}.
Combining the three replacements and setting $X=V\nAbar+\sqrt V\,\delta$, define
\begin{equation}\label{eq:SI:rescaledfunctions}
\varrho(\delta):=\sqrt V\,\widehat{\varrho}_{V\nAbar+\sqrt V\,\delta},
\qquad
\varphi(\delta):=\widehat{\varphi}_{V\nAbar+\sqrt V\,\delta},
\end{equation}
whose asymptotic forms are given in~\eqref{eq:varrhodelta} and~\eqref{eq:varphiNAsplitdelta}, respectively, we find
\begin{equation}\label{eq:SI:result}
\braket{S_A}_N
=\int_{\R}\varrho(\delta)\varphi(\delta)\,\dd\delta
 +O(V^{-1/2})\,.
\end{equation}
The error is subleading to a constant, uniformly on compact interior density ranges, including generic densities and both non-degenerate and degenerate critical densities. Density sequences approaching an occupation boundary or a singular saddle require separate analysis.

Finally, let us highlight that our error estimate of $O(V^{-1/2})$ is an upper limit and we generally expect that for $n$ away from critical points, the actual error of moving from the sum to the integral is actually exponentially suppressed.

\section{Resolution of Kronecker deltas and branch transitions}\label{ap:ResolvingKdelta}
We start this section by giving a pedagogical explanation of what is meant by resolving a Kronecker delta or a branch transition, by explaining how resolutions at different orders in $V$ interact, and by classifying the critical terms of~\eqref{eq:bvgEE} into the two kinds. We then resolve the Kronecker deltas and branch transitions of the average entanglement entropy \eqref{eq:bvgEE}, which are encoded by its indicator functions $1_\ncritset(n)$ and $1_{\ncritset^C}(n)$ and its Kronecker deltas, see proposition~\ref{prop:jumps-kdeltas}. We first find the critical points $\ncrit$ by solving the characteristic equation in appendix~\ref{ap:criticaldensity}, and we then keep our density in a regime where these contributions are non-trivial. How to choose this regime in each case is explained in appendix~\ref{ap:choosing-window}. Then after calculating $\braket{S_A}_N$ in this regime, we isolate the resolved profiles of these terms.

\subsection{What is meant by resolving Kronecker deltas and branch transitions?}\label{ap:pedagogy}

\begin{figure}[tbp]
\centering
\includegraphics{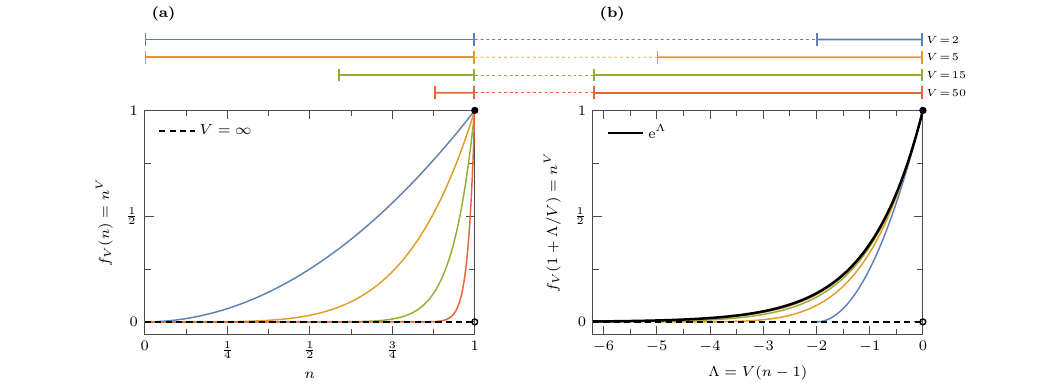}
\caption{\textit{Resolving a Kronecker delta.} \textbf{(a)}~The functions $f_V(n)=n^V$ converge pointwise, but not uniformly, to $\delta_{n,1}$ (dashed). \textbf{(b)}~In the double scaling coordinate $\Lambda=V(n-1)$, they converge to the profile $\ee^{\Lambda}$ (solid black), which resolves the Kronecker delta. The bars above the panels show, for each $V$, the window in $n$ that is mapped onto the range of $\Lambda$ shown in (b).}
\label{fig:kronecker-resolution}
\panellabel{fig:pointwise-collapse}
\panellabel{fig:double-scaling-resolution}
\end{figure}

In what follows, we will focus on a simple example that exhibits the same sort of limiting behavior resulting in Kronecker delta terms in \eqref{eq:bvgEE}: giving a pointwise limit of a Kronecker delta at a critical point. Then we will see what it means to resolve that Kronecker delta. 

Specifically, consider the sequence of functions $(f_V)_{V\in\mathbb Z_+}: f_V(n)=n^V\,,\, 0<n\leq 1$ which we plot in figure~\ref{fig:pointwise-collapse}. Now let us rewrite our sequence of functions as $f_V(n) = \ee^{V \ln n}$. Then $\lim_{V\to \infty}f_V(n) = \delta_{n,1}$ pointwise. That is, we have a critical point $\ncrit = 1$ where the exponent $\ln n$ vanishes. Taylor expanding $\ln n= (n-1) + O\bm((n-1)^2\bm)$, our sequence of functions can be written as $f_V(n) = \ee^{V [(n-1) + O\bm((1-n)^2\bm)]}$, so if we choose the scale in which $V[n-1+O\bm((1-n)^2\bm)] = O(1)$, then the exponent no longer has $V$ dependence in its scaling. Thus the limiting behavior becomes nontrivial rather than collapsing to a Kronecker delta. Consider a power-law scaling around the critical point, $n = \ncrit + \frac{\Lambda}{V^s} = 1 + \frac{\Lambda}{V^s}$ with $s>0$. Rearranging this equation for $\Lambda$ we find $\Lambda = V^s(n-1)$. So if we choose $s=1$, and keep $\Lambda \leq0$ a constant, we will be exactly in the regime where the exponent is of constant order as $V \to \infty$. The way to interpret this scaling is that $n$ is in a shrinking neighborhood of width $|\Lambda|/V$ around 1 as $V\gg1$. Rewriting the expansion in the exponent in the new variables, we find $V[n-1 + O\bm((1-n)^2\bm)] = \Lambda + O(1/V)$. So in this new fixed variable $\Lambda$, the higher-order terms vanish and the exponent $V\ln n$ behaves like $\Lambda$. Thus when we take the double scaling limit as $V \to \infty$ keeping $\Lambda$ fixed, we find the nontrivial limiting behavior
\begin{align}
    \lim_{V \to \infty} f_V(n)= \lim_{V\to \infty} \left(1+\frac{\Lambda}{V}\right)^V = \ee^{\Lambda}\,,
\end{align}
as expected; equivalently $\ee^{\Lambda+o(1)}\to \ee^\Lambda$ as $V\to \infty$. So for large $V$, in the shrinking region $n\in[1+\frac{\Lambda}{V}, 1]\subset[0,1]$, our sequence of functions is approximated by the function $\ee^{\Lambda}$. We show how well the resolved formula matches $f_V$ in figure~\ref{fig:double-scaling-resolution} compared to the pointwise limit. That is, we have found a resolution on which we can see how the sequence of functions collapses to a Kronecker delta.

A branch transition is resolved in the same way. For instance, $g_V(n)=1/(1+n^V)$ on $(0,2)$, which we plot in figure~\ref{fig:transition-collapse}, converges pointwise to $1$ for $n<1$, to $\frac12$ at $n=1$ and to $0$ for $n>1$. This is a jump whose value at the critical point is the mean of the one-sided limits, so by~\eqref{eq:jump-kronecker} it has no Kronecker delta. In the same window $n=1+\Lambda/V$ we find $g_V\to1/(1+\ee^{\Lambda})$, which interpolates between the two one-sided limits instead of returning to a single one as $|\Lambda|\to\infty$, see figure~\ref{fig:transition-double-scaling}. In the following subsections, we will resolve the Kronecker deltas and branch transitions which arise as correction terms in the average entanglement entropy~\eqref{eq:bvgEE}.

\begin{figure}[tbp]
\centering
\includegraphics{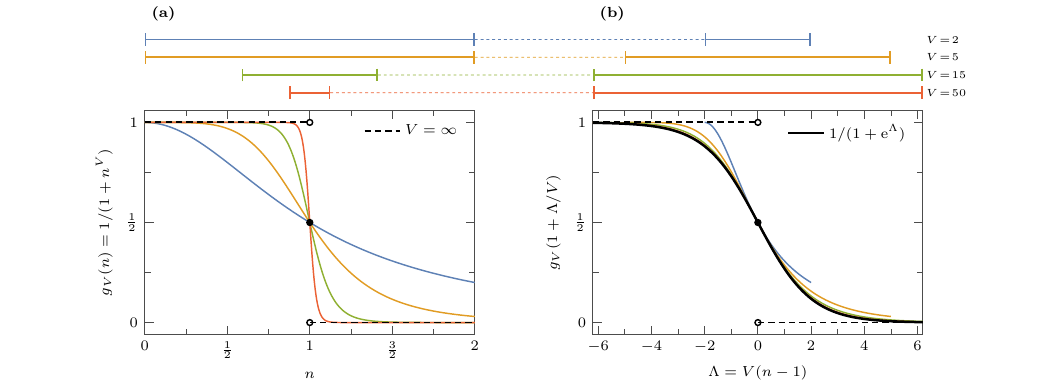}
\caption{\textit{Resolving a branch transition.} \textbf{(a)}~The functions $g_V(n)=1/(1+n^V)$ converge pointwise, but not uniformly, to a step (dashed) whose value at $n=1$ is the mean of the one-sided limits: a branch transition without a Kronecker delta. \textbf{(b)}~In the double scaling coordinate $\Lambda=V(n-1)$, they converge to the profile $1/(1+\ee^{\Lambda})$ (solid black), which interpolates between the two one-sided limits. The bars are as in figure~\ref{fig:kronecker-resolution}.}
\label{fig:transition-resolution}
\panellabel{fig:transition-collapse}
\panellabel{fig:transition-double-scaling}
\end{figure}

\subsection{Interplay from resolutions on different orders}\label{ap:interplay}
When we resolve the discontinuities in the expansion coefficients of $\braket{S_A}_N$ an additional subtlety arises. If we have a single function $f_V(n)$ as discussed in the previous section, there is a unique resolved function on the correct scale.

If we have a function $F_V(n)$ whose pointwise asymptotics is of the form
\begin{align}
    F_V(n)=V a(n) +\sqrt{V} b(n)+ c(n)+o(1)\,,
\end{align}
we naturally find an interplay between the different orders. In the window $n=\ncrit+\Lambda V^{-s}$, a coefficient that is smooth near $\ncrit$ also contributes to lower orders through its Taylor expansion. If $a$ is smooth, for instance, then for $s=\frac12$
\begin{align}\label{eq:interplay-taylor}
    V a(n)=V a(\ncrit)+\sqrt V\,\Lambda\, a'(\ncrit)+\frac12\Lambda^2 a''(\ncrit)+O(V^{-1/2})\,,
\end{align}
so the coefficients of $\sqrt V$ and of constant order in the window contain the terms $\Lambda a'(\ncrit)$ and $\frac12\Lambda^2a''(\ncrit)$, which descend from order $V$. These are the inherited terms of~\eqref{eq:inh-res}, and their superscript records the order from which they descend. In $\braket{S_A}_N$ the coefficients $a(n)$ and $b(n)$ are not smooth at $\ncrit$ themselves; the Taylor expansions instead apply to the smooth functions from which they are built, the branch coefficients $a_X$ of~\eqref{eq:aX} and the height $b_{\rm K}$ of~\eqref{eq:bK}, weighted by the resolved weights of appendix~\ref{ap:calc-avgEE-res}. The resulting inherited terms are listed in table~\ref{tab:inherited}. They vanish at $\Lambda=0$, may grow with $|\Lambda|$, and resolve no discontinuity of the pointwise coefficients.

The resolved profiles $B$ and $C$ are what remains at each order. They describe the full shape of the crossover, not only an interpolation between the pointwise values. In particular, a term that is not generated by a Taylor expansion belongs to $B$ or $C$ even if it vanishes both at $\Lambda=0$ and as $\Lambda\to\pm\infty$, such as the term proportional to $\tau$ in table~\ref{tab:resolved-kdeltas}.

Given the roots of $Y$ to leading order in $V$, the resolved profiles $B$ and $C$ take the same form at degenerate and non-degenerate critical densities. The degeneracy only enters through the Taylor expansions that produce the inherited terms, which then also appear at powers of $V$ between order $V$ and order unity, through the window, \ie the choice of $s>0$, and through the order to which the roots must be computed to get $\braket{S_A}_N^{\mathrm {res}}$ correct up to $o(1)$. All three depend on $p$, see appendices~\ref{ap:kdeltares-pert-generic} and~\ref{ap:kdeltares-pert-flat}.

\subsection{Classifying Kronecker deltas and branch transitions}\label{ap:classify-jumps}
Not every indicator function in~\eqref{eq:bvgEE} describes an isolated contribution at a critical density. To see this, recall the branch coefficients $a_X$ and $c_X$ of~\eqref{eq:aX} and~\eqref{eq:cX}, so that $a_X(n)=\bar\beta_X$. Using $r_B=1/r_A$ and $\Xi_B=-\Xi_A$, the constant-order terms of~\eqref{eq:bvgEE} at a critical density $n\neq n^*$ are
\begin{align}
    c_A-\frac14\left(\ln r_A+\frac{1-r_A}{1+r_A}\right)-\frac12\Xi_A\beta'(n)=\frac12(c_A+c_B)\,,
\end{align}
the mean of the two branches. In particular, this is why the cross term $\Xi_A$ drops out at a critical density. Hence~\eqref{eq:bvgEE} can be written equivalently as
\begin{align}\label{eq:bvgEE-branch}
    \braket{S_A}_N={}&\sum_{(X,Y)}H(\bar\beta_Y-\bar\beta_X)\left(V\bar\beta_X+c_X(n)\right)-\sqrt{\frac{2V}{\pi\kappa}}\,|\beta'(n)|\,1_{\ncritset}(n)\nonumber\\
    &+\Pi_*\,\delta_{\zeta_A(1),\zeta_B(1)}\delta_{n,n^*}+o(1)\,,
\end{align}
where the sum runs over $(X,Y)\in\set{(A,B),(B,A)}$ and $H$ is the Heaviside function with $H(0)=\frac12$, except at $n=n^*$ where the sum is replaced by $V\bbA+c_A(n^*)$.

We now classify the critical terms of~\eqref{eq:bvgEE-branch} by their branch jumps and Kronecker weights. Let $g(n)$ be one of the coefficients of~\eqref{eq:pointwise-abc}, and let $\ncrit\in\ncritset$ be such that the one-sided limits $g({\ncrit}^\pm)=\lim_{n\to{\ncrit}^\pm}g(n)$ exist. We call
\begin{subequations}\label{eq:jump-kronecker}
\begin{align}
    J[g]:={}&g({\ncrit}^+)-g({\ncrit}^-)\,,\\
    K[g]:={}&g(\ncrit)-\frac12\left(g({\ncrit}^+)+g({\ncrit}^-)\right)\,,
\end{align}
\end{subequations}
the \emph{branch jump} and the \emph{Kronecker weight} of $g$ at $\ncrit$, respectively. In the sense of section~\ref{sec:resolving-kdeltas}, $g$ has a branch transition at $\ncrit$ if and only if $J[g]\neq0$, and a Kronecker delta at $\ncrit$ if and only if $K[g]\neq0$. Equivalently, near $\ncrit$ we may write $g(n)=g_-(n)+(g_+(n)-g_-(n))H(n-\ncrit)+K[g]\,\delta_{n,\ncrit}$, where $g_\pm$ are continuous at $\ncrit$ with $g_\pm(\ncrit)=g({\ncrit}^\pm)$ and $H(0)=\frac12$. The mean of the one-sided limits is the only reference value for $K$ which does not depend on which side of $\ncrit$ is taken as the starting point, or equivalently on which species is labeled $A$. It is also the value selected by the double scaling limit of section~\ref{sec:resolving-kdeltas}: a kink at the mean of the Gaussian splits it into two equal halves, $w_0=\frac12$ at $\Lambda=0$. More precisely, for the resolved profiles $F(\Lambda)$ of~\eqref{eq:inh-res} at a given order, $F(\pm\infty)=g({\ncrit}^\pm)$ and $F(0)=g(\ncrit)$, see~\eqref{eq:BC-limits}, where for the first identity we assume, as usual in matched asymptotics, that the resolved and fixed-density expansions agree on an overlap region $V^{-s}\ll|n-\ncrit|\ll1$, so that $J[g]=F(\infty)-F(-\infty)$ and $K[g]=F(0)-\frac12(F(\infty)+F(-\infty))$ follow from the resolved profile alone. The inherited terms of~\eqref{eq:inh-res} are excluded: they arise from Taylor expanding terms of higher order in $V$, vanish at $\Lambda=0$, and may grow with $|\Lambda|$, see~\eqref{eq:flatdegencritres}.

\begin{table}[t]
\centering
\renewcommand{\arraystretch}{1.3}
\caption{\textit{Type of discontinuities of the coefficients of~\eqref{eq:bvgEE} at a critical density $\ncrit$.} $p$ is defined in~\eqref{eq:pdefn}, and -- denotes a continuous coefficient. The coefficient of $V$ is continuous. The branch jumps and Kronecker weights are given in proposition~\ref{prop:jumps-kdeltas}.}
\label{tab:discontinuities}
\begin{tabular}{@{\hspace{0.5em}} l @{\hspace{2em}} l @{\hspace{2em}} l @{\hspace{0.5em}}}
\hline
\hline
Critical density & Order $\sqrt V$ & Order unity \\
\hline
Generic, $p=1$ (non-degenerate) & Kronecker delta & Branch transition \\
Generic, $p=2$ & Kronecker delta & Kronecker delta \\
Generic, $p\geq3$ & Kronecker delta & -- \\
Flat & -- & Kronecker delta \\
\hline
\hline
\end{tabular}
\end{table}

Applying~\eqref{eq:jump-kronecker} to~\eqref{eq:bvgEE-branch} and~\eqref{eq:Theta} we find the following, summarized in table~\ref{tab:discontinuities}.
\begin{proposition}\label{prop:jumps-kdeltas}
    Let the species be distinct, $\zeta_A\neq\zeta_B$, let $\ncrit\in\ncritset$, and let $b(n)$ and $c(n)$ be the coefficients of $\sqrt V$ and of constant order in~\eqref{eq:bvgEE-branch}. The coefficient of $V$ has neither a branch transition nor a Kronecker delta at $\ncrit$. The coefficient $b$ has no branch transition, $J[b]=0$, and it has a Kronecker delta if and only if $\ncrit$ is a generic critical density, with weight $K[b]=-\sqrt{2/(\pi\kappac)}\,|\beta'(\ncrit)|$. For the coefficient $c$, let $p<\infty$ be the order of the zero of $D$ at $\ncrit$ defined in appendix~\ref{ap:choosing-window}. Then:
    \begin{enumerate}[label=(\arabic*)]
        \item At a generic critical density with $p=1$, \ie a non-degenerate one, label the species such that $A$ is selected for $n<\ncrit$. Then $c$ has a branch transition but no Kronecker delta,
        \begin{align}
            J[c]={}&-\frac12\left(\ln\rAc+\frac{1-\rAc}{1+\rAc}\right)-\Xi_A(\ncrit)\beta'(\ncrit)\,,\\
            K[c]={}&0\,.
        \end{align}
        \item At a generic critical density with $p=2$, which is degenerate with $\Xi_A(\ncrit)\neq0$, let $A$ be the species selected on both sides of $\ncrit$. Then $c$ has a Kronecker delta but no branch transition, $J[c]=0$ and $K[c]=-\frac12\Xi_A(\ncrit)\beta'(\ncrit)\neq0$.
        \item At a generic critical density with $p\geq3$, $c$ is continuous, $J[c]=K[c]=0$, with $c(\ncrit)=\frac12(\frac12-\ln2)$.
        \item At a flat critical density, label the species such that $\betaA''\geq\betaB''$. Then $c$ has a Kronecker delta but no branch transition, $J[c]=0$ and $K[c]=\Pi_*$, with $\Pi_*$ given by~\eqref{eq:Pistar}. At a degenerate flat critical density $K[c]=-\frac12$, while at a non-degenerate one
        \begin{align}\label{eq:Kflat}
            K[c]={}&\Theta(\rsc)-c_A(n^*)\nonumber\\
            ={}&P_{\rm flat}+\frac12\erfc(x_*)\left(\frac{\rsc-1}{1+\rsc}-\ln\rsc\right)\nonumber\\
            &-\ee^{-x_*^2}\sqrt{\frac{(1-\rsc)\ln(1/\rsc)}{2\pi(1+\rsc)}}\,,
        \end{align}
        where $x_*^2=\frac{1+\rsc}{1-\rsc}\,\frac{\ln(1/\rsc)}{2}$ is the argument in~\eqref{eq:Theta} and $P_{\rm flat}$ is given by~\eqref{eq:pageunresolved}.
    \end{enumerate}
\end{proposition}
\begin{proof}
    Away from $\ncrit$, exactly one term of the sum in~\eqref{eq:bvgEE-branch} is selected on each side, and all coefficients are continuous there, so the one-sided limits are the values of the selected branch at $\ncrit$. Since $\bbA=\bbB$ at $\ncrit$, the coefficient of $V$ is continuous and equal to $\beta_A\bm(\nAbar(\ncrit)\bm)$ on both sides and at $\ncrit$. The $\sqrt V$ term carries $1_{\ncritset}$, so $b({\ncrit}^\pm)=0$, hence $J[b]=0$ and $K[b]=b(\ncrit)=-\sqrt{2/(\pi\kappac)}\,|\beta'(\ncrit)|$. This is nonzero if and only if $\beta'(\ncrit)\neq0$, \ie if and only if $\ncrit$ is a generic critical density.

    (1) Since $p=1$, $D$ changes sign at $\ncrit$, so $c({\ncrit}^-)=c_A(\ncrit)$ and $c({\ncrit}^+)=c_B(\ncrit)$, while $H(0)=\frac12$ gives $c(\ncrit)=\frac12(c_A+c_B)$. Hence $K[c]=0$ and $J[c]=c_B-c_A$, which is the stated expression by $r_B=1/r_A$ and $\Xi_B=-\Xi_A$. By~\eqref{eq:Dprime}, $p=1$ if and only if $\ncrit$ is non-degenerate.

    (2) Since $p=2$, $D$ does not change sign and $c({\ncrit}^\pm)=c_A(\ncrit)$. By~\eqref{eq:Dprime} and~\eqref{eq:D2generic}, $\rAc=1$ and $\Xi_A(\ncrit)\neq0$. Hence $c_A$ and $c_B$ differ only by their cross terms, and $K[c]=\frac12(c_A+c_B)-c_A=-\frac12\Xi_A\beta'\neq0$ at $\ncrit$.

    (3) By~\eqref{eq:D2generic}, $\rAc=1$ and $\Xi_A(\ncrit)=0$, so $c_A(\ncrit)=c_B(\ncrit)=\frac12(\frac12-\ln2)$ by~\eqref{eq:cX}. Whether or not $D$ changes sign, the one-sided limits and $c(\ncrit)=\frac12(c_A+c_B)$ therefore coincide, and $J[c]=K[c]=0$.

    (4) At a non-degenerate flat critical density $p=2$, so the species selected on both sides is the same and, with the stated labeling, $r_A(n^*)=\rsc$. At a degenerate one $r_A(n)\to1$ as $n\to n^*$, so that $r_A(n^*)=\rsc=1$ and $c_A(n^*)=c_B(n^*)$, whichever branch is selected on either side. Since $\beta'(n^*)=0$, in both cases $c(n^{*\pm})=c_A(n^*)=\frac12(\frac1{1+\rsc}+\ln\frac{\rsc}{1+\rsc})$, while $c(n^*)=c_A(n^*)+\Pi_*$ by~\eqref{eq:bvgEE-branch}. Hence $J[c]=0$ and $K[c]=\Pi_*$. For $\rsc<1$, inserting~\eqref{eq:Theta} into~\eqref{eq:Pistar} and using $\erf=1-\erfc$ gives~\eqref{eq:Kflat}.
\end{proof}
In the non-degenerate flat case $p=2$, so $D$ does not change sign and $\bar\beta_A$ and $\bar\beta_B$ touch without exchanging order (see figure~\ref{fig:CritDiaTestBed}). In the degenerate flat case $p\geq3$, and for odd $p$, as in the particle-hole-like family of section~\ref{sec:PH-family}, $D$ changes sign and the branch selected at order $V$ does switch at $n^*$. There is nevertheless no branch transition at constant order, since $r_A(n)\to1$ and $\beta'(n)\to0$ as $n\to n^*$, so the constant-order coefficients $c_A$ and $c_B$ of the two branches have the same limit. In the non-degenerate flat case the logarithmic contribution to the Kronecker delta, the last two terms in~\eqref{eq:Kflat}, arises because a fraction $\erfc(x_*)$ of the Gaussian lies beyond the two kinks $\delta_\pm$, where the other branch is selected. Cases (2) and (3) are fine-tuned: case (2) requires $D(\ncrit)=0$ and $\betaA''=\betaB''$ simultaneously, and case (3) in addition $\betaA'''=\betaB'''$. We include them to emphasize that the type of contribution at constant order is decided by $p$, not by which term of~\eqref{eq:bvgEE} carries the indicator function. For odd $p\geq3$, $D$ changes sign and the branch selected at order $V$ switches at $\ncrit$, as for $p=1$, but the two branches agree at constant order, as at a degenerate flat critical density.

Branch transitions at order unity are a feature of species bipartitions: for the geometric bipartition of a single species every critical term is a Kronecker delta, see section~\ref{sec:GeometricBipartition}. Our resolution does not cover identical species, $\zeta_A=\zeta_B$, the only case in which critical densities are not isolated (proposition~\ref{prop:continuouscriticality}). This is the geometric bipartition at $f=\frac12$, where criticality is instead approached by varying $f$, and the Kronecker deltas are resolved in~\cite{Yauk:2023wbu}.

\subsection{Scaling windows and roots of \texorpdfstring{$Y$}{Y}}\label{ap:ncritAPb}
To locate the nonanalytic points of the integrand in~\eqref{eq:intsplit3ways}, consider $Y$ defined in~\eqref{eq:Ydefn}, for which $d_A/d_B=\ee^{VY}(1+O(1/V))$. Determining which of $d_A$ and $d_B$ is larger therefore reduces to determining the sign of $Y$, which is fixed up to an $O(V^{-2})$ error,
\begin{align}
    d_A>d_B \iff Y>0\,, && d_A<d_B \iff Y<0\,.\label{eq:dimensionsANDy}
\end{align}
Determining the sign of $Y$ to higher accuracy would require the next correction to~\eqref{eq:dXasymptotic}, which we shall not need.

We work in the rescaled variable $\delta=\sqrt{V}(n_A-\nAbar)$, in which $\varrho(\delta)$ is asymptotically the Gaussian $\mathcal N(0,1/\kappa)$ of~\eqref{eq:varrhodelta}. Since the integral is dominated by the saddle point $\nAbar$, only crossings $\hat\delta_0=O(1)$, \ie on the scale of the standard deviation of $\varrho(\delta)$, are relevant; a crossing further out only changes the integral by exponentially small terms. Unless stated otherwise, we further require $\hat\delta_0=\ord(1)$, meaning that $\hat\delta_0$ is strictly of constant order, so that the crossing neither escapes the Gaussian nor collapses onto its mean, $\hat\delta_0\to0$, as $V\to\infty$.

Looking at~\eqref{eq:intsplit3ways}, we see that $Y$ enters the calculation in two ways: its sign determines the regions $\set{d_A\gtrless d_B}$ over which the two logarithmic branches are integrated, and $V|Y|$ appears in the exponent of the Page term. Our goal is to probe how the entanglement entropy changes in a shrinking window around a critical density, $n=\ncrit+\Lambda V^{-s}$ with $s>0$ and fixed $\Lambda\in\R$. As we show below, the exponent $s$ must be fine-tuned, and it may differ depending on whether one wishes to resolve the split of the logarithmic branches, at $s=s_K$, or the Page term, at $s=s_P$.

Expanding~\eqref{eq:Ydefn} around $\nAbar$ with the shorthand~\eqref{eq:betashorthand}, we find
\begin{subequations}
\begin{align}
    Y(\delta)={}&\bbA-\bbB+(\bbA'+\bbB')(n_A-\nAbar)+\frac12(\bbA''-\bbB'')(n_A-\nAbar)^2+\frac16(\bbA'''+\bbB''')(n_A-\nAbar)^3\nonumber \\&+\frac1{2V}\left[\ln\frac{\bbA''}{\bbB''}+\left(\frac{\bbA'''}{\bbA''}+\frac{\bbB'''}{\bbB''}\right)(n_A-\nAbar)+ O\bm((n_A-\nAbar)^2\bm)\right]+ O\bm((n_A-\nAbar)^4\bm)+O(V^{-2})\,, \label{eq:Yexp1}\\
    ={}& \left(\bbA-\bbB+\frac1{2V} \ln\frac{\bbA''}{\bbB''}\right)+\left[\bbA'+\bbB'+\frac1{2V}\left(\frac{\bbA'''}{\bbA''}+\frac{\bbB'''}{\bbB''}\right)\right]V^{-1/2} \delta \nonumber \\ &{}+\frac12(\bbA''-\bbB'')(V^{-1/2} \delta)^2+\frac16(\bbA'''+\bbB''')(V^{-1/2} \delta)^3+ O(V^{-2})\,.\label{eq:Ypolydelta}
\end{align}
\end{subequations}
Using $\bbA'+\bbB'=2\beta'(n)$ (lemma~\ref{lem:combinedsaddle}) and $r_A=\bbA''/\bbB''$, this gives
\begin{align}\label{eq:VYexp}
    VY(\delta)={}&V(\bbA-\bbB)+2\sqrt V\beta'(n)\delta+\frac12\left[\ln r_A+(\bbA''-\bbB'')\delta^2 \right]\nonumber\\
    &+\frac{\delta}{2\sqrt V} \left[\frac{\bbA'''}{\bbA''}+\frac{\bbB'''}{\bbB''}+\frac13 (\bbA'''+\bbB''')\delta^2 \right]+O(V^{-1})\,.
\end{align}
The second line is $O(V^{-1/2})$ for $\delta=O(1)$. We show in appendix~\ref{ap:split-stability} that perturbations of this size affect neither the split of the logarithmic branches nor the Page term at constant order, so we work with
\begin{align}\label{eq:Qdef}
    Q(\delta):={}V D(n)+2\sqrt{V}\beta'(n)\delta+\frac12 g(n;\delta)\,,\qquad VY(\delta)=Q(\delta)+O(V^{-1/2})\,,
\end{align}
where
\begin{align}
    D(n)={}\bbA-\bbB\,, \qquad g(n;\delta):={} \ln r_A(n) + (\bbA''-\bbB'')\delta^2\,,
\end{align}
where $D$ is the function of appendix~\ref{ap:criticaldensity} whose zeros are the critical densities.

\subsubsection{Choosing the scaling window}\label{ap:choosing-window}
We now set $n=\ncrit+\Lambda V^{-s}$ and Taylor expand each coefficient of $Q$ around $\ncrit$ at fixed $\delta$,
\begin{subequations}
\begin{align}
    Q(\delta)={}& V \sum_{j\geq 0}\frac{D^{(j)}(\ncrit)}{j!} (n-\ncrit)^j + 2\sqrt{V}\delta \sum_{j\geq 0}\frac{\beta^{(j+1)}(\ncrit)}{j!} (n-\ncrit)^j+\frac12 \sum_{j\geq 0}\frac{g^{(j)}(\ncrit;\delta)}{j!} (n-\ncrit)^j\,,\\
    ={}& \sum_{j\geq p}D_j \Lambda^j V^{1-js} + 2\delta \sum_{j\geq \delta_{\ncrit,n^*}}B_j \Lambda^j V^{\frac12 -js}+\frac12 \sum_{j\geq 0}g_j(\delta) \Lambda^j V^{-js}\,,\label{eq:Qtaylor}
\end{align}
\end{subequations}
where $g^{(j)}$ is the $j$-th derivative with respect to $n$, and we have defined the Taylor coefficients
\begin{align}
    D_j:=\frac{D^{(j)}(\ncrit)}{j!}\,,\qquad B_j:=\frac{\beta^{(j+1)}(\ncrit)}{j!}\,,\qquad g_j(\delta):=\frac{g^{(j)}(\ncrit;\delta)}{j!}\,,
\end{align}
and the order of the first nonvanishing coefficient of $D$,
\begin{align}\label{eq:pdefn}
    p:=\min\set{j:D_j\neq0}\,.
\end{align}
Since $D(\ncrit)=0$ we have $p\geq1$, and $p<\infty$ since $D$ is analytic and, for distinct species, does not vanish identically near $\ncrit$ by proposition~\ref{prop:continuouscriticality}. Applying proposition~\ref{prop:betaX} to the combined species $AB$, we have $B_1=\beta''(\ncrit)<0$, while $B_0=\beta'(\ncrit)$ vanishes if and only if $\ncrit=n^*$. Hence the second sum in~\eqref{eq:Qtaylor} starts at $j=\delta_{\ncrit,n^*}$ with a nonvanishing coefficient. The order $p$ is controlled by the concavity ratio: by lemma~\ref{lem:combinedsaddle} and proposition~\ref{prop:monotonen},
\begin{align}\label{eq:Dprime}
    D'(n)=\beta'(n)(\nAbar'(n)-\nBbar'(n))=\beta'(n)\,\frac{1-r_A(n)}{1+r_A(n)}\,.
\end{align}
Here and in the following, a subscript $c$ denotes evaluation at $n=\ncrit$. In particular, $\betaA^{(m)}=\beta_A^{(m)}\bm(\nAbar(\ncrit)\bm)$, $\nAbc'=\nAbar'(\ncrit)=1/(1+\rAc)$ and $\nBbc'=1-\nAbc'$, together with the concavity ratios $\rAc=r_A(\ncrit)=\betaA''/\betaB''$, $\rBc=1/\rAc$ and $\rsc=\min\set{\rAc,\rBc}$, and $\kappac=\kappa(\ncrit)$. Quantities without the subscript, such as $r_A$ or $\kappa$, are evaluated at the actual density $n=N/V$. In constant-order terms, replacing one by the other only changes the result at $o(1)$. At a generic critical density, $\beta'(\ncrit)\neq0$, so $p=1$ if and only if $\ncrit$ is non-degenerate, $\betaA''\neq\betaB''$. If it is degenerate, then $\rAc=1$ and $\nAbc'=\nBbc'=\frac12$, and differentiating~\eqref{eq:Dprime} gives
\begin{align}\label{eq:D2generic}
    D_2=-\frac{\beta'(\ncrit)}{4}\,r_A'(\ncrit)=\frac{\beta'(\ncrit)}{8\betaA''}\left(\betaB'''-\betaA'''\right)=\frac{\kappac^2\,\beta'(\ncrit)}{8\betaA''}\,\Xi_A(\ncrit)\,,
\end{align}
so $p=2$ if and only if $\Xi_A(\ncrit)\neq0$, and $p\geq3$ if and only if $\rAc=1$ and $\Xi_A(\ncrit)=0$. At a flat critical density, $D_1=0$ and
\begin{align}
    D_2=\frac12\beta''(n^*)\frac{1-\rAc}{1+\rAc}=\beta''(n^*)\left(\nAbc'-\frac12\right)\,,
\end{align}
so $p=2$ if and only if $n^*$ is non-degenerate, $\betaA''\neq\betaB''$. If $n^*$ is degenerate, which is characterized by~\eqref{eq:flatdoubledegenchar}, we have $p\geq3$.

The first two sums in~\eqref{eq:Qtaylor} are led by $D_p\Lambda^pV^{1-ps}$ and $2\delta B_j\Lambda^jV^{1/2-js}$ with $j=\delta_{\ncrit,n^*}$, while the last sum is $O(1)$. Requiring these two terms to balance with $\delta=\ord(1)$ gives
\begin{align}\label{eq:sK}
    s=s_K:=\frac{1}{2(p-\delta_{\ncrit,n^*})}\,.
\end{align}
If $s<s_K$, the first term dominates and any root of $Q$ escapes the Gaussian, $|\hat\delta_0|\to\infty$. If $s>s_K$, the dependence on $\Lambda$ drops out of the leading-order equation for the roots, which either collapse onto the mean, $\hat\delta_0\to0$, or, in the non-degenerate flat case, reduce to their values at $\Lambda=0$. The window $s=s_K$ is therefore the unique window in which the crossings are strictly of constant order and depend on $\Lambda$. Since $p\geq2$ whenever $\ncrit=n^*$, $s_K$ is always well defined.

The Page term contributes at constant order only if $Q=O(1)$ on a $\delta$-interval of constant order. This requires $\sqrt V\beta'(n)=O(1)$, hence $\ncrit=n^*$ and $s\geq\frac12$, and $VD(n)=O(1)$, which then follows from $p\geq2$. For $s>\frac12$ the dependence on $\Lambda$ drops out again, so the Page term is resolved at $s=s_P:=\frac12$ in the flat critical case. At a generic critical density no such window exists, since $|Q|$ grows like $\sqrt V|\beta'(n)||\delta-\hat\delta_0|$ away from the crossing and the Page term is $O(V^{-1/2})$. The resulting windows are summarized in table~\ref{tab:windows}.
\begin{table*}[t]
\centering
\renewcommand{\arraystretch}{1.3}
\caption{\textit{Resolving windows $n=\ncrit+\Lambda V^{-s}$.} The kinks of $\varphi$ are resolved at $s=s_K$ of~\eqref{eq:sK} and the Page term at $s=s_P$. Here $p$ is the order to which $D$ vanishes at $\ncrit$.}
\label{tab:windows}
\begin{tabular*}{0.8\textwidth}{@{\hspace{0.5em}\extracolsep{\fill}}llll@{\extracolsep{0pt}\hspace{0.5em}}}
\hline
\hline
Critical density & Condition & Kinks, $s_K$ & Page, $s_P$ \\
\hline
Non-degenerate generic critical & $\ncrit\neq n^*$, $\betaA''\neq\betaB''$ & $\frac12$ & absent \\
Degenerate generic critical & $\ncrit\neq n^*$, $\betaA''=\betaB''$ & $\frac{1}{2p}$ & absent \\
Non-degenerate flat critical & $\ncrit=n^*$, $\betaA''\neq\betaB''$ & $\frac12$ & $\frac12$ \\
Degenerate flat critical & $\ncrit=n^*$, $\betaA''=\betaB''$ & $\frac{1}{2(p-1)}$ & $\frac12$ \\
\hline
\hline
\end{tabular*}
\end{table*}
Of particular note, in the degenerate flat critical case in the fourth row of table~\ref{tab:windows} we have $p\geq 3$: the kink window is strictly wider than the Page profile window. However, in appendix~\ref{ap:calc-flat-bvgEE-res} we show the wider window resolves neither a Kronecker delta nor a branch transition; for odd $p$ it only resolves the switch between the branches selected at order $V$ in~\eqref{eq:bvgEE}, \ie the kink of the volume-law coefficient. For even $p$ the same branch is selected on both sides of $n^*$, and there is no such switch.

\subsubsection{Roots of \texorpdfstring{$Y$}{Y} in the generic critical case}\label{ap:kdeltares-pert-generic}
Here $s_K=1/(2p)$ and $B_0=\beta'(\ncrit)\neq0$. Both leading terms of~\eqref{eq:Qtaylor} are of order $\sqrt V$ while the last sum is of constant order, so the crossing near the saddle is
\begin{align}\label{eq:deltaCgdeltaCPg}
    \hat\delta_0 ={}& -\frac12 \frac{\sum_{j\geq p}D_j \Lambda^j V^{1-\frac{j}{2p}}}{\sum_{j\geq 0}B_j \Lambda^j V^{\frac12 -\frac{j}{2p}}} + O(V^{-1/2})
    = \underbrace{-\frac12 \frac{\sum_{j=p}^{p+\lfloor p/2\rfloor}D_j \Lambda^j V^{1-\frac{j}{2p}}}{\sum_{j= 0}^{\lfloor p/2\rfloor}B_j \Lambda^j V^{\frac12 -\frac{j}{2p}}}}_{=: \dzg^{(p)}(\Lambda)} + o(V^{-1/4})\,.
\end{align}
We write $\dzg:=\dzg^{(p)}(\Lambda)$ for the truncated root, where g stands for generic. The truncation keeps every term whose relative size is not $o(V^{-1/4})$, which is the accuracy required in appendix~\ref{ap:split-stability}. For $p=1$, \eqref{eq:Dprime} gives $D_1=\beta'(\ncrit)(2\nAbc'-1)$, and hence
\begin{align}\label{eq:deltac1}
    \dzg^{(1)}(\Lambda)=\Lambda\left(\frac12-\nAbc'\right)=\frac{\Lambda}{2}\,\frac{\rAc-1}{\rAc+1}\,.
\end{align}
Any second root of $Q$ lies at $|\delta|\gtrsim\sqrt V$, far outside the Gaussian, and only contributes exponentially small terms.

\subsubsection{Roots of \texorpdfstring{$Y$}{Y} in the flat critical case}\label{ap:kdeltares-pert-flat}
Here $B_0=0$ and $B_1=\beta''(n^*)$.

\paragraph{Non-degenerate case} For $\betaA''\neq\betaB''$ we have $p=2$ and $s_K=s_P=\frac12$, and~\eqref{eq:Qtaylor} reduces to the quadratic
\begin{align}\label{eq:Qflat}
    Q(\delta)={}& \beta''(n^*)\left(\nAbc'-\frac12\right) \Lambda^2 + 2\delta \beta''(n^*)\Lambda+\frac12\left[\ln \rAc+(\betaA''-\betaB'')\delta^2\right]  +O(V^{-1/2})\,,
\end{align}
whose roots are
\begin{align}\label{eq:deltapm}
    \delta={}&\frac{|\beta''(n^*)| \Lambda \pm \frac12\sqrt{\Delta_Q}}{\frac12(\betaA''-\betaB'')}\,,
\end{align}
and we let $\delta_-<\delta_+$ denote the ordered roots. Using $\betaA''\nAbc'=\betaB''\nBbc'=\beta''(n^*)$ (proposition~\ref{prop:monotonen}), the discriminant simplifies to
\begin{align}\label{eq:flatdiscriminant}
    \Delta_Q:=4\beta''(n^*)^2\Lambda^2-(\betaA''-\betaB'')\left[\ln \rAc+2\beta''(n^*)\left(\nAbc'-\frac12\right)\Lambda^2\right]=\betaA''\betaB''\Lambda^2-(\betaA''-\betaB'')\ln \rAc\,,
\end{align}
which is strictly positive for every $\Lambda$, since $\betaA''\betaB''>0$, while $\betaA''-\betaB''$ and $\ln \rAc$ have opposite signs. The roots are therefore real and simple, with $|Q'(\delta_\pm)|=\sqrt{\Delta_Q}+O(V^{-1/2})$, and the $O(V^{-1/2})$ corrections in~\eqref{eq:Qflat} shift them by $O(V^{-1/2})$.

\paragraph{Degenerate case} For $\betaA''=\betaB''$ we have $\rAc=1$, so $g_0\equiv0$, and $p\geq3$. In the Page window $s=s_P=\frac12$,~\eqref{eq:Qtaylor} gives
\begin{align}\label{eq:Qdegenpage}
    Q(\delta)={}& 2\beta''(n^*)\Lambda\delta + O(V^{-1/2})\,,
\end{align}
so the crossing sits at the mean, $\hat\delta_0=o(1)$, and only the Page term is resolved; it is evaluated in~\eqref{eq:PageResDegen}. In the outer window $s=s_K=1/[2(p-1)]\leq\frac14$ with $\Lambda\neq0$,
\begin{align}
    Q(\delta)={}& \sum_{j\geq p}D_j \Lambda^j V^{1-js_K} + 2\delta \sum_{j\geq 1}B_j \Lambda^j V^{\frac12 -js_K}+O(V^{-s_K})\,,
\end{align}
where the remainder starts at $j=1$ in the last sum of~\eqref{eq:Qtaylor} since $g_0\equiv0$. Both leading terms are of order $V^{1/2-s_K}\geq V^{1/4}$, so the crossing is
\begin{align}\label{eq:deltaCfdeltaCPf}
    \hat\delta_0 ={}& -\frac12 \frac{\sum_{j\geq p}D_j \Lambda^j V^{1-js_K}}{\sum_{j\geq 1}B_j \Lambda^j V^{\frac12 -js_K}} + O(V^{-1/2})
    = \underbrace{-\frac12 \frac{\sum_{j=p}^{p-1+\lfloor p/2\rfloor}D_j \Lambda^j V^{1-js_K}}{\sum_{j= 1}^{\lfloor p/2\rfloor}B_j \Lambda^j V^{\frac12 -js_K}}}_{=: \dzf^{(p)}(\Lambda)} + o(V^{-1/4+s_K/2})\,,
\end{align}
where we write $\dzf:=\dzf^{(p)}(\Lambda)$ for the truncated root, with f for flat, to distinguish it from $\dzg$ of the generic case. The accuracy $o(V^{-1/4+s_K/2})$ is again the one required in appendix~\ref{ap:split-stability}. As in the generic case, any second root of $Q$ lies at $|\delta|\gtrsim\sqrt V$, since the coefficient of $\delta^2$ is $\frac12(\bbA''-\bbB'')=O(V^{-s_K})$.

\subsubsection{Splitting at approximate roots}\label{ap:split-stability}
\begin{figure}[t]
\centering
\includegraphics{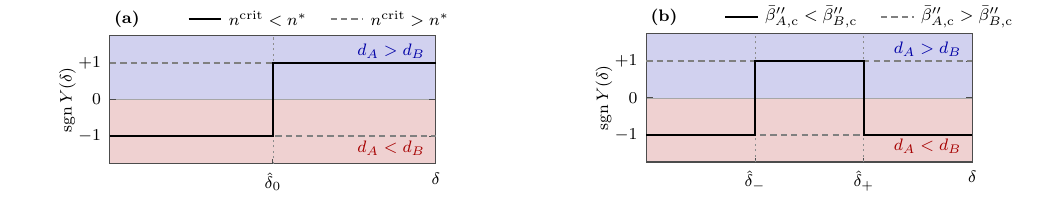}
\caption{\textit{Branch choice at criticality.} Sign of $Y(\delta)$ near its roots, which determines whether $d_A<d_B$ or $d_A>d_B$ and hence which branch of $\varphi$ is selected. \textbf{(a)}~A single root $\hat\delta_0$, as at a generic critical density and in the outer window of a degenerate flat critical density. \textbf{(b)}~Two roots $\hat\delta_\pm$, as at a non-degenerate flat critical density.}
\label{fig:branch-choice}%
\panellabel{fig:generic-branch-choice}%
\panellabel{fig:flat-branch-choice}%
\end{figure}
The branch selected on either side of the crossings follows from~\eqref{eq:dimensionsANDy} and is summarized in figure~\ref{fig:branch-choice}. We write
\begin{align}
    \varphi_A:=\ln \frac{d_{AB}}{d_A}\,,\qquad
    \varphi_B:=\ln \frac{d_{AB}}{d_B}\,,\qquad \varphi_B-\varphi_A=\ln\frac{d_A}{d_B}=VY\,,
\end{align}
and $\sigma:=\sgn\beta'(n)=\sgn(n^*-n)$, by proposition~\ref{prop:betaX} applied to the combined species, with $\sigma=+1$ for all $n$ if $\nmax=\infty$. We now show that splitting the integrals in~\eqref{eq:intsplit3ways} at the truncated roots of appendices~\ref{ap:kdeltares-pert-generic} and~\ref{ap:kdeltares-pert-flat}, instead of at the exact crossings, only changes $\braket{S_A}_N$ at $o(1)$.

\paragraph{Single crossing} In the generic critical window and in the outer window of the degenerate flat case, $Y$ has a single transverse crossing $\hat\delta_0$ within the Gaussian, where $\sigma$ is the sign of its slope. Near $\hat\delta_0$ we have $V|Y|\simeq2|\sqrt V\beta'(n)||\delta-\hat\delta_0|$, so the Page term is $O\bm(1/(\sqrt V\beta'(n))\bm)$. This is $O(V^{-1/2})$ in the generic case and $O(V^{-1/2+s_K})$ in the degenerate case, and hence $o(1)$ in both. By~\eqref{eq:dimensionsANDy}, the entanglement entropy~\eqref{eq:intsplit3ways} becomes
\begin{align}
    \braket{S_A}_N=F_\sigma(\hat\delta_0)+o(1)\,,\qquad
    F_{\sigma}(t):={}&
    \begin{cases}
        \int^t_{-\infty}\varrho\,\varphi_B\,\dd \delta +\int_t^{\infty}\varrho\,\varphi_A\,\dd \delta\,, & \sigma=+1\,, \\
        \int^t_{-\infty}\varrho\,\varphi_A\,\dd \delta +\int_t^{\infty}\varrho\,\varphi_B\,\dd \delta\,, & \sigma=-1\,,
    \end{cases}\label{eq:generic-split-integral-res}
\end{align}
for a split point $t$. Since the logarithmic part of $\varphi$ is $\min\set{\varphi_A,\varphi_B}$, we have $F_\sigma(t)\geq F_\sigma(\hat\delta_0)$ for all $t$ with equality at $t=\hat\delta_0$. Differentiating,
\begin{align}
    F_\sigma'(t)=\sigma\,\varrho(t)\,VY(t)\,,\qquad
    F_\sigma''(\hat\delta_0)=\sigma\,\varrho(\hat\delta_0)\,VY'(\hat\delta_0)=2\varrho(\hat\delta_0)\,|\sqrt V\beta'(n)|(1+o(1))\,,
\end{align}
where we used $VY'(\delta)=2\sqrt V\beta'(n)+(\bbA''-\bbB'')\delta+O(V^{-1/2})$ from~\eqref{eq:VYexp}. In the degenerate case $|\sqrt V\beta'(n)|\geq cV^{1/4}$ while $\bbA''-\bbB''=O(V^{-s_K})$. Hence $\hat\delta_0$ is a nondegenerate minimum and
\begin{align}
    F_\sigma(\hat\delta_0+\eta)-F_\sigma(\hat\delta_0)=\varrho(\hat\delta_0)\,|\sqrt V\beta'(n)|\,\eta^2(1+o(1))+O\bm(|\sqrt V\beta'(n)|\,\eta^3\bm)\,.\label{eq:Fstationary}
\end{align}
The first-order term vanishes because the two branches coincide at the crossing. Any split point within $\eta=o\bm((\sqrt V\beta'(n))^{-1/2}\bm)$ of $\hat\delta_0$ therefore reproduces $\braket{S_A}_N$ to $o(1)$. Since $|\sqrt V\beta'(n)|=|B_j\Lambda^j|V^{1/2-js_K}(1+o(1))$ with $j=\delta_{\ncrit,n^*}$, this is $\eta=o(V^{-1/4})$ in the generic case and $\eta=o(V^{-1/4+s_K/2})$ in the degenerate case. These are exactly the accuracies of the truncated roots $\dzg$ and $\dzf$ in~\eqref{eq:deltaCgdeltaCPg} and~\eqref{eq:deltaCfdeltaCPf}, respectively, so
\begin{align}
    \braket{S_A}_N=\begin{cases}
        F_\sigma(\dzg)+o(1)\,, & \ncrit\neq n^*\,,\\
        F_\sigma(\dzf)+o(1)\,, & \ncrit=n^*,\ \betaA''=\betaB''\,,
    \end{cases}\label{eq:SAasF}
\end{align}
with $\sigma=\sgn(n^*-\ncrit)$ in the generic case and $\sigma=\sgn\bm(n^*-(n^*+\Lambda V^{-s})\bm)=-\sgn\Lambda$ in the degenerate flat case.

\paragraph{Two crossings} In the non-degenerate flat window $s=\frac12$, $VY$ has two crossings $\hat\delta_\pm$ within the Gaussian, which differ by $O(V^{-1/2})$ from the simple roots $\delta_\pm$ of~\eqref{eq:deltapm}. The logarithmic part of~\eqref{eq:intsplit3ways} is $G(\hat\delta_-,\hat\delta_+)$, where for split points $t_-<t_+$
\begin{align}
    G(t_-,t_+):=\int_{-\infty}^{t_-}\varrho\,\varphi_{\rm out}\,\dd\delta+\int_{t_-}^{t_+}\varrho\,\varphi_{\rm in}\,\dd\delta+\int_{t_+}^{\infty}\varrho\,\varphi_{\rm out}\,\dd\delta\,,
\end{align}
with $(\varphi_{\rm out},\varphi_{\rm in})=(\varphi_B,\varphi_A)$ for $\betaA''<\betaB''$ and $(\varphi_A,\varphi_B)$ for $\betaA''>\betaB''$ (figure~\ref{fig:flat-branch-choice}). The first derivatives $\partial_{t_\mp}G=\pm\varrho(t_\mp)(\varphi_{\rm out}-\varphi_{\rm in})(t_\mp)$ vanish at the crossings, the mixed derivative vanishes identically, and $\partial^2_{t_\pm}G=\varrho(\hat\delta_\pm)|VY'(\hat\delta_\pm)|=\varrho(\hat\delta_\pm)\sqrt{\Delta_Q}+O(V^{-1/2})$ at the crossings. Hence
\begin{align}\label{eq:Gstationary}
    G(\hat\delta_-+\eta_-,\hat\delta_++\eta_+)-G(\hat\delta_-,\hat\delta_+)=\frac{\sqrt{\Delta_Q}}{2}\left(\varrho(\hat\delta_-)\eta_-^2+\varrho(\hat\delta_+)\eta_+^2\right)(1+O(V^{-1/2}))+O(|\eta|^3)\,,
\end{align}
and, since $\Delta_Q=O(1)$, any errors $\eta_\pm=o(1)$ in the split points change $\braket{S_A}_N$ only at $o(1)$. In particular, the leading-order roots~\eqref{eq:deltapm} suffice.

\subsection{Calculating \texorpdfstring{$\braket{S_A}_N$}{the mean entropy} in the resolved regime}\label{ap:calc-avgEE-res}

We now resolve the Kronecker deltas and branch transitions in $\braket{S_A}_N$.
As in~\cite{Yauk:2023wbu}, the indicator functions in the average entanglement entropy~\eqref{eq:bvgEE} arise from a nonuniform large-$V$ limit. At finite $V$, the transition between the two branches chosen for the integrals in~\eqref{eq:intsplit3ways} is smooth in $n$, and the indicator functions are only recovered after taking $V\gg1$ at fixed $n$. Here we instead let $n=\ncrit +\Lambda V^{-s}$ with the exponents $s$ of table~\ref{tab:windows} and take $V\gg 1$ at fixed $\Lambda$, which resolves the indicator function contributions. Near the saddle, the probability distribution $\varrho(\delta)$ takes the form~\eqref{eq:varrhodelta}, while the non-analytic function $\varphi(\delta)$ takes the form~\eqref{eq:varphiNAsplitdelta}. We evaluate the integrals~\eqref{eq:intsplit3ways} first in the generic critical regime, $\ncrit\neq n^*$, and then in the flat critical regime, $\ncrit=n^*$. A system may have both generic critical densities and a flat critical density (see section~\ref{sec:testbedmodelA}), in which case one must choose which density to resolve around. The inherited terms of~\eqref{eq:inh-res} at the non-degenerate critical densities, obtained below, are listed in table~\ref{tab:inherited}.

\begin{table*}[t]
\centering
\renewcommand{\arraystretch}{1.3}
\setlength{\extrarowheight}{3pt}
\caption{\textit{Inherited terms at the non-degenerate critical densities.} The inherited terms of~\eqref{eq:inh-res} in the window $n=\ncrit+\Lambda/\sqrt V$, obtained from~\eqref{eq:gencritres1} and~\eqref{eq:flatResAVGlog}, with the conventions of table~\ref{tab:resolved-kdeltas}. Here $b_{\rm K}$ is defined in~\eqref{eq:bK}, $w_0$ and $w_1$ in~\eqref{eq:XiArAetc}, and $w_0^*$ and $w_1^*$ in~\eqref{eq:flatweights}. In the degenerate case, the table changes and new intermediate power terms like $V^{3/4}$ arise.}
\label{tab:inherited}
\begin{tabular}{@{\hspace{0.5em}} l @{\hspace{2em}} c @{\hspace{2em}} c @{\hspace{0.5em}}}
\hline
\hline
Term & Non-degenerate generic critical & Non-degenerate flat critical \\
\hline
$B_{\rm inh}^V$ & $\Lambda\big[(1-w_0)\,a_A'(\ncrit)+w_0\,a_B'(\ncrit)\big]$ & $0$ \\
$C_{\rm inh}^V$ & $\frac12\Lambda^2\big[(1-w_0)\,a_A''(\ncrit)+w_0\,a_B''(\ncrit)\big]$ & $\frac12\Lambda^2\big[(1-w_0^*)\,a_A''(n^*)+w_0^*\,a_B''(n^*)\big]$ \\
$C_{\rm inh}^{\sqrt V}$ & $\Lambda\,b_{\rm K}'(\ncrit)\,w_1$ & $2|\beta''(n^*)|\,\Lambda\,w_1^*$ \\[2pt]
\hline
\hline
\end{tabular}
\end{table*}

\subsubsection{Calculating \texorpdfstring{$\braket{S_A}_N$}{the mean entropy} in the generic critical regime}\label{ap:resolving-generic-kdelta}
Here $\ncrit\neq n^*$, which is only a restriction if $\zeta_A(1)=\zeta_B(1)<\infty$, since otherwise $n^*$ is not critical. Then $\beta'(\ncrit)\neq0$, so the crossing is transverse in the Gaussian coordinate $\delta$ and the Page correction $-\frac12\int\varrho\,\ee^{-V|Y|}\dd\delta$ is $O(V^{-1/2})$ (appendix~\ref{ap:split-stability}). We therefore only need the logarithmic integrands, and evaluate~\eqref{eq:SAasF} in the window $n=\ncrit+\Lambda V^{-s_K}$ with $s_K=1/(2p)$.
The two integrands are expanded as follows. Here the last term in each bracket is the second-order Edgeworth correction $\mathcal{E}_2/V$ of~\eqref{eq:varrhodelta} multiplied by the leading-order term of the logarithm, which becomes $o(1)$ after integration since $\sum_{N_A}\varrho_{N_A}=1$ is exact~\eqref{eq:edgeworth-mean-zero} and $\bbA=\bbB+o(1)$ at this scale:
\begin{align}
    \varrho\varphi_B
    ={}&\varrho_{\mathrm G}(\delta)
    \left[
    \frac12\ln\frac{\beta''(n)}{\bbB''}
    +V\bbA
    +\sqrt V\,\bbB'\delta
    -\frac12\bbB''\delta^2
    \nonumber \right.\\
    &+
    \frac{\sqrt V\,\delta}{2}
    \left(
        \frac{\bbA'''}{\bbA''}
        -
        \frac{\bbB'''}{\bbB''}
    \right)
    \bbA
    +
    \frac{\sqrt V\,\delta^3}{6}
    \left(
        \bbA'''-\bbB'''
    \right)
    \bbA
    \nonumber\\
    &+
    \frac{\delta^2}{2}
    \left(
        \frac{\bbA'''}{\bbA''}
        -
        \frac{\bbB'''}{\bbB''}
    \right)
    \bbB'
    \left.+
    \frac{\delta^4}{6}
    \left(
        \bbA'''-\bbB'''
    \right)
    \bbB'
    +O(V^{-1/2})+\mathcal{E}_2(\delta)\,\bbA\right]\,,
    \label{eq:branch-logdABdB}
\end{align}
and
\begin{align}\label{eq:branch-logdABdA}
    \varrho\varphi_A
    ={}&\varrho_{\mathrm G}(\delta)
    \left[
    \frac12\ln\frac{\beta''(n)}{\bbA''}
    +V\bbB
    -\sqrt V\,\bbA'\delta
    -\frac12\bbA''\delta^2
    \nonumber \right.\\
    &+
    \frac{\sqrt V\,\delta}{2}
    \left(
        \frac{\bbA'''}{\bbA''}
        -
        \frac{\bbB'''}{\bbB''}
    \right)
    \bbB
    +
    \frac{\sqrt V\,\delta^3}{6}
    \left(
        \bbA'''-\bbB'''
    \right)
    \bbB
    \nonumber\\
    &-
    \frac{\delta^2}{2}
    \left(
        \frac{\bbA'''}{\bbA''}
        -
        \frac{\bbB'''}{\bbB''}
    \right)
    \bbA'
    \left.-
    \frac{\delta^4}{6}
    \left(
        \bbA'''-\bbB'''
    \right)
    \bbA'
    +O(V^{-1/2})+\mathcal{E}_2(\delta)\,\bbB\right]\,.
\end{align}
Evaluating the Gaussian moments on the two half-lines gives, for $\sigma=+1$, \ie $\ncrit<n^*$, which includes every case with $\nmax=\infty$,
\begin{align}\label{eq:gencritres1}
    \braket{S_A}_N^{\mathrm {g,res}}={}&V[\bbA+(\bbB-\bbA)w_0]-\sqrt{\frac{2V}{\pi\kappa}}|\beta'(n)|\,w_1\nonumber \\
    &+(1-w_0)\,c_A(n)+w_0\,c_B(n)\nonumber \\
    &-\left[\frac{1-r_A}{2(1+r_A)}+\frac{\beta'(n)}{3\kappa^2}\left(\bbA'''-\bbB'''\right)\right]\sqrt{\frac{\kappa}{2\pi}}\,\dzg\,w_1+o(1)\,,
\end{align}
where we have defined the weights
\begin{subequations}\label{eq:XiArAetc}
\begin{align}
    w_0={}&\frac12\erfc(\dzg\sqrt{\kappa/2})\,,\\
    w_1={}&\ee^{-\dzg^2\kappa/2}\,,
\end{align}
\end{subequations}
and $\dzg=\dzg^{(p)}(\Lambda)$ is given by~\eqref{eq:deltaCgdeltaCPg}, and by~\eqref{eq:deltac1} for $p=1$. For $\sigma=-1$, \ie $\ncrit>n^*$, the result follows from~\eqref{eq:gencritres1} by exchanging $A\leftrightarrow B$ in every quantity except $\dzg$, which keeps its value~\eqref{eq:deltaCgdeltaCPg}, so that the weights $w_0$ and $w_1$ are unchanged. At $\Lambda=0$ we have $\dzg=0$ and $w_0=\frac12$, so the last term vanishes and we recover the critical value of~\eqref{eq:bvgEE} at $n=\ncrit$.

At a non-degenerate generic critical density, where $p=1$, expanding the first line of~\eqref{eq:gencritres1} at $n=\ncrit+\Lambda/\sqrt V$ gives $a(\ncrit)$, the inherited terms $B_{\rm inh}^V$, $C_{\rm inh}^V$ and $C_{\rm inh}^{\sqrt V}$ of table~\ref{tab:inherited}, and the resolved profile $B$ of table~\ref{tab:resolved-kdeltas}. The corrections from evaluating the arguments of $w_0$ and of the Gaussian at $n$ rather than at $\ncrit$ cancel between its two terms at constant order. The remaining terms of~\eqref{eq:gencritres1}, evaluated at $\ncrit$ and with leading order $\delta_{0,\rm g}$, give the resolved profile $C$ of table~\ref{tab:resolved-kdeltas}.

With $A$ selected for $n<\ncrit$ as in proposition~\ref{prop:jumps-kdeltas}(1), the weight $w_0$ is the probability that the Gaussian variable $\delta$ lies beyond the kink of $\varphi$, where the branch selected for $n>\ncrit$ applies. The profile $B=b_{\rm K}(\ncrit)\,w_1$ (with $\delta_{0,\rm g}$ evaluated at leading order) is a Gaussian in $\Lambda$ with height
\begin{align}\label{eq:bK}
    b_{\rm K}(n):=-\sqrt{\frac{2}{\pi\kappa(n)}}\,|\beta'(n)|
\end{align}
at $\Lambda=0$, so it resolves the Kronecker delta of order $\sqrt V$. The profile $C$ interpolates between $c_A(\ncrit)$ and $c_B(\ncrit)$ and equals their mean at $\Lambda=0$, so it resolves the branch transition. Its last term, with
\begin{align}\label{eq:tau}
    \tau:={}&\frac{1-\rAc}{2(1+\rAc)}+\frac{\beta'(\ncrit)}{3\kappac^2}\left(\betaA'''-\betaB'''\right)\,,
\end{align}
the bracket of~\eqref{eq:gencritres1} at $\ncrit$, vanishes at $\Lambda=0$ and as $\Lambda\to\pm\infty$, so it changes neither $J[c]$ nor $K[c]$; it arises from the quadratic terms of~\eqref{eq:varphiNAsplitdelta} and from the first Edgeworth correction in~\eqref{eq:varrhodelta}. For $p\geq3$, $\rAc=1$ and $\betaA'''=\betaB'''$ by~\eqref{eq:D2generic}, so $\tau=0$ and $c_A(\ncrit)=c_B(\ncrit)$: the resolved profile $C(\Lambda;\ncrit)=\frac12(\frac12-\ln2)$ is constant, in line with proposition~\ref{prop:jumps-kdeltas}(3), and only $B$ is non-trivial.

\subsubsection{Calculating \texorpdfstring{$\braket{S_A}_N$}{the mean entropy} in the flat critical regime}\label{ap:calc-flat-bvgEE-res}
We now resolve the flat critical density $\ncrit=n^*$, which exists only if $\zeta_A(1)=\zeta_B(1)<\infty$. We first consider the Page correction term, which is not suppressed in the window $n=n^*+\Lambda V^{-1/2}$. We then calculate the logarithmic terms in the window $n=n^*+\Lambda V^{-1/2}$ for the non-degenerate case, and in both windows $s_P=\frac12$ and $s_K=1/[2(p-1)]$ for the degenerate case.

\paragraph{Page correction term}
We want to resolve the contribution
\begin{subequations}
\begin{align}
    P={}&-\frac12
    \int_{-\infty}^{\infty}
        \varrho(\delta)
        \min\Set{
            \frac{d_A-1}{d_B},
            \frac{d_B-1}{d_A}
        }
        \dd{\delta}\,, \\
        ={}&-\frac12 \int_{-\infty}^\infty\varrho(\delta)\ee^{-V|Y|}\dd\delta+o(1)\,,\\
        ={}&-\frac12 \int_{-\infty}^\infty\varrho(\delta)\ee^{-|Q(\delta)|}\dd\delta+o(1)\,,
\end{align}
\end{subequations}
where $Q$ is defined in~\eqref{eq:Qdef}. For $\betaA''\neq\betaB''$, $Q$ is the quadratic~\eqref{eq:Qflat}, whose sign changes at its ordered roots $\delta_\pm$ of~\eqref{eq:deltapm}. Following the branch choice in figure~\ref{fig:flat-branch-choice}, we split the integration into three regions,
\begin{align}
    P
    ={}&
    \begin{cases}
        -\frac12
    \left(
    \int_{-\infty}^{\delta_-}
        \varrho(\delta)\,
        \ee^{Q(\delta) }
        \dd{\delta}
    +
    \int_{\delta_-}^{\delta_+}
        \varrho(\delta)\,
        \ee^{-Q(\delta)}
        \dd{\delta}
    +
    \int_{\delta_+}^{\infty}
        \varrho(\delta)\,
        \ee^{Q(\delta)}
        \dd{\delta}
    \right)
    +o(1)\,, &\betaA''<\betaB''\,,\\
    -\frac12
    \left(
    \int_{-\infty}^{\delta_-}
        \varrho(\delta)\,
        \ee^{-Q(\delta) }
        \dd{\delta}
    +
    \int_{\delta_-}^{\delta_+}
        \varrho(\delta)\,
        \ee^{Q(\delta)}
        \dd{\delta}
    +
    \int_{\delta_+}^{\infty}
        \varrho(\delta)\,
        \ee^{-Q(\delta)}
        \dd{\delta}
    \right)
    +o(1)\,, & \betaA''>\betaB''\,.
    \end{cases}\label{eq:flat-page-split-nondegen}
\end{align}
Completing the square in each region, using $\betaA''\nAbc'=\betaB''\nBbc'=\beta''(n^*)$, we obtain for $\betaA''>\betaB''$ (otherwise exchange $A\leftrightarrow B$)
\begin{align}\label{eq:near-flat-page-correction}
    P(\Lambda)={}&-\frac14\ee^{|\beta''(n^*)|\Lambda^2/2}\Big\{\sqrt{\tfrac{1+\rAc}{2}}\Big[\erf\bm(\sqrt{-\betaA''}(\delta_++\nAbc'\Lambda)\bm)-\erf\bm(\sqrt{-\betaA''}(\delta_-+\nAbc'\Lambda)\bm)\Big]\nonumber\\
    &\hspace{17mm}+\sqrt{\tfrac{1+\rBc}{2}}\Big[2-\erf\bm(\sqrt{-\betaB''}(\delta_+-\nBbc'\Lambda)\bm)+\erf\bm(\sqrt{-\betaB''}(\delta_--\nBbc'\Lambda)\bm)\Big]\Big\}+o(1)\,.
\end{align}
At criticality, $\Lambda=0$, the roots are $\delta_\pm=\pm\sqrt{\ln(1/\rsc)/|\betaA''-\betaB''|}$, and~\eqref{eq:near-flat-page-correction} simplifies to
\begin{align}\label{eq:pageunresolved}
    P_{\rm flat}=-\frac12\erfc\!\left(\sqrt{\frac{\ln(\frac1{\rsc})}{1-{\rsc}}} \right)\sqrt{\frac{1+\frac1{\rsc}}{2}} -\frac12\erf\!\left(\sqrt{\frac{\ln(\frac1{\rsc})}{\frac1{\rsc}-1}} \right)\sqrt{\frac{1+\rsc}{2}}\,.
\end{align}

For $\betaA''=\betaB''$, the Page term is resolved in the window $n=n^*+\Lambda V^{-1/2}$, where $Q$ is given by~\eqref{eq:Qdegenpage}, so that
\begin{align}\label{eq:PageResDegen}
    P(\Lambda) ={}& -\frac12 \int_{\R} \varrho(\delta) \ee^{-2|\beta''(n^*)\Lambda \delta|}\dd \delta+o(1)
    =-\frac12 \operatorname{erfcx}\bm( \sqrt{2/\kappa }\, |\beta''(n^*)\Lambda| \bm)+o(1)\,,
\end{align}
where $\operatorname{erfcx}(x)=\ee^{x^2}\erfc(x)$ is the scaled complementary error function. At criticality we find $P(0)=-\frac12$. In the outer window $s_K$ the Page term is $o(1)$ (appendix~\ref{ap:split-stability}).

\paragraph{Logarithmic term}
We first consider the non-degenerate case $\betaA''\neq \betaB''$, for which the branch choice of figure~\ref{fig:flat-branch-choice} gives
\begin{align}
    \braket{S_A}_N^{\rm {f,res}}-P ={}&
    \begin{cases}
        \int_{-\infty}^{\delta_-}\varrho\,\varphi_B\,\dd \delta + \int_{\delta_-}^{\delta_+}\varrho\,\varphi_A\,\dd \delta+\int^{\infty}_{\delta_+}\varrho\,\varphi_B\,\dd \delta+o(1)\,, &\betaA''<\betaB''\,, \\
   \int_{-\infty}^{\delta_-}\varrho\,\varphi_A\,\dd \delta + \int_{\delta_-}^{\delta_+}\varrho\,\varphi_B\,\dd \delta+\int^{\infty}_{\delta_+}\varrho\,\varphi_A\,\dd \delta+o(1)\,, & \betaA''>\betaB''\,,
    \end{cases}
\end{align}
where by~\eqref{eq:Gstationary} the leading-order roots~\eqref{eq:deltapm} suffice. Inserting~\eqref{eq:branch-logdABdA} and~\eqref{eq:branch-logdABdB}, in which the second-order Edgeworth corrections $\mathcal{E}_2/V$ again integrate to $o(1)$, and using the moments of the Gaussian, we find up to an $o(1)$ remainder
\begin{align}\label{eq:flatResAVGlog}
    \braket{S_A}_N^{\mathrm {f,res}}-P={}&
    \begin{cases}
   V\bbB-\Lambda |\beta''(n^*)|[2w^*_1+\Lambda(\nAbc'-\frac12)w^*_0]\,, &  \betaA''<\betaB''\,,\\
       V\bbA+\Lambda |\beta''(n^*)|[2w^*_1+\Lambda(\nAbc'-\frac12)w^*_0]\,, &  \betaA''>\betaB''\,,
    \end{cases}\nonumber\\
    &+\frac12\left(\frac{1+(\rsc-1)w^*_2}{1+\rsc}-\ln(1+1/\rsc)-w^*_0\ln \rsc\right)\,,
\end{align}
where we have defined the weights
\begin{subequations}\label{eq:flatweights}
\begin{align}
    w^*_0={}&\frac12\erfc(-\delta_-\sqrt{\kappa/2})+\frac12\erfc(\delta_+\sqrt{\kappa/2})\,,\\
    w_1^*={}&\frac{\ee^{-\delta_+^2\kappa/2}-\ee^{-\delta_-^2\kappa/2}}{\sqrt{2\pi\kappa}}\,,\\
    w_2^*={}&w_0^*+\sqrt\frac{\kappa}{2\pi}(\delta_+ \ee^{-\delta_+^2\kappa/2}-\delta_- \ee^{-\delta_-^2\kappa/2})\,.
\end{align}
\end{subequations}
For $\betaA''>\betaB''$, $A$ is selected on both sides of $n^*$, $\varphi$ has two kinks within the Gaussian, at the roots $\delta_-<\delta_+$ of~\eqref{eq:Qflat}, and $B$ is selected outside them. The weights $w_0^*$, $w_1^*$ and $w_2^*$ are the probability of this outer region and, up to normalization, the first and second moments of $\delta$ over it. Expanding $V\bbA$ around $n^*$, the first line of~\eqref{eq:flatResAVGlog} gives $a(n^*)$ and the inherited terms $C_{\rm inh}^V$ and $C_{\rm inh}^{\sqrt V}$ of table~\ref{tab:inherited}, while its second line together with $P(\Lambda)$ gives the resolved profile $C$ of table~\ref{tab:resolved-kdeltas}. Taking $\Lambda=0$ we find the constant-order term for the flat critical case,
\begin{align}\label{eq:Theta}
        \Theta(\rsc) ={}&\frac12\erf\! \left(\sqrt{\frac{1+{\rsc}}{1-{\rsc}}\,\frac{\ln(\frac1{\rsc})}{2}} \right)
    \left(\frac{1}{1+{\rsc}}-\ln(1+\frac1{\rsc})\right)\nonumber \\
    &+ \frac12\erfc\! \left(\sqrt{\frac{1+{\rsc}}{1-{\rsc}}\,\frac{\ln(\frac1{\rsc})}{2}} \right)\left(\frac{1}{1+\frac1{\rsc}}-\ln(1+{\rsc})\right)\nonumber \\
    &-\exp\!\left(-\frac{1+{\rsc}}{1-{\rsc}}\frac{\ln(\frac1{\rsc})}{2}\right)\sqrt{\frac{1-{\rsc}}{1+{\rsc}}\,\frac{\ln(\frac1{\rsc})}{2\pi}} +P_{\rm flat}\,.
\end{align}
Hence $C(0;n^*)=\Theta(\rsc)$: the resolved Page term $P$ resolves $P_{\rm flat}$, and the remaining terms of $C$ resolve the logarithmic contribution to the Kronecker delta~\eqref{eq:Kflat}. Although $b$ vanishes near $n^*$, the term of order $\sqrt V$ survives at constant order as $C_{\rm inh}^{\sqrt V}$. Note that $\Theta$ is not defined for $\rsc=1$, \ie in the degenerate case $\betaA''=\betaB''$, which we treat now.

\paragraph{Degenerate flat critical case}
First consider the window $n=n^*+\Lambda V^{-1/2}$, in which the Page term is resolved. Since $p\geq3$, we have $VD(n)=O(V^{-1/2})$, so $V\bbA=V\bbB+o(1)$, as well as $\sqrt V\beta'(n)=\beta''(n^*)\Lambda+O(V^{-1/2})$ and $r_A(n)=1+O(V^{-1/2})$. By~\eqref{eq:Qdegenpage}, $d_A<d_B$ if and only if $\Lambda\delta>0$. Inserting~\eqref{eq:varphiNAsplitdelta} into~\eqref{eq:intsplit3ways} and using~\eqref{eq:PageResDegen}, we find
\begin{subequations}\label{eq:flatdegencritres}
\begin{align}
    \braket{S_A}_N^{\mathrm {f,degen,res},s_P}={}&V\bbB-\frac12\left(\ln 2-\frac12+ \operatorname{erfcx}\bm( \sqrt{2/\kappa }\, |\beta''(n^*)\Lambda| \bm)\right)-2\Lambda|\beta''(n^*)|\int_{\set{d_B>d_A}} \varrho(\delta)\delta\,\dd \delta+o(1)\,,\label{eq:flatdegencritres1}\\
    ={}&V\bbA-\frac12\left(\ln 2-\frac12+ \operatorname{erfcx}\bm( \sqrt{2/\kappa }\, |\beta''(n^*)\Lambda| \bm)\right)-2\Lambda|\beta''(n^*)|\int_{\set{d_B>d_A}} \varrho(\delta)\delta\,\dd \delta+o(1)\,,\label{eq:flatdegencritres2}
\end{align}
\end{subequations}
where we see manifest invariance under subsystem exchange. The term proportional to $\Lambda$ arises from the Taylor expansion of a term of higher order in $V$ and is not required to resolve any Kronecker delta: the only Kronecker delta arising in this case is from the Page correction term. In the notation of~\eqref{eq:inh-res}, this term is the inherited term $C_{\rm inh}^{\sqrt V}=-\sqrt{2/(\pi\kappac)}\,|\beta''(n^*)\Lambda|$, expanding $V\bbB$ gives $C_{\rm inh}^V=\frac14\beta''(n^*)\Lambda^2$, and the remaining terms give the resolved profile, the flat-critical entry of table~\ref{tab:resolved-kdeltas} at $\rsc=1$. At criticality we find
\begin{align} \label{eq:Further-degenerateCrit}
    \braket{S_A}_N^{\mathrm{f,degen}}=V\bbA-\frac12\left(\ln 2+\frac12\right)+o(1)\,.
\end{align}
This is the limit $\rsc\to1$ of the non-degenerate result $V\bbA+\Theta(\rsc)$, since the argument of $\erf$ and $\erfc$ in~\eqref{eq:Theta} tends to $1$, the last term vanishes, and $P_{\rm flat}\to-\frac12$.

Now consider the outer window $n=n^*+\Lambda V^{-s_K}$ with $s_K=1/[2(p-1)]$, $p\geq3$ and $\Lambda\neq0$. There is neither a Kronecker delta nor a branch transition at constant order to resolve here: the Page term is $o(1)$, and the constant-order indicator terms of~\eqref{eq:bvgEE} vanish since $r_A(n)=1+o(1)$. We evaluate~\eqref{eq:SAasF} with $\sigma=-\sgn\Lambda$ as in the generic case, so that the result is~\eqref{eq:gencritres1} with $\dzg\to\dzf$. The terms involving $\Xi_A$ and the third derivatives are now $o(1)$, since $\beta'(n)=O(V^{-s_K})$ and $\sqrt VD(n)=O(V^{-s_K})$, and the remaining constant terms reduce to $\frac12(\frac12-\ln2)$, since $r_A(n)=1+o(1)$. For $\Lambda<0$, \ie $\sigma=+1$, we find
\begin{align}
    \braket{S_A}_N^{\mathrm{f,degen,res},s_K} ={}& V\left[\bbA+(\bbB-\bbA)w_0\right]-\sqrt{\frac{2V}{\pi\kappa}}\,|\beta'(n)|\,w_1+\frac12\left(\frac12-\ln 2\right)+o(1)\,,\label{eq:flatdegenouter}
\end{align} %
where $w_0=\frac12\erfc(\dzf\sqrt{\kappa/2})$ and $w_1=\ee^{-\kappa\dzf^2/2}$ are the weights~\eqref{eq:XiArAetc} with $\dzg\to\dzf$, and $\dzf=\dzf^{(p)}(\Lambda)$ is given by~\eqref{eq:deltaCfdeltaCPf}. For $\Lambda>0$, \ie $\sigma=-1$, we exchange $A\leftrightarrow B$ in every quantity except $\dzf$, $w_0$ and $w_1$, as in the generic case. In this regime we resolve neither a Kronecker delta nor a branch transition, but, for odd $p$, the switch between the branches (species $A$ or species $B$) selected at order $V$ by the entanglement entropy~\eqref{eq:bvgEE}, \ie the kink of the volume-law coefficient. For even $p$, $D$ does not change sign, so the same branch is selected for $\Lambda\to\pm\infty$ and the volume-law coefficient has no kink at $n^*$. We cannot take $\Lambda=0$ here, so this window does not rediscover the behavior at criticality in~\eqref{eq:Further-degenerateCrit}.

\bibliography{references}
\typeout{get arXiv to do 4 passes: Label(s) may have changed. Rerun}
\end{document}